\documentclass[twocolumn,american,floatfix, table, dvipsnames,superscriptaddress]{revtex4-1}
\usepackage[T1]{fontenc}
\usepackage[latin9]{inputenc}
\usepackage{color}
\usepackage{babel}
\usepackage{booktabs}
\usepackage{mathrsfs}
\usepackage{amsmath}
\usepackage{amssymb}
\usepackage{graphicx}
\usepackage{esint}
\usepackage[unicode=true,
 bookmarks=false,
 breaklinks=true,pdfborder={0 0 1},backref=false,colorlinks=true]
 {hyperref}
\hypersetup{
 pdfcreator={},pdfproducer={LaTeX with hyperref},linkcolor=blue,anchorcolor=blue,citecolor=blue,filecolor=red,menucolor=red,pagecolor=red,urlcolor=blue,pdfstartview=FitV,pdfhighlight=/I,pdfpagelayout=TwoColumns,hypertexnames=true}

\makeatletter

\providecommand{\tabularnewline}{\\}

\usepackage{lmodern}
\usepackage{mathptmx, newtxtext, newtxmath, xspace}

\usepackage[table, dvipsnames]{xcolor}
\definecolor{colorA}{cmyk}{0,0,0,0.05}
\definecolor{colorB}{cmyk}{0.14,0.04,0,0}
\definecolor{colorC}{cmyk}{0.02,0.0799,0,0}
\definecolor{colorD}{cmyk}{0.099,0.14,0,0}

\usepackage{natbib}
\setcitestyle{numbers}

\newcommand{\mathbbm}[1]{\text{\usefont{U}{bbm}{m}{n}#1}}

\makeatother

\begin{document}
\title{Cascade of vestigial orders in two-component superconductors: nematic,
ferromagnetic, $s$-wave charge-4e, and $d$-wave charge-4e states}
\author{Matthias Hecker }
\affiliation{School of Physics and Astronomy, University of Minnesota, Minneapolis
55455 MN, USA}
\author{Roland Willa }
\affiliation{Institute for Condensed Matter, Karlsruhe Institute of Technology,
Karlsruhe 76131, Germany}
\affiliation{Institute of Systems Engineering, School of Engineering, HES-SO Valais-Wallis,
Sion, Switzerland}
\author{Jörg Schmalian }
\affiliation{Institute for Condensed Matter, Karlsruhe Institute of Technology,
Karlsruhe 76131, Germany}
\affiliation{Institute for Quantum Materials and Technologies, Karlsruhe Institute
of Technology, Karlsruhe 76131, Germany}
\author{Rafael M. Fernandes}
\affiliation{School of Physics and Astronomy, University of Minnesota, Minneapolis
55455 MN, USA}
\date{\today }
\begin{abstract}
Electronically ordered states that break multiple symmetries can melt
in multiple stages, similarly to liquid crystals. In the partially-melted
phases, known as vestigial phases, a bilinear made out of combinations
of the multiple components of the primary order parameter condenses.
Multi-component superconductors are thus natural candidates for vestigial
order, since they break both the $U(1)$-gauge and also time-reversal
or lattice symmetries. Here, we use group theory to classify all possible
real-valued and complex-valued bilinears of a generic two-component
superconductor on a tetragonal or hexagonal lattice. While the more
widely investigated real-valued bilinears correspond to vestigial
nematic or ferromagnetic order, the little explored complex-valued
bilinears correspond to a vestigial charge-$4e$ condensate, which
itself can have an underlying $s$-wave, $d_{x^{2}-y^{2}}$-wave,
or $d_{xy}$-wave symmetry. To properly describe the fluctuating regime
of the superconducting Ginzburg-Landau action and thus access these
competing vestigial phases, we employ both a large-$N$ and a variational
method. We show that while vestigial order can be understood as a
weak-coupling effect in the large-$N$ approach, it is akin to a moderate-coupling
effect in the variational method. Despite these distinctions, both
methods yield similar results in wide regions of the parameter space
spanned by the quartic Landau coefficients. Specifically, we find
that the nematic and ferromagnetic phases are the leading vestigial
instabilities, whereas the various types of charge-$4e$ order are
attractive albeit subleading vestigial channels. The only exception
is for the hexagonal case, in which the nematic and $s$-wave charge-$4e$
vestigial states are degenerate. We discuss the limitations of our
approach, as well as the implications of our results for the realization
of exotic charge-$4e$ states in material candidates.
\end{abstract}
\maketitle

\section{Introduction}

The vast majority of phase transitions can be understood employing
Landau's concept of a symmetry-breaking order parameter $\eta$ that
acquires a non-zero expectation value $\langle\eta\rangle\neq0$ below
a transition temperature $T_{c}$ \textendash{} or below a threshold
value of a non-thermal tuning parameter. Fluctuations above $T_{c}$,
encoded in the expectation value of the bilinear $\eta^{2}$, can
also be captured within the Ginzburg-Landau formalism via methods
that go beyond mean-field, thus describing observable phenomena such
as para-conductivity and diamagnetic fluctuations above the onset
of superconductivity \citep{Aslamazov1968,Aslamasov1968b,Schmidt1968}.
The versatility of the Ginzburg-Landau approach in describing disparate
systems without a detailed knowledge of the microscopic interactions
involved makes it a powerful phenomenological method to assess the
properties of various correlated materials.

When the order parameter has mutiple components, $\boldsymbol{\eta}=(\eta_{1},\eta_{2},\dots)$,
as is the case in magnetic and superconducting phases, the quantity
$\left|\boldsymbol{\eta}\right|^{2}$ is just one of many bilinears
of the general form $\eta_{i}M_{ij}\eta_{j}$, each of which describes
a different fluctuation mode (summation over repeated indices is implied).
The matrices $M_{ij}$ defining these bilinears are constrained by
the symmetries of the system, much alike the types of order parameter
$\boldsymbol{\eta}$ allowed are themselves restricted by the same
symmetries. While $\langle\left|\boldsymbol{\eta}\right|^{2}\rangle$
is always non-zero, since the bilinear $\eta_{i}\delta_{ij}\eta_{j}$
transforms trivially under the symmetry operations of the system (i.e.
it does not break any symmetries of the system), the other bilinears
$\eta_{i}M_{ij}\eta_{j}$ generally have non-trivial transformation
properties. In these cases, $\left\langle \eta_{i}M_{ij}\eta_{j}\right\rangle $
can be interpreted as a symmetry-breaking composite order parameter
which only acquires a non-zero value below a temperature $T^{*}\geq T_{c}$.
When $T^{*}$ is larger than $T_{c}$, there is a range of temperatures
$T_{c}<T<T^{*}$ in which $\left\langle \eta_{i}M_{ij}\eta_{j}\right\rangle \neq0$
but $\langle\boldsymbol{\eta}\rangle=0$, which defines a so-called
vestigial phase \citep{Nie2013,Fradkin2015,Fernandes2019}. This denomination
is motivated by the fact that the symmetries broken by $\left\langle \eta_{i}M_{ij}\eta_{j}\right\rangle \neq0$
correspond to a subset of the symmetries broken when the primary order
parameter acquires a non-zero value, $\langle\boldsymbol{\eta}\rangle\neq0$,
i.e. the vestigial phase is a partially-melted version of the primary
phase. Evidently, vestigial order is a fluctuation phenomena, whose
description requires methods that go beyond the mean-field approximation.
More broadly, vestigial phases are an example of intertwined orders
\citep{Fradkin2015}, as the bilinear $\left\langle \eta_{i}M_{ij}\eta_{j}\right\rangle $
and the order parameter $\langle\boldsymbol{\eta}\rangle$ are intrinsically
connected, notwithstanding the fact that they describe ordered states
with different broken symmetries (for a recent review on vestigial
orders, see Ref. \citep{Fernandes2019}). 

The idea of vestigial order is a natural generalization to electronic
systems of concepts typically employed in studies of liquid crystals.
As such, vestigial electronic nematicity has been widely investigated
in a number of correlated materials \citep{Kivelson98}, such as iron
pnictides \citep{Hu_Kivelson2008,Xu2008,Fernandes2012}, cuprates
\citep{Nie2013,Nie2017,Mukhopadhyay2019}, heavy-fermion compounds
\citep{Seo2020}, and twisted moiré systems \citep{Cao2021}. Beyond
that, the framework of vestigial phases has also been employed to
describe a broad range of phenomena observed or proposed to occur
in diverse settings \citep{Berg2009,Agterberg2011,Chern2012,Wang_Chubukov,Orth2015,Galitski2015,Fernandes2016_vestigial,Fischer2016,Zhang2017,Wang2017,Hecker2018,Borisov2019,Christensen2019,Little2020,Willa2020,Agterberg2020,PLee_PDW,Fernandes2021,Jian2021,Konig2021,Banerjee2022,Drouin2022,Strockoz2022,Scheurer2023},
from Bose-Einstein condensates of ultracold atoms in optical cavities
\citep{Demler2017} to valence-bond solid phases in quantum magnets
\citep{Sandvik2020}. It is important to emphasize that the concept
of a partially ordered state in itself is not new, as it dates back
to investigations of order-by-disorder phenomena in frustrated magnets
\citep{Villain77,Fradkin1978,Henley89,Chandra90,Korshunov06}. A more
extensive literature search reveals that, already in 1988, Golubovi\'{c}
and Kosti\'{c} employed a large-$N$ approach to a Ginzburg-Landau
model for incommensurate density-waves to show that partially-ordered
states, which they even dubbed ``nematiclike,'' are naturally expected
to emerge \citep{Golubovic1988}. To the best of our knowledge, this
work, which has somehow remained unknown to much of the community
(including us), obtains for the first time some of the results that
would emerge again in more recent large-$N$ studies of vestigial
nematicity in cuprates and pnictides \citep{Hu_Kivelson2006,Hu_Kivelson2008,Fernandes2012,Nie2013,Si2016}.
Finally, we stress that our analysis of vestigial order is performed
at finite temperatures. Corresponding investigations for the ground
state behavior were performed in Refs. \citep{Fernandes2012} and
\citep{Fernandes2013}. At $T=0$ it is also possible that vestigial
phases are affected by topological terms in the collective order-parameter
field theory \citep{Rampp2022}.

In this paper, we revisit the issue of vestigial orders of superconductors
described by two gap functions related by a symmetry of the lattice,
which we consider to be either tetragonal or hexagonal/trigonal. Quite
generally, systems with multiple coupled condensates provide a fertile
ground for composite orders above $T_{c}$ \citep{Bojesen2013,Bojesen2014}.
While it has been known that real-valued composite order parameters
describing nematicity or ferromagnetism can condense before the onset
of two-component superconductivity \citep{Fischer2016,Hecker2018,Venderbos_Fernandes,Fernandes2019,Willa2020},
the possibility of vestigial phases characterized by complex-valued
composite order parameters, which generally describe charge-$4e$
condensates \citep{Berg2009,Babaev2010,Agterberg2020}, has not been
as systematically investigated. Recent works focusing primarily on
nematic superconductors in hexagonal lattices have shown that charge-$4e$
order can indeed be stabilized as a vestigial phase \citep{Fernandes2021,Jian2021,Zeng2021}.
Here, we first apply the group-theoretical formalism introduced in
Ref. \citep{Fernandes2019} to the case of two-component $p$-wave,
$d$-wave, and $f$-wave superconductors in a lattice with either
fourfold or sixfold/threefold rotational symmetry. The superconducting
ground states in these cases are either nematic or chiral as it can
be obtained from a straightforward mean-field minimization of the
Ginzburg-Landau action. By going beyond the real-valued nematic and
ferromagnetic bilinears discussed in Ref. \citep{Fernandes2019},
which transform trivially within the $U(1)$ group but non-trivially
within the lattice point group, our group-theoretical classification
of complex-valued bilinears reveals not only $s$-wave charge-$4e$
composite order parameters (which transform non-trivially within $U(1)$
but trivially within the point group), but also even more exotic $d_{xy}$-wave
and $d_{x^{2}-y^{2}}$-wave charge-$4e$ orders, which transform non-trivially
within both the $U(1)$ and the point groups. Interestingly, inside
each of the chiral and nematic superconducting ground states, at least
one real-valued bilinear and one complex-valued bilinear are simultaneously
non-zero, which suggests the possibility of both vestigial nematic/ferromagnetic
order and vestigial charge-$4e$ order, thus expanding the list of
systems where the elusive quartet condensate can potentially be found
\citep{Nozieres1982,Korshunov1985,Nozieres1998,Volovik1992,Wu2005,Aligia2005,Berg2009,Radzihovsky2009,Babaev2010,Agterberg2011,Radzihovsky2011,Moon2012,Agterberg2020,HongYao2017,Fernandes2021,Jian2021,Zeng2021,Gnezdilov_Wang2022}.

To further investigate this possibility, we solve the Ginzburg-Landau
action for the two-component superconductor via two different approaches
that go beyond the mean-field approximation: the large-$N$ method
and the variational method. While both have been widely employed to
investigate vestigial phases arising from the condensation of real-valued
composite order parameters \citep{Golubovic1988,Hu_Kivelson2006,Hu_Kivelson2008,Fernandes2012,Nie2013,Si2016,Fischer2016,Nie2017,Yip2022},
they have not been systematically used to study the competition between
vestigial nematic/ferromagnetic and charge-$4e$ orders. We find that
treating all possible vestigial orders on an equal footing is not
possible within the large-$N$ method. This is due to the Fierz identities
relating the different bilinears, which introduce an unavoidable ambiguity
in the decoupling scheme of the quartic terms via Hubbard-Stratonovich
auxiliary fields. By imposing some physically-motivated but \emph{ad
hoc }restrictions on the decoupling procedure, we obtain the leading
and subleading vestigial instabilities in the parameter space spanned
by the ratios between the quartic coefficients of the Ginzburg-Landau
action. Quite generally, we find in every region of the phase diagram
attraction in one real-valued bilinear channel and one complex-valued
bilinear channel, indicating the possibility of a ``cascade'' of
vestigial phases. Despite the viability of both channels to form vestigial
phases, the leading vestigial instabilities are those associated with
the real-valued bilinears, corresponding to nematic and ferromagnetic
orders, whereas the vestigial instabilities associated with the charge-$4e$
states are subleading. The only exception is for the case of a hexagonal
nematic superconducting state, where vestigial nematicity and vestigial
$s$-wave charge-$4e$ order are degenerate \textendash{} as previously
reported in Ref. \citep{Fernandes2021}.

In distinction to the large-$N$ method, which is controlled yet limited
to the regime of many order-parameter components, the uncontrolled
variational approach is able to describe the proper number of order-parameter
components and hence allows us to treat on an equal footing all possible
composite order parameters. The key limitation of the variational
approach is that it only describes weak, Gaussian fluctuations. The
results obtained from the variational approach largely mirror those
obtained within the large-$N$ approach, with one important difference.
There are regions in the parameter space without any viable vestigial
phase, i.e. the instability channels associated with the real-valued
bilinears and complex-valued bilinears are all repulsive. In fact,
in the variational approach, the vestigial channels are only attractive
when the Landau coefficients of the squared non-trivial bilinears
are at least comparable in magnitude with the Landau coefficient of
the squared trivial bilinear, a behavior we dub ``moderate coupling.''
In contrast, in the large-$N$ approach, a vestigial instability is
present regardless of how small the coefficients of the squared non-trivial
bilinears are, which we identify as a ``weak-coupling'' behavior.
Nevertheless, despite these differences, the large-$N$ and variational
methods give the same results outside of these parameter-space regions
in which the variational method gives no (or only one) vestigial instability.
We also discuss under which conditions a vestigial instability implies
a vestigial phase. In doing so, we find that the most natural extension
of the variational ansatz that also includes the possibility of superconducting
order does not work properly, which exposes a possible limitation
of the variational approach. Finally, we explore the implications
of our results for various candidate two-component superconductors,
such as doped $\mathrm{Bi_{2}Se_{3}}$, twisted bilayer graphene,
$\mathrm{UPt_{3}}$, $\mathrm{UTe_{2}}$, $\mathrm{Sr_{2}RuO_{4}}$,
$\mathrm{URu_{2}Si_{2}}$, $\mathrm{KV_{3}Sb_{5}}$, 4Hb-$\mathrm{TaS_{2}}$,
and $\mathrm{CaSn_{3}}$. We also discuss ways in which the subleading
charge-$4e$ vestigial instability can be uncovered in these and other
systems.

This paper is organized as follows: in Sec. \ref{sec:Classification-of-bilinears}
we employ a group-theoretical formalism and classify all possible
real-valued and complex-valued bilinears of two-component superconductors
in systems with point groups $\mathsf{D_{4h}}$ or $\mathsf{D_{6h}}$.
In Sec. \ref{sec:Ginzburg_Landau}, we introduce the Ginzburg-Landau
actions associated with these superconducting degrees of freedom,
and re-derive the mean-field phase diagram in the parameter space
spanned by the quartic Landau coefficients. The properties and hierarchy
of leading and subleading vestigial instabilities of this model are
obtained via a large-$N$ approach in Sec. \ref{sec:Large--approach}
and a variational approach in Sec. \ref{sec:Variational-Approach}.
Section \ref{sec:Discussion-and-conclusions} is devoted to a comprehensive
discussion of the results and to the conclusions. Details about the
group-theoretical formalism are presented in Appendix \ref{App:Real-and-complex},
whereas Appendices \ref{App:variational-details}, \ref{App:One-component-superconductor-in}
and \ref{App:Two-component-superconductor} contain further details
about the variational approach.

\section{Classification of bilinears\label{sec:Classification-of-bilinears}}

\begin{table*}[t] 	
\setlength{\tabcolsep}{3.2pt}  	
\setlength{\arrayrulewidth}{.1em} 	
\renewcommand{\arraystretch}{1.7}  	
\newcommand{\colA}[1]{\multicolumn{1}{ >{\columncolor{colorA!100}[\tabcolsep]} c}{#1}}  
\newcommand{\colB}[1]{\multicolumn{1}{ >{\columncolor{colorB!100}[\tabcolsep]} c}{#1}}  
\newcommand{\colC}[1]{\multicolumn{1}{ >{\columncolor{colorC!100}[\tabcolsep]} c}{#1}}  
\newcommand{\colD}[1]{\multicolumn{1}{ >{\columncolor{colorD!100}[\tabcolsep]} c}{#1}} 	
\begin{tabular}{ccccc} 
\hline  
$\mathsf{D_{4h}}$ & $A_{1g}$ & $B_{1g}$ & $B_{2g}$ & $A_{2g}$ 
\\[0.1em]  \hline  
$\Gamma_{0}^{U}\big|_{s}$ & \colA{$\Psi^{A_{1g}}$} & \colB{$\Psi^{B_{1g}}$} & \colB{$\Psi^{B_{2g}}$} & \colB{\textemdash{}} 
\\[0.1em]  \hline  
$\Gamma_{0}^{U}\big|_{a}$ & \colA{\textemdash{}} & \colB{\textemdash{}} & \colB{\textemdash{}} & \colB{$\Psi^{A_{2g}}$} 
\\[0.1em]  \hline
$\Gamma_{+2}^{U}$ & \colC{$\psi^{A_{1g}}$} & \colD{$\psi^{B_{1g}}$} & \colD{$\psi^{B_{2g}}$} & \colD{\textemdash{}} 
\\[0.1em]  \hline 
$\Gamma_{-2}^{U}$ & \colC{$\bar{\psi}^{A_{1g}}$} & \colD{$\bar{\psi}^{B_{1g}}$} & \colD{$\bar{\psi}^{B_{2g}}$} & \colD{\textemdash{}} 
\\[0.1em]  \hline 
\end{tabular}
$\hspace{0.0em}$ 
\renewcommand{\arraystretch}{1.3} 
\begin{tabular}{rl} 
$\Psi^{B_{1g}}$: & $d_{x^2\!-\!y^2}$-nematic \\
$\Psi^{B_{2g}}$: & $d_{xy}$-nematic \\
$\Psi^{A_{2g}}$: & ferromagnetic \\
$\psi^{A_{1g}}$: & $s$-wave charge-$4e$ \\
$\psi^{B_{1g}}$: & $d_{x^2\!-\!y^2}$-wave charge-$4e$ \\
$\psi^{B_{2g}}$: & $d_{xy}$-wave charge-$4e$ 
\end{tabular}
$\hfill$
\renewcommand{\arraystretch}{1.7}
\begin{tabular}{cccc} 
\hline  
$\mathsf{D_{6h}}$ & $A_{1g}$ & $E_{2g}$ & $A_{2g}$ 
\\[0.1em]  \hline 
$\Gamma_{0}^{U}\big|_{s}$ & \colA{$\Psi^{A_{1g}}$} & \colB{$\boldsymbol{\Psi}^{E_{2g}}$} & \colB{\textemdash{}} 
\\[0.1em]  \hline 
$\Gamma_{0}^{U}\big|_{a}$ & \colA{\textemdash{}} & \colB{\textemdash{}} & \colB{$\Psi^{A_{2g}}$} 
\\[0.1em]  \hline 
$\Gamma_{+2}^{U}$ & \colC{$\psi^{A_{1g}}$} & \colD{$\boldsymbol{\psi}^{E_{2g}}$} & \colD{\textemdash{}} 
\\[0.1em]  \hline 
$\Gamma_{-2}^{U}$ & \colC{$\bar{\psi}^{A_{1g}}$} & \colD{$\bar{\boldsymbol{\psi}}^{E_{2g}}$} & \colD{\textemdash{}} 
\\[0.1em]  \hline 
\end{tabular}
$\hspace{0.0em}$ 
\renewcommand{\arraystretch}{1.3} 
\begin{tabular}{ll} 
$\boldsymbol{\Psi}^{E_{2g}}$: & \hspace{-0.0em}$\begin{pmatrix}d_{x^2\!-\!y^2}\\ d_{xy}\end{pmatrix}$-nematic \\
$\Psi^{A_{2g}}$: & ferromagnetic \\
$\psi^{A_{1g}}$: & $s$-wave charge-$4e$ \\
$\boldsymbol{\psi}^{E_{2g}}$: & $\begin{pmatrix}d_{x^2\!-\!y^2}\\ d_{xy}\end{pmatrix}$-wave charge-$4e$ \\
\end{tabular}
\caption{The set of $N_{\Gamma}=10$ non-zero bilinear components associated with the composite orders of the two-component superconducting order parameter $\boldsymbol{\Delta}=(\Delta_{1},\Delta_{2})$ in the cases of a tetragonal lattice (Eq. \ref{eq:case1},  left panel) and of a hexagonal lattice (Eq. \ref{eq:case2},  right panel). The rows and columns of these ``multiplication tables" correspond,  respectively,  to the bilinear decompositions in the  $U(1)$ gauge sector and in the point-group lattice sector (\ref{eq:U(1)})-(\ref{eq:D3d}).  The explicit expressions for the bilinears are given in Eq.  (\ref{eq:bilinears_exp}).  Next to each multiplication table,  we also identify the composites according to the type of vestigial order they promote once condensed.\label{tab:bil_components}} 
\end{table*} 

Our starting point is a two-component superconducting (SC) order parameter,
$\boldsymbol{\Delta}=(\Delta_{1},\Delta_{2})$, which transforms as
a two-dimensional irreducible representation (IR) of the point-groups
$\mathsf{D_{4h}}$ or $\mathsf{D_{6h}}$, which describe tetragonal
and hexagonal lattices, respectively. In particular, we have:
\begin{align}
\mathrm{tetragonal\,(\mathsf{D_{4h}})} & : & \mathrm{IR}(\boldsymbol{\Delta}) & =E_{g}\;\mathrm{or}\;E_{u},\label{eq:case1}\\
\mathrm{hexagonal\,(\mathsf{D_{6h}})} & : & \mathrm{IR}(\boldsymbol{\Delta}) & =E_{1g},\;E_{2g},\;E_{1u},\mathrm{or}\;E_{2u}.\label{eq:case2}
\end{align}
This parametrization can describe the $m_{l}=\pm1$ singlet $(d_{xz},d_{yz})$-wave
state ($E_{g}$ or $E_{1g}$); the $m_{l}=\pm1$ triplet $(p_{x},p_{y})$-wave
state ($E_{u}$ or $E_{1u}$); the $m_{l}=\pm2$ singlet $(d_{x^{2}-y^{2}},d_{xy})$-wave
state ($E_{2g}$); the $m_{l}=\pm2$ triplet $(f_{x^{2}z-y^{2}z},f_{xyz})$-wave
state ($E_{2u}$). These states have been proposed in a variety of
materials, such as the tetragonal-lattice $\mathrm{Sr_{2}RuO_{4}}$
\citep{Luke1998,Jia2006} and $\mathrm{URu_{2}Si_{2}}$ \citep{Kasahara2009,Balicas2013,Schemm2015},
the trigonal-lattice $A_{\mathrm{x}}\mathrm{Bi_{2}Se_{3}}$ with $A=\mathrm{Cu},\,\mathrm{Nb},\,\mathrm{Sr}$
\citep{Fu2014,Matano2015,Pan2016,Asaba2016,Venderbos2016a,Hecker2018},
the hexagonal-lattice $\mathrm{UPt_{3}}$ \citep{Sauls1994,Schemm2014,Avers2020},
and the triangular-moiré superlattice twisted bilayer graphene \citep{Venderbos_Fernandes,SZLin2018,Kozii2019,Chichinadze2020,Scheurer2020,Cao2021}.

In this section, we apply the group-theoretical method outlined in
Ref. \citep{Fernandes2019} to comprehensively identify all bilinear
combinations formed out of $\boldsymbol{\Delta}$. While previous
works have focused only on real-valued bilinears, corresponding to
nematic and ferromagnetic vestigial order \citep{Fernandes2019,Fischer2016,Willa2020},
here we show that there is an entire family of complex-valued bilinears
corresponding to different types of charge-4e superconductivity, of
which the results of Ref. \citep{Fernandes2021} are a special case.
A summary of the results of this section is presented in Table \ref{tab:bil_components},
with the bilinears defined in Eqs. (\ref{eq:real_bilinears})-(\ref{eq:compl_bilinears}).

Consider a general order parameter parameter $\boldsymbol{\eta}=(\eta_{1},\dots,\eta_{\dim\Gamma})$
that transforms according to the IR $\Gamma$ of the group $\mathscr{G}$.
The corresponding bilinear components, denoted by $C^{\ell}$, with
$\ell=1,\dots,N_{\Gamma}$, can be expressed as 
\begin{align}
C^{\ell} & =\boldsymbol{\eta}^{T}M^{\ell}\boldsymbol{\eta}.\label{eq:C_l}
\end{align}
Here, $M^{\ell}$ is a $\dim\Gamma\times\dim\Gamma$ matrix. Since
the component $C^{\ell}$ (\ref{eq:C_l}) is a scalar, the matrix
$M^{\ell}$ has to be symmetric {[}$(M^{\ell})^{T}=M^{\ell}${]},
which reduces the total number of bilinear components from $\dim^{2}\Gamma$
to $N_{\Gamma}=\frac{1}{2}\dim\Gamma\left(1\!+\dim\Gamma\right)$.
Naturally, these components can be grouped into the IRs of the group
$\mathcal{G}$ according to the product decomposition $\Gamma\otimes\Gamma=\Gamma_{0}\oplus\Gamma_{1}\oplus\dots$.
The explicit bilinear components (\ref{eq:C_l}) are deduced from
the transformation properties of $\boldsymbol{\eta}$ as shown in
details in App. \ref{App:Real-and-complex}.

For many condensed-matter systems of interest, the symmetry group
itself is a product of two groups, $\mathcal{G}=\mathcal{G}_{\mathrm{int}}\otimes\mathcal{G}_{s}$,
where $\mathcal{G}_{s}$ is a space group and $\mathcal{G}_{\mathrm{int}}$
is a continuous internal group. Indeed, this is the case for magnetic
materials with negligible spin-orbital coupling, where $\mathcal{G}_{\mathrm{int}}=SU(2)$
corresponds to spin-rotational symmetry, or superconductors, where
$\mathcal{G}_{\mathrm{int}}=U(1)$ is the gauge symmetry. In these
cases, it suffices to classify the matrices associated with the two
subspaces separately and then simply multiply them, enforcing the
resulting matrix $M^{\ell}$ to be symmetric. Generally, the resulting
bilinear components can be categorized into four sectors according
to their subspace transformation properties. (i) They are fully symmetry-preserving,
i.e. transform trivially under the operations of both subgroups. (ii-iii)
They break symmetries related to only one subgroup, i.e. they transform
non-trivially within one group but trivially within the other subgroup.
(iv) They break symmetries related to both subgroups, i.e. they transform
non-trivially under the operations of both subgroups. 

For our cases of interest, Eqs. (\ref{eq:case1}) and (\ref{eq:case2}),
it is sufficient to consider the point groups $\mathcal{G}_{p}=\mathsf{D_{4h}}$
or $\mathcal{G}_{p}=\mathsf{D_{6h}}$ rather than the space group
$\mathcal{G}_{s}$, since we only consider cases where superconductivity
is a uniform order that does not break translational symmetry. As
explained in App. \ref{App:Real-and-complex}, a (complex) superconducting
order parameter transforms effectively as a ``Nambu'' doublet $\hat{\boldsymbol{\Delta}}=(\Delta,\bar{\Delta})$
under the $U(1)$ symmetry operations, i.e. according to a two-dimensional
representation. We denote this representation as $\Gamma_{\Delta}=\Gamma_{+1}^{U}\oplus\Gamma_{-1}^{U}$,
with $\Gamma_{m}^{U}$ denoting the IRs of the $U(1)$ group and $m\in\{0,\pm1,\pm2,\dots\}$.
More generally, if an order parameter transforms as $\Gamma_{+m}^{U}\oplus\Gamma_{-m}^{U}$,
it corresponds to a condensate with charge $2me$, whose condensation
lowers the continuous $U(1)$ gauge symmetry to a discrete $Z_{m}$
symmetry. Thus, in this notation, the two-component superconductor
$\boldsymbol{\Delta}=(\Delta_{1},\Delta_{2})$ associated with the
lattice IR $E_{i}$ transforms effectively as the four-component Nambu
vector $\hat{\boldsymbol{\Delta}}=(\boldsymbol{\Delta},\bar{\boldsymbol{\Delta}})$
according to $\Gamma=\Gamma_{\Delta}\otimes E_{i}$. Here, the subscript
$i$ is defined as $i=\left\{ g,u\right\} $ for $\mathsf{D_{4h}}$
and $i=\left\{ 1g,\,2g,\,1u,\,2u\right\} $ for $\mathsf{D_{6h}}$.
To identify the bilinear components, we rewrite the product representation
separately in the two subsectors, namely, the gauge sector and the
latice sector, $\Gamma\otimes\Gamma=(\Gamma_{\Delta}\otimes\Gamma_{\Delta})\otimes(E_{i}\otimes E_{i})$.
Individually, they decompose according to 
\begin{align}
U(1) & : & \Gamma_{\Delta}\otimes\Gamma_{\Delta} & =\left[\Gamma_{0}^{U}\oplus\big(\Gamma_{+2}^{U}\oplus\Gamma_{-2}^{U}\big)\right]_{s}\oplus\left[\Gamma_{0}^{U}\right]_{a},\label{eq:U(1)}\\
\mathsf{D_{4h}} & : & E_{i}\otimes E_{i} & =\left[A_{1g}\oplus B_{1g}\oplus B_{2g}\right]_{s}\oplus\left[A_{2g}\right]_{a},\label{eq:D4h}\\
\mathsf{D_{6h}} & : & E_{i}\otimes E_{i} & =\left[A_{1g}\oplus E_{2g}\right]_{s}\oplus\left[A_{2g}\right]_{a},\label{eq:D3d}
\end{align}
where the subscripts $s,\,a$ denote the channels associated with
symmetric and antisymmetric matrices, respectively (details in App.
\ref{App:Real-and-complex}). For a one-component superconductor,
regardless of how it transforms in the lattice sector, the bilinears
are always trivial within the point group. As a result, the $N_{\Gamma_{\Delta}}=3$
bilinear components comprise the trivial combination $|\Delta|^{2}$
and the doublet $(\Delta^{2},\bar{\Delta}^{2})$, which transform
according to $\Gamma_{0}^{U}$ and $\left(\Gamma_{+2}^{U}\oplus\Gamma_{-2}^{U}\right)$,
respectively. While the former corresponds to superconducting fluctuations,
which are present at any temperature, the latter corresponds to a
charge-$4e$ superconducting order parameter, which, as discussed
above, lowers the continuous $U(1)$ gauge symmetry to a discrete
$Z_{2}$ one.

Going back to our two-component superconductors described by Eqs.
(\ref{eq:case1}) and (\ref{eq:case2}), the symmetric/antisymmetric
matrices resulting from the decompositions (\ref{eq:U(1)})-(\ref{eq:D3d})
allow us to identify the $N_{\Gamma}=10$ bilinear components shown
in Table \ref{tab:bil_components} in a straightforward way. To make
the notation more transparent, real-valued bilinears (i.e. which transform
trivially within $U(1)$) are denoted by $\Psi^{n}$, whereas complex-valued
bilinears (i.e. which transform according to $\Gamma_{\pm2}^{U}$)
are denoted by $(\psi^{n},\bar{\psi}^{n})$; in both case, $n$ denotes
the point-group IR of the bilinear. In the table, the four combinations
of bilinears mentioned above are highlighted with different colors:
a bilinear that is trivial in both the gauge and lattice sectors is
highlighted in gray; a bilinear that is trivial in the gauge sector
and non-trivial in the lattice sector is highlighted in blue; a bilinear
that is non-trivial in the gauge sector and trivial in the lattice
sector is highlighted in pink; and a bilinear that is non-trivial
in both gauge and lattice sectors is highlighted in purple.

Explicitly, for the $\mathsf{D_{4h}}$ case (\ref{eq:case1}), one
obtains the trivial combination $\Psi^{A_{1g}}=\boldsymbol{\Delta}^{\dagger}\tau^{0}\boldsymbol{\Delta}$
with Pauli matrices $\tau^{i}$ ; the three real-valued non-trivial
bilinears
\begin{align}
\Psi^{B_{1g}} & =\boldsymbol{\Delta}^{\dagger}\tau^{z}\boldsymbol{\Delta}, & \Psi^{B_{2g}} & =\boldsymbol{\Delta}^{\dagger}\tau^{x}\boldsymbol{\Delta}, & \Psi^{A_{2g}} & =\boldsymbol{\Delta}^{\dagger}\tau^{y}\boldsymbol{\Delta},\label{eq:real_bilinears}
\end{align}
and the three complex-valued bilinears
\begin{align}
\psi^{A_{1g}} & =\boldsymbol{\Delta}^{T}\tau^{0}\boldsymbol{\Delta}, & \psi^{B_{1g}} & =\boldsymbol{\Delta}^{T}\tau^{z}\boldsymbol{\Delta}, & \psi^{B_{2g}} & =\boldsymbol{\Delta}^{T}\tau^{x}\boldsymbol{\Delta}.\label{eq:compl_bilinears}
\end{align}
Writing them explicitly in terms of the two SC components $\boldsymbol{\Delta}=(\Delta_{1},\Delta_{2})$
yields:

\begin{align}
\Psi^{A_{1g}} & =|\Delta_{1}|^{2}+|\Delta_{2}|^{2}, & \psi^{A_{1g}} & =\Delta_{1}^{2}+\Delta_{2}^{2},\nonumber \\
\Psi^{A_{2g}} & =\mathsf{i}\left(\bar{\Delta}_{2}\Delta_{1}-\bar{\Delta}_{1}\Delta_{2}\right),\nonumber \\
\Psi^{B_{1g}} & =|\Delta_{1}|^{2}-|\Delta_{2}|^{2}, & \psi^{B_{1g}} & =\Delta_{1}^{2}-\Delta_{2}^{2},\nonumber \\
\Psi^{B_{2g}} & =\bar{\Delta}_{1}\Delta_{2}+\bar{\Delta}_{2}\Delta_{1}, & \psi^{B_{2g}} & =2\Delta_{1}\Delta_{2}.\label{eq:bilinears_exp}
\end{align}

Among the real-valued composite order parameters (\ref{eq:real_bilinears}),
$\Psi^{A_{2g}}$ corresponds to an out-of-plane ferromagnetic moment
that breaks time reversal symmetry, while $\Psi^{B_{1g}}$ and $\Psi^{B_{2g}}$
describe electronic nematicity that breaks the tetragonal symmetry
of the lattice by making, respectively, the two cartesian axes inequivalent
($B_{1g}$ or $d_{x^{2}-y^{2}}$-nematic) or the two diagonals inequivalent
($B_{2g}$ or $d_{xy}$-nematic). As for the complex-valued composites
in (\ref{eq:compl_bilinears}), all of them correspond to a type of
charge-$4e$ superconductivity, as explained above. They can be further
classified according to how they transform upon the point-group operations.
Thus, $\psi^{A_{1g}}$ is an $s$-wave charge-$4e$ superconductor
while $\psi^{B_{1g}}$ and $\psi^{B_{2g}}$ are, respectively, $d_{x^{2}-y^{2}}$-wave
and $d_{xy}$-wave charge-$4e$ superconductors. For later convenience,
we introduce the groups of point-group IRs associated with the real-valued
(\ref{eq:real_bilinears}) and the complex-valued (\ref{eq:compl_bilinears})
non-trivial bilinears,
\begin{align}
\mathbb{G}_{\mathbb{R}} & =\{A_{2g},B_{1g},B_{2g}\}, & \mathbb{G}_{\mathbb{C}} & =\{A_{1g},B_{1g},B_{2g}\},\label{eq:IR_groups}
\end{align}
 as well as the full real group $\mathbb{G}_{\mathbb{R}}^{0}=\{A_{1g},\mathbb{G}_{\mathbb{R}}\}$
containing also the trivial lattice IR.

In the $\mathsf{D_{6h}}$ case (\ref{eq:case2}), the only modication
required is to combine the two $B$-channel bilinears into a single
$E_{2g}$-channel bilinear:

\begin{align}
\boldsymbol{\Psi}^{E_{2g}} & =\big(\Psi^{B_{1g}},\,-\Psi^{B_{2g}}\big)=\big(\boldsymbol{\Delta}^{\dagger}\tau^{z}\boldsymbol{\Delta},\,-\boldsymbol{\Delta}^{\dagger}\tau^{x}\boldsymbol{\Delta}\big),\nonumber \\
\boldsymbol{\psi}^{E_{2g}} & =\big(\psi^{B_{1g}},\,-\psi^{B_{2g}}\big)=\big(\boldsymbol{\Delta}^{T}\tau^{z}\boldsymbol{\Delta},\,-\boldsymbol{\Delta}^{T}\tau^{x}\boldsymbol{\Delta}\big).\label{eq:psi_E2g}
\end{align}
To make the presentation more transparent, hereafter we will only
derive the results for the $\mathsf{D_{4h}}$ case (\ref{eq:case1})
and just mention the replacements necessary to recover the results
for the $\mathsf{D_{6h}}$ case (\ref{eq:case2}). 

\section{Superconducting phase diagram: Ginzburg-Landau theory \label{sec:Ginzburg_Landau}}

To keep the analysis general, in this paper we use a Ginzburg-Landau
formalism to obtain the phase diagrams of the two-component superconducting
order parameters $\boldsymbol{\Delta}=(\Delta_{1},\Delta_{2})$ of
Eqs. (\ref{eq:case1}) and (\ref{eq:case2}). Since the values of
the Landau coefficients depend on the microscopic model, we will consider
the entire parameter space spanned by the Landau coefficients. Our
only restriction is that the free energy is bounded, i.e. that the
underlying superconducting transition is second-order. The Ginzburg-Landau
expansion for the superconducting action can be expressed as (see,
for instance, Ref. \citep{Sigrist1991}) 
\begin{align}
\mathcal{S} & =\int_{\mathsf{x}}r_{0}\,\boldsymbol{\Delta}^{\dagger}\boldsymbol{\Delta}\;+\;\mathcal{S}^{\mathrm{grad}}+\;\mathcal{S}^{\mathrm{int}}\,,\label{eq:action1}
\end{align}
where the variable $\mathsf{x}=(\tau,\boldsymbol{r})$ comprises both
imaginary time and position, and the $\boldsymbol{\Delta}(\mathsf{x})$
dependence is left implicit. The quadratic coefficient is $r_{0}=a_{0}(T-T_{0})$
with $a_{0}>0$ and $T_{0}>0$ denoting the bare superconducting transition.
The gradient ($\mathcal{S}^{\mathrm{grad}}$) and interaction ($\mathcal{S}^{\mathrm{int}}$)
contributions to the action depend on the point-group symmetry. For
the $\mathsf{D_{4h}}$ tetragonal case (\ref{eq:case1}), the interaction
term is given, in terms of the bilinears (\ref{eq:real_bilinears}),
by
\begin{align}
\mathcal{S}^{\mathrm{int}} & =\int_{\mathsf{x}}\Big[u\,\big(\Psi^{A_{1g}}\big)^{2}+v\,\big(\Psi^{A_{2g}}\big)^{2}+w\,\big(\Psi^{B_{1g}}\big)^{2}\Big],\label{eq:S_int}
\end{align}
where the interaction parameters $u$, $v$, and $w$ have to satisfy
$u>0$, $v>-u$ and $w>-u$ in order for the action to be bounded.
Note that the interaction term can only have squared bilinears as
it must transform trivially. The gradient term, which is essential
to account for order parameter fluctuations, is more conveniently
expressed in momentum space, 
\begin{align}
\mathcal{S}^{\mathrm{grad}} & =\frac{T}{V}\sum_{k}\boldsymbol{\Delta}_{k}^{\dagger}\left(f_{\boldsymbol{k}}^{A_{1g}}\tau^{0}\!+\!f_{\boldsymbol{k}}^{B_{1g}}\tau^{z}\!+\!f_{\boldsymbol{k}}^{B_{2g}}\tau^{x}\right)\boldsymbol{\Delta}_{k},\label{eq:S_grad}
\end{align}
where we defined the Fourier transform $\boldsymbol{\Delta}_{k}=\frac{T}{V}\int d\mathsf{x}\:\boldsymbol{\Delta}(\mathsf{x})e^{-ik\mathsf{x}}$
with the volume $V$ and the variable $k=(\omega_{n},\boldsymbol{k})$
comprising bosonic Matsubara frequency $\omega_{n}=2\pi nT$ and momentum
$\boldsymbol{k}$. In the continuum limit, the gradient functions
in (\ref{eq:S_grad}) are given by:

\begin{align}
f_{\boldsymbol{k}}^{A_{1g}} & =\mathsf{d}_{0}(k_{x}^{2}+k_{y}^{2})+\mathsf{d}_{z}k_{z}^{2}, & f_{\boldsymbol{k}}^{B_{1g}} & =\mathsf{d}_{1}(k_{x}^{2}-k_{y}^{2}),\nonumber \\
f_{\boldsymbol{k}}^{B_{2g}} & =\mathsf{d}_{2}2k_{x}k_{y},\label{eq:f_gradient}
\end{align}
where $\mathsf{d}_{i}$ are stiffness coefficients. 

The $\mathsf{D_{6h}}$ hexagonal case (\ref{eq:case2}) is obtained
by setting $w=0$ and $\mathsf{d}_{2}=\mathsf{d}_{1}$ in Eqs. (\ref{eq:S_int})
and (\ref{eq:S_grad}), respectively. Note that for other point groups
that have threefold rotational symmetry, additional gradient terms
may be allowed. For instance, for the trigonal $\mathsf{D_{3d}}$
group, which describes the lattice symmetries of $A_{\mathrm{x}}\mathrm{Bi_{2}Se_{3}}$,
the two extra terms, $\mathsf{d}_{3}2k_{y}k_{z}$ and $\mathsf{d}_{3}2k_{x}k_{z}$,
must be added to $f_{\boldsymbol{k}}^{B_{1g}}$ and $f_{\boldsymbol{k}}^{B_{2g}}$,
respectively, as explained in Ref. \citep{Hecker2018}. 

\begin{figure}
\includegraphics[scale=0.9]{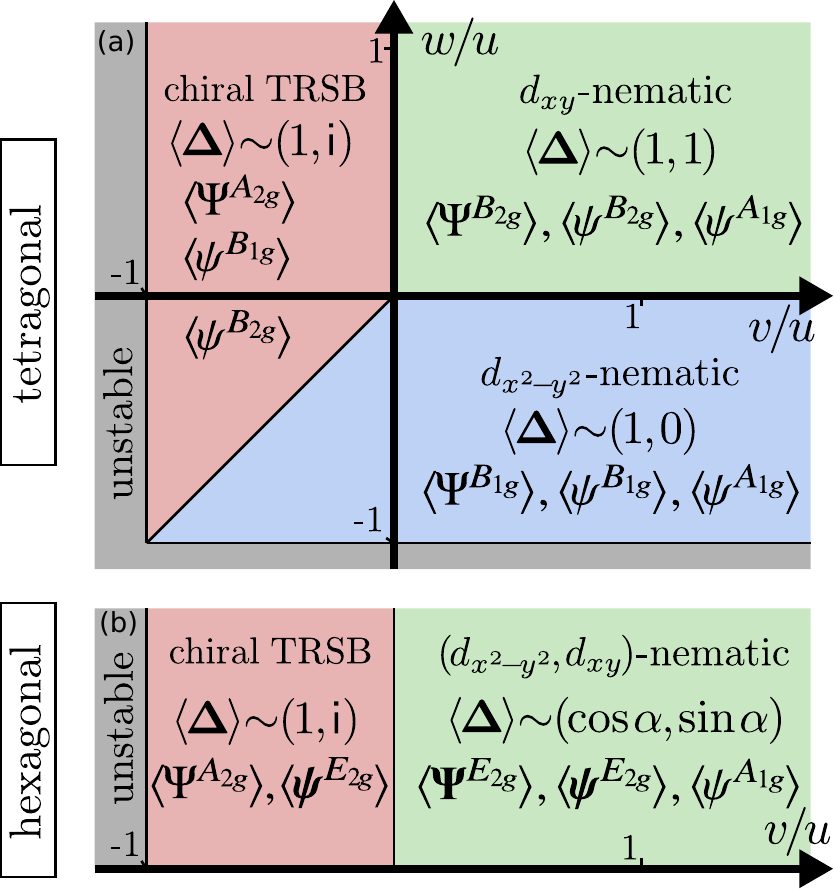}

\caption{Mean-field phase diagram, in the parameter space spanned by the quartic
Landau coefficients of Eq. (\ref{eq:action1}), of the two-component
superconducting pairing states of the (a) tetragonal case $\mathsf{D_{4h}}$
(\ref{eq:case1}) and (b) hexagonal case $\mathsf{D_{6h}}$ (\ref{eq:case2}).
In each phase, we show the relationship between the two components
of the superconducting order parameter, $\boldsymbol{\Delta}=(\Delta_{1},\Delta_{2})$,
as well as all non-zero bilinears defined in Eqs. (\ref{eq:real_bilinears})-(\ref{eq:compl_bilinears}).
Each phase is labeled by the additional symmetry that they break besides
the $U(1)$ gauge symmetry, which could be time-reversal symmetry
(TRS) or a point-group symmetry. In the regions labeled ``unstable''
(dark-gray), the action is unbounded. \label{fig:Mean-field-phase-diagram}}
\end{figure}

Before discussing the emergence of vestigial phases, we do a mean-field
calculation to obtain the superconducting phase diagram of the Ginzburg-Landau
action in Eq. (\ref{eq:action1}). The results are well known \citep{Sigrist1991}:
minimization of the action gives three distinct states, as shown in
Fig. \ref{fig:Mean-field-phase-diagram}(a). For $w,v>0$ the superconducting
ground state is $\langle\boldsymbol{\Delta}\rangle\sim(1,1)$, which
we denote as the $d_{xy}$-nematic (or $B_{2g}$-nematic) SC state.
It not only breaks the $U(1)$ gauge symmetry, but also the fourfold
rotational symmetry of $\mathsf{D_{4h}}$ by making the two diagonals
inequivalent. Indeed, substituting $\langle\boldsymbol{\Delta}\rangle\sim(1,1)$
into the real-valued bilinear expressions (\ref{eq:real_bilinears}),
we find that only the $d_{xy}$-nematic (or $B_{2g}$-nematic) composite
order parameter $\langle\Psi^{B_{2g}}\rangle$ is non-zero in this
phase. As for the complex-valued bilinears in Eq. (\ref{eq:compl_bilinears}),
two charge-$4e$ composite order parameters are non-zero in this phase,
namely, the $s$-wave $\langle\psi^{A_{1g}}\rangle$ and the $d_{xy}$-wave
$\langle\psi^{B_{2g}}\rangle$. 

In the region $w<\min(0,v)$ of the parameter space spanned by the
quartic Landau coefficients, the superconducting ground state is $\langle\boldsymbol{\Delta}\rangle\sim(1,0)$,
which corresponds to a $d_{x^{2}-y^{2}}$-nematic (or $B_{1g}$-nematic)
superconductor. In this case, tetragonal symmetry is broken due to
the inequivalence between the horizontal and vertical Cartesian axes.
The corresponding non-zero composite order parameters are the $d_{x^{2}-y^{2}}$-nematic
(or $B_{1g}$-nematic) $\langle\Psi^{B_{1g}}\rangle$, the $s$-wave
charge-$4e$ $\langle\psi^{A_{1g}}\rangle$, and the $d_{x^{2}-y^{2}}$-wave
charge-$4e$ $\langle\psi^{B_{1g}}\rangle$. Finally, in the region
$v<\min(0,w)$, the ground state is the chiral superconductor $\langle\boldsymbol{\Delta}\rangle\sim(1,\mathsf{i})$.
This is a time-reversal symmetry-breaking (TRSB) phase that respects
all symmetries of the tetragonal lattice. The associated composite
order parameters are the ferromagnetic $\langle\Psi^{A_{2g}}\rangle$,
the $d_{x^{2}-y^{2}}$-wave charge-$4e$ $\langle\psi^{B_{1g}}\rangle$,
and the $d_{xy}$-wave charge-$4e$ $\langle\psi^{B_{2g}}\rangle$.

While it is straightforward to verify which non-trivial bilinears
are non-zero by simply substituting the superconducting solutions
in Eqs. (\ref{eq:real_bilinears}) and (\ref{eq:compl_bilinears}),
valuable insight can be obtained directly from the interaction action,
Eq. (\ref{eq:S_int}). Indeed, we can loosely interpret the interaction
action, which is quartic in the superconducting order parameters,
as an effective action that is quadratic in the bilinears. This suggests,
for instance, that $w<0$ should favor the condensation of $\langle\Psi^{B_{1g}}\rangle$,
whereas $v<0$ should promote $\langle\Psi^{A_{2g}}\rangle\neq0$.
At first sight, this oversimplified analysis would seem to suggest
that no composite orders would condense when $w,v>0$. To see why
this is not the case, we use the fact that

\begin{equation}
\big(\Psi^{A_{1g}}\big)^{2}=\big(\Psi^{B_{1g}}\big)^{2}+\big(\Psi^{B_{2g}}\big)^{2}+\big(\Psi^{A_{2g}}\big)^{2},\label{eq:aux_Fierz}
\end{equation}
to rewrite the interaction action as:

\begin{equation}
\mathcal{S}^{\mathrm{int}}=\int_{\mathsf{x}}\left[\left(u+w\right)\big(\Psi^{A_{1g}}\big)^{2}+\left(v-w\right)\big(\Psi^{A_{2g}}\big)^{2}-w\,\big(\Psi^{B_{2g}}\big)^{2}\right].\label{eq:alt_Sint}
\end{equation}
Thus, $w>0$ should favor the condensation of $\langle\Psi^{B_{2g}}\rangle$.
Equation (\ref{eq:aux_Fierz}) is an example of a so-called Fierz
identity. The complete list of Fierz identities in our case is:

\begin{align}
\big(\Psi^{A_{1g}}\big)^{2} & =\sum_{n\in\mathbb{G}_{\mathbb{R}}}\left(\Psi^{n}\right)^{2}, & \big(\Psi^{A_{1g}}\big)^{2} & =\left|\psi^{B_{1g}}\right|^{2}+\big(\Psi^{B_{2g}}\big)^{2},\nonumber \\
\big(\Psi^{A_{1g}}\big)^{2} & =\left|\psi^{A_{1g}}\right|^{2}+\big(\Psi^{A_{2g}}\big)^{2}, & \big(\Psi^{A_{1g}}\big)^{2} & =\left|\psi^{B_{2g}}\right|^{2}+\big(\Psi^{B_{1g}}\big)^{2}.\label{eq:Fierz_ids_D4h}
\end{align}
The Fierz identities imply that the representation of the interaction
action (\ref{eq:S_int}) in terms of the three bilinear channels $\Psi^{A_{1g}}$,
$\Psi^{A_{2g}}$ and $\Psi^{B_{1g}}$ is not unique. On the contrary,
by inserting these identities in the interaction action (\ref{eq:S_int}),
it can be represented in terms of infinitely many combinations of
composite bilinears. Note that the mean-field results are insensitive
to this choice of representation of the quartic action.

The superconducting phase diagram in the case of a two-component order
parameter defined on the $\mathsf{D_{6h}}$ lattice according to Eq.
(\ref{eq:case2}) is obtained by setting $w=0$ in Eq. (\ref{eq:S_int})
and then minimizing the action. There are two distinct ground states,
as shown in Fig. \ref{fig:Mean-field-phase-diagram}(b). For $v<0$,
we obtain the chiral state $\langle\boldsymbol{\Delta}\rangle\sim(1,\mathsf{i})$
and non-zero composite order parameters $\langle\Psi^{A_{2g}}\rangle$
and $\langle\boldsymbol{\psi}^{E_{2g}}\rangle$. For $v>0$, the superconducting
ground state is the nematic one, $\langle\boldsymbol{\Delta}\rangle\sim(\cos\alpha,\sin\alpha)$,
which is accompanied by the non-zero composite order parameters $\langle\boldsymbol{\Psi}^{E_{2g}}\rangle$,
$\langle\psi^{A_{1g}}\rangle$ and $\langle\boldsymbol{\psi}^{E_{2g}}\rangle$.
Note that the continuous degeneracy of the nematic superconducting
state, indicated by the angle $\alpha$, is an artifact of stopping
the Ginzburg-Landau at fourth order. Addition of the sixth order term
$\mathcal{S}_{6}=u_{6}\Psi^{E_{2g},1}\Big[\big(\Psi^{E_{2g},1}\big)^{2}-3\big(\Psi^{E_{2g},2}\big)^{2}\Big]$
reduces the degeneracy to threefold, as one would have expected for
a system with threefold rotational symmetry \citep{Sigrist1991,Fu2014,Venderbos_Fernandes}.
For completeness, we also list the Fierz identities for the $\mathsf{D_{6h}}$
case, noting that $\left|\boldsymbol{\Psi}^{E_{2g}}\right|^{2}=\big(\Psi^{E_{2g},1}\big)^{2}+\big(\Psi^{E_{2g},2}\big)^{2}$,

\begin{align}
\big(\Psi^{A_{1g}}\big)^{2} & \!=\big(\Psi^{A_{2g}}\big)^{2}\!+\left|\boldsymbol{\Psi}^{E_{2g}}\right|^{2}\!, & 2\big(\Psi^{A_{1g}}\big)^{2} & =\left|\boldsymbol{\Psi}^{E_{2g}}\right|^{2}\!+\left|\boldsymbol{\psi}^{E_{2g}}\right|^{2}\!,\nonumber \\
\big(\Psi^{A_{1g}}\big)^{2} & \!=\big(\Psi^{A_{2g}}\big)^{2}\!+\left|\psi^{A_{1g}}\right|^{2}\!.\label{eq:Fierz_ids_D6h}
\end{align}

The results of this section reveal that each superconducting ground
state is accompanied by several non-zero non-trivial bilinears. For
instance, in the case of the $\mathsf{D_{4h}}$ lattice, each superconducting
state has one real-valued (corresponding to nematicity or ferromagnetism)
and two complex-valued (corresponding to charge-$4e$ superconductivity)
non-zero bilinears. In a mean-field approach, these composite order
parameters condense simultaneously and along with superconductivity.
However, as we discuss in the next sections, fluctuations may allow
them to condense before the onset of superconductivity, giving rise
to vestigial phases.

\section{Vestigial-orders phase diagram: large-$N$ approach\label{sec:Large--approach}}

As discussed in the previous section, it is necessary to go beyond
the mean-field approximation in order to assess whether a vestigial
phase, characterized by a non-zero composite order $\left\langle \Psi^{n}\right\rangle \neq0$
or $\left\langle \psi^{n}\right\rangle \neq0$, can emerge before
the onset of superconductivity, i.e. $\langle\boldsymbol{\Delta}\rangle=0$.
Fluctuations can be accounted for via different approaches. For instance,
renormalization-group (RG) methods have been widely used to study
actions of the form (\ref{eq:action1}) \citep{Millis2010,Qi_Xu2009,Fernandes2012,Christensen2018}.
An interesting outcome of a $\left(4-\varepsilon\right)$ RG calculation
is that the mean-field phase boundaries obtained in Fig. \ref{fig:Mean-field-phase-diagram}
remain unchanged \citep{Christensen2018}. However, to the best of
our knowledge, a consistent RG scheme that treats the vestigial orders
on an equal footing as the primary order has yet to be established. 

Here, we focus on the large-$N$ approach, which has been widely employed
to search for vestigial phases \citep{Golubovic1988,Hu_Kivelson2006,Hu_Kivelson2008,Fernandes2012,Nie2013,Si2016}.
The underlying assumption is that the number of order parameter components
can be extended from a given value $N_{0}$ ($N_{0}=2$ in the present
case) to an arbitrary $N\gg1$. The key point is that, in the $N\rightarrow\infty$
limit, the partition function associated with the action (\ref{eq:action1}),
$\mathcal{Z}=\int D\boldsymbol{\Delta}\,\exp\left(-S\left[\boldsymbol{\Delta}\right]\right)$,
can be computed exactly. The procedure consists of performing Hubbard-Stratonovich
transformations of the quartic terms of the action by introducing
appropriate auxiliary fields. The resulting action is quadratic in
the $\boldsymbol{\Delta}$ fields, and the calculation of the partition
function reduces to the straightforward evaluation of a Gaussian functional
integral. The remaining functional integral over the auxiliary fields
can be evaluated exactly in the $N\rightarrow\infty$ limit via the
saddle-point method. This leads to self-consistent equations for the
uniform auxiliary fields, which can then be solved to determine whether
any symmetry-breaking auxiliary field can condense still in the fluctuating
regime of the $\boldsymbol{\Delta}$ field (for more details on the
large-$N$ method, see Refs. \citep{Golubovic1988,Hu_Kivelson2006,Hu_Kivelson2008,Fernandes2012,Nie2013,Si2016}).
Note that while this method is controlled in the small parameter $1/N$,
there is no guarantee that the physical case $N_{0}$ is captured
by this expansion. While the large-$N$ method has been previously
used to investigate vestigial phases, the focus has been extensively
on the real bilinears only. Our goal in this section is to extend
this method to also include the complex bilinears.

By writing the quartic action in the form of Eq. (\ref{eq:S_int}),
it is natural to use the bilinears $\Psi^{A_{1g}}$, $\Psi^{A_{2g}}$,
and $\Psi^{B_{1g}}$ to introduce the Hubbard-Stratonovich fields.
The resulting self-consistent equations, however, will not bring any
information about the complex bilinears $\psi^{n}$. Self-consistent
equations for the latter could be obtained by exploiting the Fierz
identities (\ref{eq:Fierz_ids_D4h}). But the issue is that these
identities can be used to generate an infinite number of bilinear
representations for the quartic terms of the action, rendering a simultaneous
analysis of all possible vestigial orders intractable. This issue
is well known in Hartree-Fock-like solutions of interacting fermionic
Hamiltonians, since one only has access to the particular channels
in which the interactions are decomposed. Despite these shortcomings,
there is still useful information about the vestigial orders that
can be obtained from the large-$N$ approach, as we discuss below.
Note that some of the results obtained in this section recover results
previously reported elsewhere \citep{Golubovic1988,Hu_Kivelson2006,Hu_Kivelson2008,Fernandes2012,Nie2013,Si2016}.

\subsection{Derivation of the self-consistent equations}

We start by considering an arbitrary representation of the interaction
action in terms of the bilinears of $\boldsymbol{\Delta}$: 
\begin{align}
\mathcal{S}^{\mathrm{int}} & =\frac{1}{N}\int_{\mathsf{x}}\Big[\sum_{n\in\mathbb{G}_{\mathbb{R}}^{0}}U_{n}\,(\Psi^{n})^{2}+\sum_{n\in\mathbb{G}_{\mathbb{C}}}u_{n}\,|\psi^{n}|^{2}\Big]\,.\label{eq:S_int_largeN}
\end{align}
Here, the coefficients $U_{n}$ and $u_{n}$, which we dub interaction
parameters, are functions of the quartic Landau coefficients $u$,
$v$, and $w$ that depend on the particular representation chosen.
For example, for the representation shown in Eq. (\ref{eq:S_int}),
there are only three non-zero parameters: $U_{A_{1g}}=u$, $U_{A_{2g}}=v$
and $U_{B_{1g}}=w$. Quite generally, only a subset of all possible
$U_{n}$ and $u_{n}$ will be non-zero for a given representation.
Nevertheless, to keep the formalism general, we will keep them undetermined
for now. We perform the Hubbard-Stratonovich transformations
\begin{align}
1 & =\prod_{n\in\mathbb{G}_{\mathbb{R}}^{0}}\!\int\!\mathcal{D}\Phi^{n}\,\text{exp}\Big[\frac{N}{4U_{n}}\!\int_{\mathsf{x}}\!\Big(\Phi^{n}-\frac{2U_{n}}{N}\Psi^{n}\Big)^{2}\Big]\,,\label{eq:HS_real}\\
1 & =\prod_{n\in\mathbb{G}_{\mathbb{C}}}\!\int\!\mathcal{D}\!\left(\phi^{n},\bar{\phi}^{n}\right)\,\text{exp}\Big[\frac{N}{4u_{n}}\!\int_{\mathsf{x}}\Big|\phi^{n}-\frac{2u_{n}}{N}\psi^{n}\Big|^{2}\Big]\,,\label{eq:HS_complex}
\end{align}
to decouple the interaction action, which results in the introduction
of the auxiliary bosonic fields $\Phi^{n}$ and $\phi^{n}$. The action
then becomes
\begin{align}
\mathcal{S}_{N} & =N\mathcal{S}_{0}+\frac{1}{2}\frac{V}{T}\sum_{k,k^{\prime}}\hat{\boldsymbol{\Delta}}_{k}^{\dagger}\,\,\mathcal{G}_{k,k^{\prime}}^{-1}\,\,\hat{\boldsymbol{\Delta}}_{k^{\prime}}\,,\label{eq:SN}
\end{align}
where we introduced the momentum-space Nambu vector $\hat{\boldsymbol{\Delta}}_{k}=\left(\boldsymbol{\Delta}_{k},\bar{\boldsymbol{\Delta}}_{-k}\right)$.
Here, $\mathcal{S}_{0}$ depends only quadratically on the auxiliary
fields 
\begin{align}
\mathcal{S}_{0} & =-\sum_{n\in\mathbb{G}_{\mathbb{R}}^{0}}\frac{V}{4TU_{n}}\sum_{k}|\Phi_{k}^{n}|^{2}-\sum_{n\in\mathbb{G}_{\mathbb{C}}}\frac{V}{4Tu_{n}}\sum_{k}|\phi_{k}^{n}|^{2},\label{eq:SC}
\end{align}
and the Nambu Green's function is given by
\begin{align}
\mathcal{G}_{k,k^{\prime}}^{-1} & =2r_{0}\delta_{kk^{\prime}}M^{A_{1g}}+2\sum_{n\in\mathbb{G}_{\mathbb{R}}^{0}}\left[f_{\boldsymbol{k}}^{n}\delta_{kk^{\prime}}+\Phi_{k-k^{\prime}}^{n}\right]M^{n}\nonumber \\
 & \quad+\sum_{n\in\mathbb{G}_{\mathbb{C}}}\left[\phi_{k-k^{\prime}}^{n}(m^{n})^{\dagger}+\mathrm{h.c.}\right].\label{eq:lN_Greens_fcn}
\end{align}
where we defined the $M^{n}$, $m^{n}$ matrices (see Appendix \ref{App:Real-and-complex}):

\begin{align}
M^{A_{1g}} & =\tau^{0}\sigma^{0}/2, & m^{A_{1g}} & =\tau^{0}\sigma^{-}, & M^{A_{2g}} & =\tau^{y}\sigma^{z}/2,\nonumber \\
M^{B_{1g}} & =\tau^{z}\sigma^{0}/2, & m^{B_{1g}} & =\tau^{z}\sigma^{-},\nonumber \\
M^{B_{2g}} & =\tau^{x}\sigma^{0}/2, & m^{B_{2g}} & =\tau^{x}\sigma^{-},\label{eq:M_m-1}
\end{align}
with $\sigma^{\pm}=(\sigma^{x}\pm\mathsf{i}\sigma^{y})/2$. The form
factors $f_{\boldsymbol{k}}^{n}$ are given by Eq. (\ref{eq:f_gradient})
and $f_{\boldsymbol{k}}^{A_{2g}}=0$. Since the action (\ref{eq:SN})
is Gaussian in the superconducting field $\boldsymbol{\Delta}$, the
corresponding integration in the partition function can be carried
out, leading to an effective action that depends only on the auxiliary
fields
\begin{align}
\mathcal{S}_{\mathrm{eff}} & =N\Big\{\mathcal{S}_{0}+\frac{1}{4}\text{Tr}\log\left(\mathcal{G}^{-1}\right)\Big\}\,,\label{eq:S_eff}
\end{align}
where we dropped an unimportant constant. Importantly, the expectation
values of the bilinear combinations $\Psi^{n}$, $\psi^{n}$ are directly
proportional to the expectation values of the auxiliary fields via
\begin{align}
\langle\Psi^{n}\rangle & =\frac{N}{2U_{n}}\langle\Phi^{n}\rangle_{\Phi}, & \langle\psi^{n}\rangle & =\frac{N}{2u_{n}}\langle\phi^{n}\rangle_{\Phi}.\label{eq:expect_Phi}
\end{align}
Therefore, we identify the auxiliary fields as composite order parameters.
Note that we carefully distinguish the usual expectation value obtained
by integrating over $\int\!\mathcal{D}\!\left(\boldsymbol{\Delta},\bar{\boldsymbol{\Delta}}\right)$
and the expectation value with respect to the auxiliary fields $\langle\mathcal{O}\rangle_{\Phi}=\mathcal{Z}_{\Phi}^{-1}\int\!\mathcal{D}\!\left(\Phi^{n},\phi^{n},\bar{\phi}^{n}\right)\mathcal{O}\exp\left[-\mathcal{S}_{\mathrm{eff}}\right]$,
where $\mathcal{Z}_{\Phi}$ is the contribution to the partition function
that depends on the auxiliary fields only. 

In the limit $N\rightarrow\infty$, the prefactor $N$ in Eq. (\ref{eq:S_eff})
justifies a saddle-point analysis. Technically, this means that we
expand the effective action (\ref{eq:S_eff}) up to second order around
the homogeneous field values $\Phi_{0}^{n}$ and $\phi_{0}^{n}$ that
extremize $\mathcal{S}_{\mathrm{eff}}$:

\begin{align}
\mathcal{S}_{\mathrm{eff}} & \approx\mathcal{S}_{\mathrm{eff}}\big|_{0}+\frac{1}{2}\sum_{j,j^{\prime}}\int_{\mathsf{x},\mathsf{x}^{\prime}}\frac{\partial^{2}\mathcal{S}_{\mathrm{eff}}}{\partial X_{j}(\mathsf{x})\partial X_{j^{\prime}}(\mathsf{x}^{\prime})}\Big|_{0}\quad\times\nonumber \\
 & \qquad\qquad\qquad\,\left(X_{j}(\mathsf{x})-X_{j0}\right)\left(X_{j^{\prime}}(\mathsf{x}^{\prime})-X_{j^{\prime}0}\right).\label{eq:S_eff_expans}
\end{align}
For the sake of compactness, we used a single variable $\boldsymbol{X}=\big(\Phi^{n},\phi^{n},\bar{\phi}^{n}\big)$
to denote all composite fields. The Gaussian form (\ref{eq:S_eff_expans})
allows for a direct evaluation of the expectation values (\ref{eq:expect_Phi}),
yielding, to leading order, 
\begin{align}
\langle\Phi^{n}\rangle_{\Phi} & =\Phi_{0}^{n}, & \langle\phi^{n}\rangle_{\Phi} & =\phi_{0}^{n}.\label{eq:expect_Phi2}
\end{align}
Thus, Eqs. (\ref{eq:expect_Phi2}) and (\ref{eq:expect_Phi}) imply
that the homogeneous fields ($\Phi_{0}^{n}$, $\phi_{0}^{n}$) are
the composite order parameters associated with the non-trivial superconducting
bilinear combinations (\ref{eq:real_bilinears})-(\ref{eq:compl_bilinears}).
By definition, the homogeneous values $\Phi_{k}^{n}=\Phi_{0}^{n}\delta_{k,0}$
and $\phi_{k}^{n}=\phi_{0}^{n}\delta_{k,0}$ are given by $\frac{\partial\mathcal{S}_{\mathrm{eff}}}{\partial X_{j}}\big|_{0}=0$,
yielding the saddle-point equations 
\begin{align}
r_{0}-R_{0} & =-2U_{A_{1g}}\Pi^{A_{1g}},\label{eq:sadd_lN_1}\\
\Phi_{0}^{n} & =2U_{n}\Pi^{n}, & \!\!n & \in\mathbb{G}_{\mathbb{R}},\label{eq:sadd_lN_2}\\
\phi_{0}^{n} & =2u_{n}\pi^{n}, & \!\!n & \in\mathbb{G}_{\mathbb{C}}.\label{eq:sadd_lN_3}
\end{align}
Here, we have introduced the renormalized mass parameter $R_{0}\equiv r_{0}+\Phi_{0}^{A_{1g}}$,
as well as the integrals 
\begin{align}
\Pi^{n} & =\frac{T}{2V}\!\sum_{k}\!\mathrm{tr}\left[\mathcal{G}_{k,k}M^{n}\right], & \pi^{n} & =\frac{T}{2V}\!\sum_{k}\!\mathrm{tr}\left[\mathcal{G}_{k,k}m^{n}\right],\label{eq:Pi_n-1}
\end{align}
with $n\in\mathbb{G}_{\mathbb{R}}^{0}$ and $n\in\mathbb{G}_{\mathbb{C}}$,
respectively. The solution of the coupled saddle-point equations (\ref{eq:sadd_lN_1})-(\ref{eq:sadd_lN_3})
determines the renormalized mass $R_{0}$ and the set of composite
order parameters $\big(\Phi_{0}^{n},\phi_{0}^{n}\big)$ for a given
reduced temperature $r_{0}$. It is important to notice that equations
(\ref{eq:sadd_lN_2})-(\ref{eq:sadd_lN_3}) can only have a non-zero
solution for $\big(\Phi_{0}^{n},\phi_{0}^{n}\big)$ if the corresponding
interaction parameter is negative ($U_{n},u_{n}<0$), i.e. if that
particular vestigial-order channel is attractive. 

\subsection{Hierarchy of vestigial orders\label{subsec:largeN-instabilities}}

The onset of a non-zero $\big(\Phi_{0}^{n},\phi_{0}^{n}\big)$ via
a continuous transition in the regime where the primary order parameter
vanishes implies the existence of a vestigial phase. Of course, if
the composite transition is first-order, it may trigger a simultaneous
transition in the superconducting channel; we will get back to this
point in Sec. \ref{subsec:Vestigial-instabilities-versus}. To identify
the leading vestigial instability, we determine the highest critical
temperature associated with each composite order parameter. This is
achieved by computing the respective composite-order susceptibilities,
which can be evaluated in a straightforward way by means of the expansion
(\ref{eq:S_eff_expans}):
\begin{align}
\chi_{\Psi^{n}}^{\!}(R_{0}) & =\frac{\chi_{\Psi^{n}}^{(0)}}{1+2U_{n}\chi_{\Psi^{n}}^{(0)}}, & \chi_{\psi^{n}}^{\!}(R_{0}) & =\frac{\chi_{\psi^{n}}^{(0)}}{1+2u_{n}\chi_{\psi^{n}}^{(0)}}.\label{eq:susys}
\end{align}
Here, we defined the bare susceptibilities as 
\begin{align}
\chi_{\Psi^{n}}^{(0)} & =-\frac{\partial\Pi^{n}}{\partial\Phi_{0}^{n}}\big|_{\Phi_{0}^{n}=\phi_{0}^{n}=0}=\frac{T}{V}\sum_{k}\mathrm{tr}\left[\mathcal{G}_{k}^{0}M^{n}\mathcal{G}_{k}^{0}M^{n}\right],\label{eq:largeN_Chi0}\\
\chi_{\psi^{n}}^{(0)} & =-\frac{\partial\pi^{n}}{\partial\phi_{0}^{n}}\big|_{\Phi_{0}^{n}=\phi_{0}^{n}=0}=\!\frac{T}{2V}\!\sum_{k}\!\mathrm{tr}\left[\mathcal{G}_{k}^{0}m^{n}\mathcal{G}_{k}^{0}(m^{n})^{\dagger}\right]\!,\label{eq:largeN_chi0}
\end{align}
with $n\in\mathbb{G}_{\mathbb{R}}$ and $n\in\mathbb{G}_{\mathbb{C}}$,
respectively. The Green's function in the disordered regime is given
by $\mathcal{G}_{k}^{0}=\mathcal{G}_{k,k}\left[R_{0},\Phi_{0}^{n}=\phi_{0}^{n}=0\right]$. 

\begin{figure*}
\includegraphics[scale=0.8]{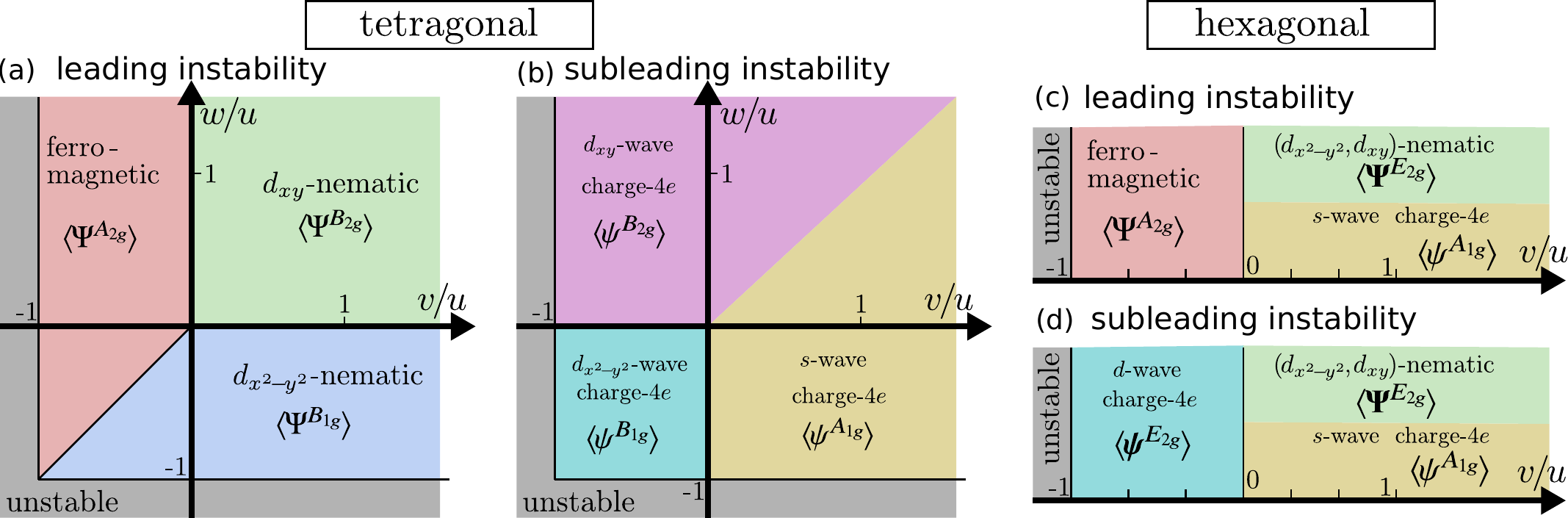}

\caption{Large-$N$ phase diagrams for the leading and subleading vestigial-order
instabilities associated with a primary two-component superconducting
phase in a system with tetragonal $\mathsf{D_{4h}}$ symmetry (panels
a,b) and hexagonal $\mathsf{D_{6h}}$ symmetry (panels c,d). As in
the mean-field phase diagram of Fig. \ref{fig:Mean-field-phase-diagram},
the parameter space is that spanned by the quartic Landau coefficients
of Eq. (\ref{eq:action1}). The vestigial charge-$4e$ instabilities
are always subleading with respect to either the vestigial nematic
or the vestigial ferromagnetic instability, except in the region $v>0$
of the hexagonal phase diagram, where the vestigial $s$-wave charge-$4e$
phase and the vestigial nematic phase are degenerate.\label{fig:largeN-phase-diagram}}
\end{figure*}

To make the analysis more transparent, we simplify the gradient terms
in Eq. (\ref{eq:S_grad}) by setting $f_{\boldsymbol{k}}^{B_{1g}}=f_{\boldsymbol{k}}^{B_{2g}}=0$,
which is equivalent to assuming that the superconducting fluctuations
are isotropic in the $(k_{x},\,k_{y})$ plane. In this case, $\mathcal{G}_{k}^{0}=G_{\boldsymbol{k}}^{0}\tau^{0}\sigma^{0}$
with $G_{\boldsymbol{k}}^{0}=\left[R_{0}+\mathsf{d}_{0}(k_{x}^{2}+k_{y}^{2})+\mathsf{d}_{z}k_{z}^{2}\right]^{-1}$,
and all bare susceptibilities (\ref{eq:largeN_Chi0})-(\ref{eq:largeN_chi0})
become identical:

\begin{equation}
\chi_{\Psi^{n}}^{(0)}=\chi_{\psi^{n}}^{(0)}=\frac{T_{0}}{V}\!\sum_{\boldsymbol{k}}\Big(G_{\boldsymbol{k}}^{0}\Big)^{2}=\frac{T_{0}}{8\pi d_{0}\sqrt{d_{z}}}\,R_{0}^{-1/2},\label{eq:chi_0}
\end{equation}
where, in the spirit of the Ginzburg-Landau expansion, we replaced
$T$ by $T_{0}$. By solving Eq. (\ref{eq:sadd_lN_1}) for $\Phi_{0}^{n}=\phi_{0}^{n}=0$,
we can find how the renormalized mass $R_{0}=R_{0}\left(r_{0}\right)>0$
depends on the reduced temperature $r_{0}$. The key point is that
$R_{0}$ vanishes at the (bare) superconducting transition temperature,
which we denote by $r_{c}$, and increases monotonically as a function
of the reduced temperature for $r_{0}>r_{c}$. Therefore, $\chi_{\Psi^{n}}^{(0)},\,\chi_{\psi^{n}}^{(0)}\rightarrow+\infty$
at the superconducting transition, which in turn implies that any
negative $U_{n},u_{n}$ will cause the susceptibility of the corresponding
composite order parameter to diverge before the onset of long-range
superconducting order, see Eqs. (\ref{eq:susys}). The reduced temperature
$r_{c}^{*}$ for which the divergence takes place is given by $R_{0}\left(r_{c}^{*}\right)=R_{0}^{*}$
with

\begin{equation}
R_{0}^{*}\equiv T_{0}^{2}U_{n}^{2}\,\big/\,\big(16\pi^{2}d_{0}^{2}d_{z}\big)\,.\label{eq:R0_rc}
\end{equation}
Of course, a similar expression holds for $u_{n}$. Since $R_{0}$
is a monotonically increasing function of $r_{0}$, it follows that
$r_{c}^{*}>r_{c}$. In fact, $r_{c}^{*}$ can be found by substituting
Eq. (\ref{eq:R0_rc}) in the self-consistent equation (\ref{eq:sadd_lN_1}): 

\begin{align}
r_{c}^{*} & =r_{c}+R_{0}^{*}+\frac{T_{0}U_{A_{1g}}}{2\pi\mathsf{d}_{0}\sqrt{\mathsf{d}_{z}}}\sqrt{R_{0}^{*}}\label{eq:rc_star0}\\
 & =R_{0}^{*}-\frac{T_{0}U_{A_{1g}}}{2\pi\sqrt{\mathsf{d}_{0}\mathsf{d}_{z}}}\left(\Lambda-\sqrt{R_{0}^{*}/\mathsf{d}_{0}}\right),\label{eq:rc_star}
\end{align}
where we explicitly inserted $r_{c}=-2U_{A_{1g}}\Pi^{A_{1g}}\big|_{R_{0}=0}$
and the in-plane momentum cutoff $\Lambda\gg\sqrt{R_{0}^{*}/\mathsf{d}_{0}}$.
Using Eqs. (\ref{eq:R0_rc}) and (\ref{eq:rc_star}), we can gain
insight into which vestigial channel has the highest critical temperature
by determining the most negative interaction parameter, which we denote
by $\mathtt{U}_{\mathrm{min}}\equiv\mathrm{min}\left\{ U_{n},u_{n}\right\} $,
since the most negative interaction parameter will correspond to the
largest $R_{0}^{*}$. The issue is that the set $\left\{ U_{n},u_{n}\right\} $
is not unique, as it depends on which combination of Fierz identities
(\ref{eq:Fierz_ids_D4h}) is used to rewrite the interaction action
(\ref{eq:S_int}) in terms of bilinears. In fact, there are infinite
many $\left\{ U_{n},u_{n}\right\} $ sets, which makes the analysis
of determining the leading vestigial-phase instability within the
large-$N$ approach intractable. To proceed, we exploit the mean-field
results obtained in the previous section to impose reasonable restrictions
on the $\left\{ U_{n},u_{n}\right\} $ sets. We first divide the phase
diagram in three regions, corresponding to each of the three mean-field
superconducting ground states shown in Fig. \ref{fig:Mean-field-phase-diagram}(a).
For each region, we only allow representations of the interaction
action (\ref{eq:S_int}) in terms of the three bilinears that acquire
non-zero values for that ground state. Moreover, we only replace a
given bilinear by a combination of other bilinears according to the
Fierz identities.

This procedure yields a small number of $\left\{ U_{n},u_{n}\right\} $
sets. For instance, in the region of the phase diagram bounded by
$v,w>0$, where the $d_{xy}$-nematic superconducting state $\langle\boldsymbol{\Delta}\rangle\sim(1,1)$
is the ground state, we find three sets $\mathcal{U}=\left\{ U_{A_{1g}},\,U_{B_{2g}},\,u_{A_{1g}},\,u_{B_{2g}}\right\} $
given by:

\begin{align}
\mathcal{U}_{1}= & \left\{ u+v,\,-w,\,w-v,\,0\right\} , & \mathcal{U}_{2}= & \left\{ u+w,\,-v,\,0,\,v-w\right\} ,\nonumber \\
\mathcal{U}_{3}= & \left\{ u+v+w,\,0,\,-v,\,-w\right\} .
\end{align}
They can be rewritten in a more compact form by introducing a parameter
$\epsilon\in\big\{0,1,\frac{w}{w-v}\big\}$:

\begin{align}
U_{A_{1g}} & =u+\text{\ensuremath{\underbar{\ensuremath{\epsilon}}}}v+\epsilon w, & U_{B_{2g}} & =-\text{\ensuremath{\underbar{\ensuremath{\epsilon}}}}w-\epsilon v,\nonumber \\
u_{A_{1g}} & =\text{\ensuremath{\underbar{\ensuremath{\epsilon}}}}(w-v), & u_{B_{2g}} & =\epsilon(v-w),
\end{align}
where $\text{\ensuremath{\underbar{\ensuremath{\epsilon}}}}\equiv1-\epsilon$.
A straightforward comparison of the minimum values of the three sets
in the $v,w>0$ range gives $\mathtt{U}_{\mathrm{min}}=-w\in\{U_{B_{2g}},\,u_{B_{2g}}\}$
for $w>v$ and $\mathtt{U}_{\mathrm{min}}=-v\in\{U_{B_{2g}},\,u_{A_{1g}}\}$
for $w<v$. At first sight, these results seem to suggest that the
vestigial $d_{xy}$-nematic phase is degenerate with the vestigial
$d_{xy}$-wave ($s$-wave) charge-$4e$ phase for $w>v$ ($w<v$).
However, this is not the case because $U_{A_{1g}}$ is different in
the two situations. Consider for concreteness $w>v$: while $\mathtt{U}_{\mathrm{min}}=U_{B_{2g}}=-w$
is from the set $\mathcal{U}_{1}$, for which $U_{A_{1g}}=u+v$, $\mathtt{U}_{\mathrm{min}}=u_{B_{2g}}=-w$
is from the set $\mathcal{U}_{3}$, for which $U_{A_{1g}}=u+v+w$.
Now, as shown in Eq. (\ref{eq:rc_star}), the vestigial-phase transition
temperature $r_{c}^{*}$ depends on $U_{A_{1g}}$: the larger $U_{A_{1g}}$
is, the smaller $r_{c}^{*}$ is. Therefore, because $U_{A_{1g}}=u+v+w$
from set $\mathcal{U}_{3}$ is larger than $U_{A_{1g}}=u+v$ from
set $\mathcal{U}_{1}$, the $d_{xy}$-nematic phase is the leading
vestigial instability of the system, while the $d_{xy}$-wave charge-$4e$
phase is the sub-leading vestigial instability. The situation is analogous
for $w<v$, where the $d_{xy}$-nematic phase is the leading vestigial
instability and the $s$-wave charge-$4e$ phase, the subleading one.
We verified the validity of this semi-quantitative argument via a
direct computation of $r_{c}^{*}$.

The other regions of the phase diagram can be analyzed in a similar
fashion. In the region $w<\min(0,v)$, whose mean-field ground state
is the $d_{x^{2}-y^{2}}$-nematic superconducting state $\langle\boldsymbol{\Delta}\rangle\sim(1,0)$,
the three relevant sets of interaction parameters are parametrized
by:

\begin{align}
U_{A_{1g}} & \!=u+\epsilon v, & U_{B_{1g}} & \!=w-\text{\ensuremath{\underbar{\ensuremath{\epsilon}}}}v, & u_{A_{1g}} & \!=-\epsilon v, & u_{B_{2g}} & \!=\text{\ensuremath{\underbar{\ensuremath{\epsilon}}}}v,\label{eq:sets_B1g}
\end{align}
where $\epsilon\in\big\{0,1,\frac{v-w}{v}\big\}$. Finally, in the
$v<\min(0,w)$ region, associated with the TRSB chiral superconducting
ground state $\langle\boldsymbol{\Delta}\rangle\sim(1,\mathsf{i})$,
the sets of interaction parameters are given by:
\begin{align}
U_{A_{1g}} & \!=u+\epsilon w, & U_{A_{2g}} & \!=v-\text{\ensuremath{\underbar{\ensuremath{\epsilon}}}}w, & u_{B_{2g}} & \!=-\epsilon w, & u_{B_{1g}} & \!=\text{\ensuremath{\underbar{\ensuremath{\epsilon}}}}w,\label{eq:sets_chiral}
\end{align}
with $\epsilon$ acquiring the values $\big\{0,1,\frac{w-v}{w}\big\}$.
The resulting phase diagrams in Fig. \ref{fig:largeN-phase-diagram}(a)-(b)
for the leading and subleading vestigial instabilities are obtained
by computing the maximum transition temperature of each vestigial
phase, considering the sets of interaction parameters $\left\{ U_{n},u_{n}\right\} $
given above. In all cases, the real-valued composite order parameters
give the leading vestigial instability and the complex-valued ones,
the subleading vestigial phases. 

The same analysis can be performed for the case of a two-component
superconducting order parameter in a hexagonal system with point group
$\mathsf{D_{6h}}$, whose mean-field phase diagram was shown in Fig.
\ref{fig:Mean-field-phase-diagram}(b). In the $v<0$ region, for
which the mean-field ground state is the chiral superconductor, there
are two relevant sets of interaction parameters:

\begin{equation}
\left\{ U_{A_{1g}}=u,\,U_{A_{2g}}=v\right\} ,\quad\left\{ U_{A_{1g}}=u-v,\,u_{E_{2g}}=v\right\} \,.
\end{equation}
Similar to the tetragonal case, in the hexagonal case the real-valued
composite order parameter \textendash{} the ferromagnetic $\left\langle \Psi_{A_{2g}}\right\rangle $
\textendash{} wins over the complex-valued composite order parameter
\textendash{} the $d$-wave charge-$4e$ $\left\langle \boldsymbol{\psi}^{E_{2g}}\right\rangle $.
In the phase-diagram region where the nematic superconductor is the
mean-field ground state, $v>0$, there are three sets:

\begin{align}
\left\{ U_{A_{1g}}=u+v,\,U_{E_{2g}}=-v\right\}  & ,\quad\left\{ U_{A_{1g}}=u+v,\,u_{A_{1g}}=-v\right\} \,,\nonumber \\
\left\{ U_{A_{1g}}=u-v,\,u_{E_{2g}}=v\right\} \,.
\end{align}
Interestingly, in this regime, there is a degeneracy between the vestigial
nematic and $s$-wave charge-$4e$ states, as previously reported
in Ref. \citep{Fernandes2021}. The resulting phase diagrams for the
leading and subleading vestigial instabilities are shown in Fig. \ref{fig:largeN-phase-diagram}(c)-(d).

\section{Vestigial-orders phase diagram: variational approach\label{sec:Variational-Approach}}

The inability of the large-$N$ approach to treat on an equal footing
all possible composite order parameters motivates us to consider in
this section an alternative method: the variational approach. Although
such an approach, which has been previously employed to study real-valued
composite orders \citep{Fischer2016,Nie2017,Yip2022}, is uncontrolled,
it allows one to determine the leading and subleading vestigial instabilities
of the system of the action (\ref{eq:action1}) in a much less biased
way. The method is based on a trial action $\mathcal{S}_{0}$ that
contains the variational parameters. The underlying principle relies
on the general (convexity) inequality \citep{Moshe2003}
\begin{align}
\langle e^{-A}\rangle & \geq e^{-\langle A\rangle}.\label{eq:convex_ineq}
\end{align}
To apply it to our problem, we rewrite the partition function as

\begin{align}
\mathcal{Z} & =\int\mathcal{D}(\boldsymbol{\Delta},\bar{\boldsymbol{\Delta}})\,e^{-\mathcal{S}}=\mathcal{Z}_{0}\,\left\langle e^{-(\mathcal{S}-\mathcal{S}_{0})}\right\rangle _{0},\label{eq:part_func1}
\end{align}
where $\langle\mathcal{O}\rangle_{0}$ denotes the expectation value
with respect to $\mathcal{S}_{0}$,
\begin{align}
\langle\mathcal{O}\rangle_{0} & \equiv\frac{1}{\mathcal{Z}_{0}}\int\mathcal{D}(\boldsymbol{\Delta},\bar{\boldsymbol{\Delta}})\,\,\mathcal{O}\,\,e^{-\mathcal{S}_{0}},\label{eq:expect_value_A}
\end{align}
and $\mathcal{Z}_{0}\equiv\int\mathcal{D}(\boldsymbol{\Delta},\bar{\boldsymbol{\Delta}})\,e^{-\mathcal{S}_{0}}$
is the partition function of the trial action. Applying the relationship
(\ref{eq:convex_ineq}) to Eq. (\ref{eq:part_func1}) leads to the
inequality $F\leq F_{v}$ between the actual free energy $F=-T\log\mathcal{Z}$
and the variational free energy
\begin{align}
F_{v} & =-T\log\mathcal{Z}_{0}+T\langle\mathcal{S}-\mathcal{S}_{0}\rangle_{0}.\label{eq:var_free_energy}
\end{align}
The remaining task is to minimize the variational free energy $F_{v}$
with respect to the variational parameters to find the optimal solution
under the constraints imposed on the trial action $\mathcal{S}_{0}$.
The success of this method crucially depends on the chosen ansatz
for $\mathcal{S}_{0}$.

\subsection{Gaussian variational ansatz\label{subsec:Variational-Ansatz}}

One commonly used variational ansatz is a Gaussian trial action \citep{Moshe2003,Fischer2016,Nie2017}.
In this work, we employ the most general form of this ansatz by introducing
a variational parameter to each of the possible bilinear combinations
(\ref{eq:real_bilinears})-(\ref{eq:compl_bilinears}). In particular,
the trial action is given by
\begin{align}
\mathcal{S}_{0} & =\frac{1}{2}\frac{V}{T}\sum_{k}\hat{\boldsymbol{\Delta}}_{k}^{\dagger}\,\mathcal{G}_{k}^{-1}\,\hat{\boldsymbol{\Delta}}_{k},\label{eq:trial_action}
\end{align}
where, as in the previous section, we introduced $\hat{\boldsymbol{\Delta}}_{k}=(\boldsymbol{\Delta}_{k},\bar{\boldsymbol{\Delta}}_{-k})$,
with $\boldsymbol{\Delta}_{k}=(\Delta_{1k},\Delta_{2k})$. The trial
Green's function has a similar form as Eq. (\ref{eq:lN_Greens_fcn}),
with $R_{0}\equiv r_{0}+\Phi^{A_{1g}}$ and $f_{\boldsymbol{k}}^{B_{1g}}=f_{\boldsymbol{k}}^{B_{2g}}=0$:

\begin{align}
\mathcal{G}_{k}^{-1} & =2\Big(\!R_{0}+\!f_{\boldsymbol{k}}^{A_{1g}}\!\Big)M^{A_{1g}}+2\!\!\sum_{n\in\mathbb{G}_{\mathbb{R}}}\!\!\!\Phi^{n}M^{n}+\!\!\sum_{n\in\mathbb{G}_{\mathbb{C}}}\!\!\!\left(\bar{\phi}^{n}m^{n}\!+\mathrm{H.c.}\right).\label{eq:var_Greens_fcn}
\end{align}

It is important to highlight the differences with respect to the large-$N$
approach in Sec. \ref{sec:Large--approach}: in that case, the quantities
$\Phi^{n}$, $\phi^{n}$ were auxiliary bosonic fields introduced
via a Hubbard-Stratonovich transformation of the interaction action
(\ref{eq:S_int}), which in turn depended on the representation of
the latter in terms of the bilinears. As a result, only a few fields
could be introduced simultaneously, which was the main limitation
of the large-$N$ approach. On the other hand, in the variational
approach, because $\Phi^{n}$, $\phi^{n}$ are variational parameters,
we can introduce all of them simultaneously. The matrices $M^{n}$,
$m^{n}$ are those defined in Eq. (\ref{eq:M_m-1}).

Having set up the Gaussian trial action (\ref{eq:trial_action}),
it is straightforward to derive the variational free energy (\ref{eq:var_free_energy});
details are presented in Appendix \ref{App:variational-details}.
We obtain the free energy density $f_{v}=F_{v}/V$ (up to an unimportant
constant)
\begin{align}
f_{v} & =\!\frac{T}{2V}\!\!\sum_{k}\!\mathrm{tr}\log\left(\mathcal{G}_{k}^{-1}\right)+2\left[r_{0}\!-\!R_{0}\!+\!U_{A_{1g}}\Pi^{A_{1g}}\right]\!\Pi^{A_{1g}}\nonumber \\
 & \quad-\!2\!\!\sum_{n\in\mathbb{G}_{\mathbb{R}}}\left[\Phi^{n}\!-\!U_{n}\Pi^{n}\right]\Pi^{n}-\!\!\!\sum_{n\in\mathbb{G}_{\mathbb{C}}}\!\!\!\left[\left(\phi^{n}\!-\!u_{n}\pi^{n}\right)\bar{\pi}^{n}\!+\mathrm{c.c.}\right],\label{eq:f_phi}
\end{align}
where the integrals $\Pi^{n}$, $\pi^{n}$ are the same as those defined
in Eq. (\ref{eq:Pi_n-1}); we repeat their definitions here for the
sake of clarity:
\begin{align}
\Pi^{n} & =\frac{T}{2V}\sum_{k}\mathrm{tr}\left[\mathcal{G}_{k}M^{n}\right], & \pi^{n} & =\frac{T}{2V}\sum_{k}\mathrm{tr}\left[\mathcal{G}_{k}m^{n}\right].\label{eq:Pi_n}
\end{align}
In contrast to the large-$N$ approach, here the interaction parameters
$\left\{ U_{n},u_{n}\right\} $ are all simultaneously non-zero and
unambiguously defined: 
\begin{align}
U_{A_{1g}} & =3u+v+w, & u_{A_{1g}} & =u-v+w, & U_{A_{2g}} & =u+3v-w,\nonumber \\
U_{B_{1g}} & =u-v+3w, & u_{B_{1g}} & =u+v+w,\nonumber \\
U_{B_{2g}} & =u-v-w, & u_{B_{2g}} & =u+v-w.\label{eq:int_parameters}
\end{align}

\begin{figure*}
\includegraphics[scale=0.8]{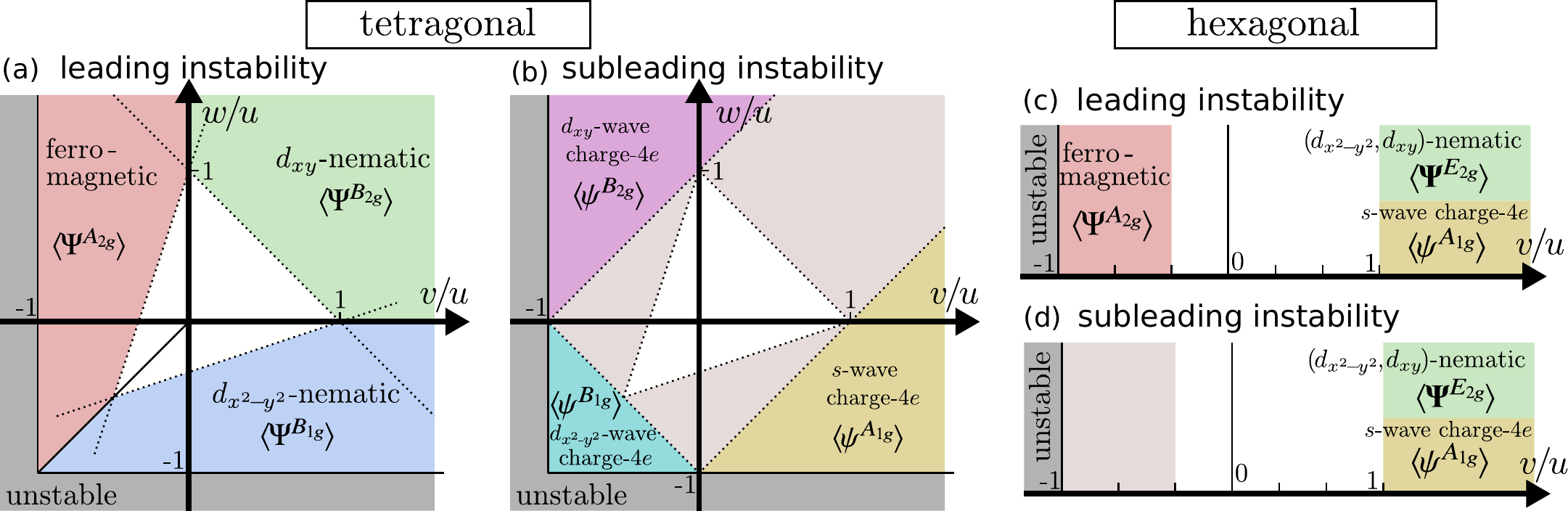}

\caption{Variational phase diagrams for the leading and subleading vestigial-order
instabilities associated with a primary two-component superconducting
phase in a system with tetragonal $\mathsf{D_{4h}}$ symmetry (panels
a,b) and hexagonal $\mathsf{D_{6h}}$ symmetry (panels c,d). The parameter
space is that spanned by the quartic Landau coefficients of Eq. (\ref{eq:action1}).
The only difference with respect to the large-$N$ phase diagrams
of Fig. \ref{fig:largeN-phase-diagram} is the existence of regions
where there is no vestigial instability (white region) or no subleading
vestigial instability (light-gray region). Outside of these regions,
the variational and large-$N$ results agree with each other. \label{fig:variational-phase-diagram}}
\end{figure*}

\subsection{Free-energy minimum\label{subsec:Free-energy-minimum}}

We now proceed to minimizing the free energy density $f_{v}$ in Eq.
(\ref{eq:f_phi}) with respect to the variational parameters, which
we collectively parametrize as $X_{i}\in\left\{ R_{0},\Phi^{n},\phi^{n},\bar{\phi}^{n}\right\} $,
where the vector $\boldsymbol{X}$ has dimension $L=\dim\mathbb{G}_{\mathbb{R}}^{0}+2\dim\mathbb{G}_{\mathbb{C}}$.
It is convenient to interpret the free energy as an implicit function
of the integrals $\Pi^{n}$ and $\pi^{n}$, which are themselves functions
of $X_{i}$, i.e. $\Pi^{n}\left(R_{0},\Phi^{n},\phi^{n},\bar{\phi}^{n}\right)$
and $\pi^{n}\left(R_{0},\Phi^{n},\phi^{n},\bar{\phi}^{n}\right)$.
As we show in Appendix \ref{App:variational-details}:

\begin{align}
\frac{\partial f_{v}}{\partial X_{i}}\Big|_{\Pi^{n},\pi^{n}} & =0\,.\label{eq:partial_fv}
\end{align}
As a result, the full derivative is given by
\begin{align}
\frac{df_{v}}{dX_{i}} & =\!\!\!\sum_{n\in\mathbb{G}_{\mathbb{R}}^{0}}\!\!\!V_{n}\frac{\partial\Pi^{n}}{\partial X_{i}}+\!\!\sum_{n\in\mathbb{G}_{\mathbb{C}}}\!\!\Big(v_{n}\frac{\partial\pi^{n}}{\partial X_{i}}\!+\!\bar{v}_{n}\frac{\partial\bar{\pi}^{n}}{\partial X_{i}}\Big),\label{eq:dfdx}
\end{align}
where we defined $V_{n}\equiv\frac{\partial f_{v}}{\partial\Pi^{n}}$
and $v_{n}\equiv\frac{\partial f_{v}}{\partial\pi^{n}}$. An explicit
evaluation gives:
\begin{align}
\!\!V_{A_{1g}} & =2\left(r_{0}\!-\!R_{0}\right)\!+\!4U_{A_{1g}}\Pi^{A_{1g}},\!\!\!\!\label{eq:VA1g}\\
\!\!V_{n} & =-2\Phi^{n}+4U_{n}\Pi^{n}, & \!\!n & \in\mathbb{G}_{\mathbb{R}},\label{eq:Vn}\\
\!\!\bar{v}_{n} & =-\phi^{n}+2u_{n}\pi^{n}, & \!\!n & \in\mathbb{G}_{\mathbb{C}}.\label{eq:vn}
\end{align}
Upon introducing the $L$-dimensional vectors $\boldsymbol{\mathcal{V}}$
and $\boldsymbol{\mathcal{P}}$ with $\mathcal{V}_{i}\in\left\{ V_{A_{1g}},V_{n},v^{n},\bar{v}^{n}\right\} $
and $\mathcal{P}_{i}\in\left\{ \Pi^{A_{1g}},\Pi^{n},\pi^{n},\bar{\pi}^{n}\right\} $,
Eqs. (\ref{eq:dfdx}) can be expressed as a matrix equation

\begin{equation}
\frac{df_{v}}{dX_{i}}=\sum_{j=1}^{L}\hat{\mathcal{P}}_{ij}\,\mathcal{V}_{j},\label{eq:dfdx_2}
\end{equation}
with the matrix elements $\hat{\mathcal{P}}_{ij}=\frac{\partial\mathcal{P}_{j}}{\partial X_{i}}$.
Because the matrix $\hat{\mathcal{P}}$ is generically non-singular,
i.e. $\det\hat{\mathcal{P}}\neq0$, the linear set of equations (\ref{eq:dfdx_2})
is only solved by the trivial solution $\boldsymbol{\mathcal{V}}=0$,
which is equivalent to:
\begin{align}
r_{0}-R_{0} & =-2U_{A_{1g}}\Pi^{A_{1g}},\label{eq:var_1}\\
\Phi^{n} & =2U_{n}\Pi^{n}, & \!\!n & \in\mathbb{G}_{\mathbb{R}},\label{eq:var_2}\\
\phi^{n} & =2u_{n}\pi^{n}, & \!\!n & \in\mathbb{G}_{\mathbb{C}}.\label{eq:var_3}
\end{align}

Recall that $\Pi^{n}$ and $\pi^{n}$ are functions of all variational
fields $R_{0},\Phi^{n},\phi^{n}$. These are the self-consistent equations
that determine the variational free-energy minimum. Although they
have the same functional form as the large-$N$ equations (\ref{eq:sadd_lN_1})-(\ref{eq:sadd_lN_3}),
the key difference is that the interaction parameters are unambiguously
determined by Eq. (\ref{eq:int_parameters}). We end this section
by noting that the variational parameters are indeed the composite
order parameters. Upon a direct computation of the bilinear expectation
values, we find (see Appendix \ref{App:variational-details}), 
\begin{align}
\langle\Psi_{q=0}^{n}\rangle & =2\Pi^{n}\overset{\text{(\ref{eq:var_2})}}{=}\frac{1}{U_{n}}\Phi^{n}, & \langle\psi_{q=0}^{n}\rangle & =2\pi^{n}\overset{\text{(\ref{eq:var_3})}}{=}\frac{1}{u_{n}}\phi^{n},\label{eq:expc_Cn_zetan_main}
\end{align}
for $n\in\mathbb{G}_{\mathbb{R}}$ and $n\in\mathbb{G}_{\mathbb{C}}$,
respectively. These expressions agree with those obtained for the
large-$N$ approach, Eq. (\ref{eq:expect_Phi}), upon setting $N=2$.

\subsection{Hierarchy of vestigial instabilities\label{subsec:Var_leading-instability}}

It is now straightforward to find the leading and subleading vestigial
instabilities of the system by linearizing the self-consistent variational
equations (\ref{eq:var_1})-(\ref{eq:var_3}) in the composite fields
$\Phi^{n},\phi^{n}$ independently. To leading order, the integral
expansions become $\Pi^{n}\approx-\chi_{\Psi^{n}}^{(0)}\Phi^{n}$
and $\pi^{n}\approx-\chi_{\psi^{n}}^{(0)}\phi^{n}$ where the bare
susceptibilities $\chi_{\Psi^{n}}^{(0)}$ and $\chi_{\psi^{n}}^{(0)}$
are defined in Eqs. (\ref{eq:largeN_Chi0}) and (\ref{eq:largeN_chi0}).
Substituting these expressions in the self-consistent equations leads
to the following instability condition in a given channel:

\begin{equation}
0=1+2U_{n}\chi_{\Psi^{n}}^{(0)}\:,\quad0=1+2u_{n}\chi_{\psi^{n}}^{(0)}\,,\label{eq:inst_conds_car}
\end{equation}
where $\chi_{\Psi^{n}}^{(0)}=\chi_{\psi^{n}}^{(0)}$ is given by the
same expression as in Eq. (\ref{eq:chi_0}). With the instability
conditions (\ref{eq:inst_conds_car}) and the first variational self-consistent
equation (\ref{eq:var_1}) having the same functional structure as
in the previous section, we also recover the same critical reduced
temperature $r_{c}^{*}$:
\begin{align}
r_{c}^{*} & =R_{0}^{*}-\frac{T_{0}U_{A_{1g}}}{2\pi\sqrt{\mathsf{d}_{0}\mathsf{d}_{z}}}\left(\Lambda-\sqrt{R_{0}^{*}/\mathsf{d}_{0}}\right),\label{eq:rc_star-1}
\end{align}
with $R_{0}^{*}\equiv T_{0}^{2}U_{n}^{2}\,\big/\,\big(16\pi^{2}d_{0}^{2}d_{z}\big)$,
and similar for $u_{n}$, cf. Eq. (\ref{eq:R0_rc}).

While the expression for the critical reduced temperature (\ref{eq:rc_star-1})
is the same as in the large-$N$ approach {[}Eq. (\ref{eq:rc_star}){]},
we emphasize two important differences: (i) the interaction parameters
$U_{n},u_{n}$ are unambiguously defined for all channels in Eq. (\ref{eq:int_parameters})
and (ii) the interaction parameter $U_{A_{1g}}$ is the same for all
channels. Consequently, we can find out the leading and subleading
vestigial instabilities by identifying the smallest and second smallest
negative interaction parameters $U_{n},u_{n}$.

The resulting phase diagram for the leading vestigial instability
is presented in Fig. \ref{fig:variational-phase-diagram}(a). When
compared to the large-$N$ phase diagram of Fig. \ref{fig:largeN-phase-diagram}(a),
the key difference is that the variational phase diagram displays
a region near the origin where no vestigial order is present {[}white
region in Fig. \ref{fig:variational-phase-diagram}(a){]}. This result
was previously found in Ref. \citep{Fischer2016} and is also consistent
with the findings of Ref. \citep{Yip2022}. It can be understood by
taking the $v,\,w\rightarrow0$ limit in Eq. (\ref{eq:int_parameters}),
which yields $U_{n}=u_{n}=u>0$, implying that all vestigial channels
are repulsive. In contrast, none of the $U_{n}$, $u_{n}$ sets obtained
in the large-$N$ approach had a contribution from $u$, which is
the coefficient of the squared trivial bilinear $\left(\Psi^{A_{1g}}\right)^{2}\propto\big(\left|\Delta_{1}\right|^{2}+\left|\Delta_{2}\right|^{2}\big)^{2}$
in the original action (\ref{eq:S_int}). As such, $u$ penalizes
large-amplitude superconducting fluctuations. In the variational approach,
such an energy penalty must be overcome by the energy gain of condensing
a non-trivial bilinear, which depends on combinations of $v$ and
$w$. Consequently, there are threshold values for the interaction
parameters $U_{n}$, $u_{n}$ below which no vestigial order emerges. 

This is an important qualitative distinction between the large-$N$
and variational results: in the former case, vestigial order is a
weak-coupling effect, in the sense that it emerges for any $\left|v\right|,\left|w\right|\ll u$,
whereas in the latter case it is a moderate-coupling effect, as it
requires $\left|v\right|,\left|w\right|\sim u$. Outside the white
region of the phase diagram of Fig. \ref{fig:variational-phase-diagram}(a),
the large-$N$ and variational phase diagrams predict the same leading
instabilities, which are all related to the condensation of real-valued
composite order parameters. The fact that two different methods give
the same results in these regions provides strong support for the
emergence of vestigial phases in these parameter ranges. 

The subleading vestigial instabilities can be readily obtained by
computing the second smallest negative interaction parameters $U_{n}$,
$u_{n}$ from Eq. (\ref{eq:int_parameters}) in the $\left(v/u,\,w/u\right)$
parameter space. Fig. \ref{fig:variational-phase-diagram}(b) shows
the resulting phase diagram. Besides the white region near the origin
where no vestigial instability can take place, there is a wider light-gray
region in which the system displays no subleading vestigial instability.
Outside of these regions, the phase diagram agrees with that obtained
in the large-$N$ approach {[}Fig. \ref{fig:largeN-phase-diagram}(b){]},
consisting of complex-valued charge-$4e$ composite order parameters
with different angular momentum. 

Extension of this analysis to the case of a $\mathsf{D_{6h}}$ hexagonal
two-component superconductor parameterized by Eq. (\ref{eq:case2})
is straightforward. In this case, the interaction parameters are given
by: 

\begin{align}
U_{A_{1g}} & =3u+v, & u_{A_{1g}} & =u-v, & U_{A_{2g}} & =u+3v,\nonumber \\
U_{E_{2g}} & =u-v, & u_{E_{2g}} & =u+v.\label{eq:int_parameters_D6h}
\end{align}

The phase diagrams corresponding to the leading and subleading vestigial
instabilities are shown in Figs. \ref{fig:variational-phase-diagram}(c)-(d).
Similarly to the $\mathsf{D_{4h}}$ tetragonal case, there are regions
of the phase diagram in which no vestigial channel is attractive (white
region) or only one vestigial channel is attractive (light-gray region).
Outside of these regions, the phase diagrams agree with those obtained
with the large-$N$ approach, see Figs. \ref{fig:largeN-phase-diagram}(c)-(d).
Interestingly, there is no subleading vestigial instability on the
$v<0$ side of the phase diagram, where only the real-valued ferromagnetic
composite order parameter can condense. On the $v>0$ side, the vestigial
nematic instability is always degenerate with the vestigial $s$-wave
charge-$4e$ instability, since $U_{E_{2g}}=u_{A_{1g}}$. Such a degeneracy,
which was also present in the large-$N$ approach, has been attributed
in Ref. \citep{Fernandes2021} to a hidden discrete symmetry of the
Ginzburg-Landau action that permutes operators in the gauge and in
the lattice sectors.

\subsection{Vestigial instabilities \emph{versus} vestigial phases \label{subsec:Vestigial-instabilities-versus}}

It is important to emphasize that the phase diagrams in Figs. \ref{fig:largeN-phase-diagram}
and \ref{fig:variational-phase-diagram} show the parameter regimes
in which there are attractive vestigial instabilities, which onset
at a (reduced) temperature $r_{c}^{*}$ that is larger than the superconducting
transition (reduced) temperature in the absence of vestigial order,
$r_{c}$. This is a necessary but not sufficient condition to ensure
the emergence of a vestigial phase preceding the primary superconducting
phase. The reason is because of the feedback effect of the condensation
of the composite order parameter on the superconducting fluctuations,
which renormalizes the superconducting transition temperature to larger
values, $\tilde{r}_{c}>r_{c}$. Thus, a vestigial phase characterized
by $\left\langle \Psi^{n}\right\rangle \neq0$ or $\left\langle \psi^{n}\right\rangle \neq0$
while $\langle\boldsymbol{\Delta}\rangle=0$ requires $r_{c}^{*}>\tilde{r}_{c}$. 

Within the variational approach, it would seem straightforward to
consider a modified ansatz with $\hat{\boldsymbol{\Delta}}_{k}$ replaced
by $\hat{\boldsymbol{\Delta}}_{k}-\hat{\boldsymbol{\delta}}$ in the
trial action (\ref{eq:trial_action}), where $\boldsymbol{\delta}=(\delta_{1},\delta_{2})$
denotes the superconducting variational parameter. The issue is that,
even for a simple one-component superconductor, which does not have
any vestigial orders, such a variational ansatz gives a first-order
superconducting transition. For completeness, this analysis is presented
in Appendix \ref{App:One-component-superconductor-in}; the formulas
for the two-component case are given in Appendix \ref{App:Two-component-superconductor}.
The bottom line is that this unphysical result indicates that the
modified trial action is not appropriate to describe the onset of
superconductivity, let alone the joint onset of superconducting and
composite orders. Additional work will be necessary to design an appropriate
ansatz. We note that a non-mean-field first-order superconducting
transition was also found in the seminal work \citep{Halperin1974},
where gauge-field fluctuations were considered within a large-$N$
approach. It was later realized that this effect holds only for type-I
superconductors \citep{Dasgupta1981,Kleinert1982,Mo2002,Kleinert2006}.
Whether these results are related to the issues encountered in the
variational approach remains to be determined.

Despite this shortcoming, one can still assess whether the condition
$r_{c}^{*}>\tilde{r}_{c}$ is self-consistently satisfied by the variational
equations (\ref{eq:var_1})-(\ref{eq:var_3}), which are identical
to the large-$N$ equations (\ref{eq:sadd_lN_1})-(\ref{eq:sadd_lN_3}).
In this formulation, $\tilde{r}_{c}$ is signaled by the vanishing
of one of the eigenvalues of the Green's function (\ref{eq:var_Greens_fcn})
{[}or, equivalently, (\ref{eq:lN_Greens_fcn}){]} evaluated at zero
momentum, i.e. $\mathrm{det}\,\mathcal{G}_{k=0}^{-1}\left(\tilde{r}_{c}\right)=0$.
That condition ensures that the superconducting susceptibility is
divergent. When only one of the composite order parameters condenses,
say $\Phi^{n}$, the latter condition is met when $R_{0}=\left|\Phi^{n}\right|$.
Therefore, as long as the vestigial phase transition at $r_{c}^{*}$
is second order, i.e. $\Phi^{n}\left(r_{0}\rightarrow\left(r_{c}^{*}\right)^{-}\right)\rightarrow0$,
the vestigial instability will not trigger a simultaneous superconducting
instability, implying that a vestigial phase emerges. Even if the
vestigial phase transition is first-order, a vestigial phase appears
as long as the jump of the composite order parameter is not too large,
$\left|\Delta\Phi^{n}\right|<R_{0}^{*}$, with $R_{0}^{*}$ given
by Eq. (\ref{eq:R0_rc}). The determination of whether the vestigial
phase transition is second-order or first-order requires solving the
non-linear equations (\ref{eq:var_1})-(\ref{eq:var_3}). While a
systematic analysis of this problem is beyond the scope of our work,
important insight can be gained from previous studies of the equivalent
large-$N$ equations (\ref{eq:sadd_lN_1})-(\ref{eq:sadd_lN_3}). 

For the tetragonal $\mathsf{D_{4h}}$ case, the large-$N$ equations
for a single composite order parameter were analyzed in detail in
Ref. \citep{Fernandes2012} in the context of magnetically-driven
nematicity and, before that, in Refs. \citep{Golubovic1988,Hu_Kivelson2008}.
The outcome of the coupled vestigial and primary transitions was found
to depend not only on the quartic Landau coefficients, but also on
stiffness coefficients $\mathsf{d}_{0},\,\mathsf{d}_{z}$. Essentially,
systems that are more anisotropic, i.e. with $\mathsf{d}_{z}/\mathsf{d}_{0}\ll1$,
tend to display vestigial phases over wider parameter ranges. 

A small modification of the model leads to more ``universal'' results,
in the sense that they depend only on the ratio between the quartic
Landau coefficients. In this modified version of the model, the anisotropic
gradient term $f_{\boldsymbol{k}}^{A_{1g}}$ in Eq. (\ref{eq:f_gradient})
is replaced by an isotropic term $f_{\boldsymbol{k}}^{A_{1g}}=\mathsf{d}_{0}\boldsymbol{k}^{2}$,
but the dimensionality of the system $d$ is allowed to assume fractional
values $2\leq d\leq3$. As shown in Ref. \citep{Fernandes2012} (see
also Ref. \citep{Golubovic1988}), for a given vestigial instability
with attractive effective interaction $U_{n}<0$ or $u_{n}<0$, there
are three different regimes for the coupled vestigial and superconducting
phase transitions, which we denote here as: (i) type-I split transitions,
in which case the vestigial and superconducting instabilities are
split and second-order; (i) type-II split transitions, in which case
the vestigial and superconducting instabilities are split but one
of them is first-order; (iii) simultaneous transition, in which case
there is a single first-order vestigial plus superconducting transition.
The system's regime depends only on the ratio $\left|U_{n}\right|/U_{A_{1g}}$
and the dimensionality $d$ according to \citep{Fernandes2012}:

\begin{align}
\frac{\left|U_{n}\right|}{U_{A_{1g}}} & <3-d, &  & \textrm{type-I}\;\textrm{split},\label{eq:1st_2nd_A}\\
3-d<\frac{\left|U_{n}\right|}{U_{A_{1g}}} & <\frac{6-2d}{6-d}, &  & \textrm{type-II\;\textrm{split}},\label{eq:1st_2nd_B}\\
\frac{\left|U_{n}\right|}{U_{A_{1g}}} & >\frac{6-2d}{6-d}, &  & \textrm{simultaneous}\;\textrm{transitions}.\label{eq:1st_2nd_C}
\end{align}
\begin{figure}
\includegraphics[scale=0.675]{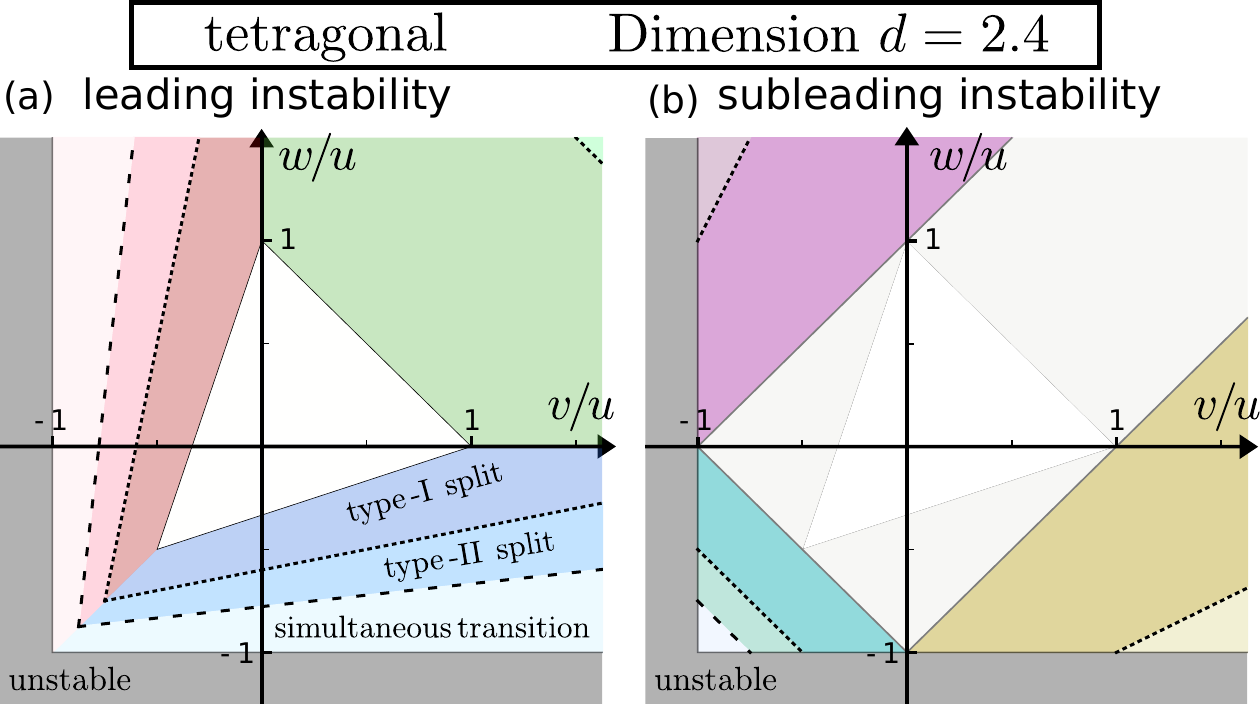}

\caption{Variational phase diagrams of Fig. \ref{fig:variational-phase-diagram}
for the tetragonal $\mathsf{D_{4h}}$ case with the phase boundaries
$\left|U_{n}\right|/U_{A_{1g}}=3-d$ (dotted line) and $\left|U_{n}\right|/U_{A_{1g}}=(6-2d)/(6-d)$
(dashed line) that determine the three different regimes for the coupled
vestigial and superconducing transitions (see main text): two split
second-order transitions (labeled type-I split); two split transitions
with one of them first-order (labeled type-II split); one simultaneous
first-order transition. For concreteness, here we set $d=2.4$. \label{fig:1st_2nd_order}}
\end{figure}

Of course, similar expressions hold for $\left|u_{n}\right|/U_{A_{1g}}$.
The key point is that a vestigial phase only exists in the type-I
split and type-II split regimes. In Fig. \ref{fig:1st_2nd_order},
we include the phase boundaries set by Eqs. (\ref{eq:1st_2nd_A})-(\ref{eq:1st_2nd_C})
separating these three regimes in the variational phase diagram of
Fig. \ref{fig:variational-phase-diagram}, for the case $d=2.4$.
Clearly, there is a wide region in parameter space where the vestigial
instability leads to a vestigial phase. Note that, upon increasing
the dimensionality $d$, the lines move closer to the origin, which
decreases the area of the phase diagram where vestigial phases exist.
In the fully isotropic case $d=3$, vestigial phases are absent, as
noted in Refs. \citep{Golubovic1988,Fernandes2012}.

For the hexagonal $\mathsf{D_{6h}}$ case, the vestigial nematic transition
is that of a 3-state Potts-model, which is first-order above its upper
critical dimension $d_{\mathrm{upper}}\lesssim3$. This problem was
analyzed in Ref. \citep{Hecker2018} for a system with lower trigonal
point-group symmetry $\mathsf{D_{3d}}$, which has the additional
stiffness coefficient $\mathsf{d}_{3}$ discussed below Eq. (\ref{eq:S_grad}).
A wide regime where vestigial nematic order emerges was reported for
a sufficiently anisotropic system. Note also that even in the dark-gray
regions of the phase diagrams in Figs. \ref{fig:largeN-phase-diagram}
and \ref{fig:variational-phase-diagram}, where the bare superconducting
transition is itself first-order, it is in principle possible for
a vestigial phase to be stabilized. However, our formalism does not
allow us to access these regions.

More broadly, the fact that there is more than one attractive vestigial
channel suggests that it is in principle possible for the system to
have sequential vestigal instabilities, giving rise to a cascade of
vestigial phases. The aforementioned issues with the modified Gaussian
variational ansatz that includes a non-zero superconducting order
parameter make a quantitative analysis challenging. On a qualitative
level, it is interesting to note that, in the cases studied here,
there is always a symmetry-allowed trilinear coupling between one
real-valued and two complex-valued bilinears. Specifically, for the
$\mathsf{D_{4h}}$ case, there are three such trilinear couplings
with coefficients $\tilde{\lambda}_{i}$, 

\begin{align}
\tilde{\mathcal{S}}_{1} & =\tilde{\lambda}_{1}\int_{\mathsf{x}}\Psi^{B_{1g}}\left(\psi^{A_{1g}}\bar{\psi}^{B_{1g}}+\bar{\psi}^{A_{1g}}\psi^{B_{1g}}\right),\label{eq:S1-1}\\
\tilde{\mathcal{S}}_{2} & =\tilde{\lambda}_{2}\int_{\mathsf{x}}\Psi^{B_{2g}}\left(\psi^{A_{1g}}\bar{\psi}^{B_{2g}}+\bar{\psi}^{A_{1g}}\psi^{B_{2g}}\right),\label{eq:S2-1}\\
\tilde{\mathcal{S}}_{3} & =\tilde{\lambda}_{3}\int_{\mathsf{x}}\Psi^{A_{2g}}\,\mathsf{i}\left(\bar{\psi}^{B_{2g}}\psi^{B_{1g}}-\bar{\psi}^{B_{1g}}\psi^{B_{2g}}\right),\label{eq:S3-1}
\end{align}
whereas for the $\mathsf{D_{6h}}$ case there is only one:
\begin{align}
\tilde{\mathcal{S}}_{1} & =\tilde{\lambda}_{1}\int_{\mathsf{x}}\boldsymbol{\Psi}^{E_{2g}}\cdot\left(\psi^{A_{1g}}\bar{\boldsymbol{\psi}}^{E_{2g}}+\bar{\psi}^{A_{1g}}\boldsymbol{\psi}^{E_{2g}}\right).\label{eq:trilin_coup_Eg}
\end{align}
Because of these couplings, if the complex-valued composite order
parameter associated with the subleading instability condenses inside
the leading vestigial phase, it will necessarily trigger a non-zero
complex-valued composite order parameter associated with the third
channel. However, as Fig. \ref{fig:variational-phase-diagram}(b)
or Eq. (\ref{eq:int_parameters_D6h}) shows, this third channel is
always repulsive. Due to this parasitic effect, the system would incur
an energy penalty if the subleading attractive complex-valued order
parameter were to condense inside the primary vestigial phase. 

\section{Discussion and conclusions \label{sec:Discussion-and-conclusions}}

In summary, we employed a group-theoretical formalism to classify
and investigate all possible vestigial orders that emerge in two-component
superconductors in systems with fourfold or sixfold/threefold rotational
symmetry. Our focus was to treat on an equal footing the widely investigated
real-valued ferromagnetic or nematic bilinears and the little-explored
complex-valued bilinears that describe $s$-wave, $d_{x^{2}-y^{2}}$-wave,
and $d_{xy}$-wave charge-$4e$ condensates. The large-$N$ and variational
calculations that we performed reveal that the real-valued vestigial-order
instabilities are always the leading ones, although the complex-valued
vestigial-order channels are attractive over wide regions of the parameter
space spanned by the quartic coefficients of the phenomenological
Ginzburg-Landau action. Only in the particular case of a hexagonal
system with quartic coefficient $v>0$ we found degenerate nematic
and charge-$4e$ vestigial instabilities, as first pointed out in
Ref. \citep{Fernandes2021}. In all other cases, the charge-$4e$
composite order was found to be subleading with respect to the ferromagnetic
or nematic composite orders. 

Our systematic comparison between the controlled large-$N$ method
with the uncontrolled variational method revealed important caveats
of both approaches in their ability to describe vestigial phases.
The large-$N$ method does not offer a way to treat on an equal footing
all possible real-valued and complex-valued bilinears, whereas the
variational method faces difficulties to account for the instability
of the primary superconducting order parameter. Notwithstanding these
shortcomings, we found wide regions in the parameter-space where the
hierarchy of instabilities obtained from both methods agreed with
each other, giving us confidence on the reliability of these findings.
One important qualitative difference between the two methods is that,
in the large-$N$ approach, the emergence of vestigial orders is a
weak-coupling effect, in that the Landau coefficients of the non-trivial
squared bilinears can be much smaller than the coefficient of the
trivial squared bilinear. On the other hand, in the variational approach,
it is a moderate-coupling effect, in that the non-trivial coefficients
need to be comparable to the trivial coefficient in order for the
vestigial channels to become attractive. This difference stems from
the distinct ways in which large-amplitude fluctuations are energetically
penalized in each scenario. Because both methods have intrinsic limitations
\textendash{} the physical $N$ in our problem is not large and the
variational action ansatz is arbitrary \textendash{} it will be interesting
to exactly solve this model via Monte Carlo simulations to elucidate
the validity of each approach. Previous Monte Carlo calculations on
a related Ginzburg-Landau model seem to be qualitatively consistent
with the large-$N$ results \citep{Vicari2005}.

We note that, by setting $f_{\boldsymbol{k}}^{B_{1g}}=f_{\boldsymbol{k}}^{B_{2g}}=0$
in Eq. (\ref{eq:S_grad}), the phase diagrams obtained in this work
neglected the in-plane anisotropic gradient terms that are allowed
in the Ginzburg-Landau action. Inclusion of these terms is expected
to cause minor changes in the phase boundaries. The most important
impact would be on the degeneracy between the nematic and charge-$4e$
orders in the $v>0$ region of the phase diagram of the hexagonal
system. Interestingly, in the large-$N$ approach of Ref. \citep{Fernandes2021},
these additional gradient terms were shown to remove the degeneracy
by actually favoring the charge-$4e$ vestigial phase. A more important
effect not considered here, which has been little explored in the
broader context of superconducting vestigial orders, is the coupling
to the electromagnetic gauge fields. These should be particularly
relevant for nearly-2D systems, where phase fluctuations can play
a more important role than amplitude fluctuations. A recent work showed
that the phase boundaries of the mean-field phase diagrams shown in
Fig. \ref{fig:Mean-field-phase-diagram} are fundamentally changed
when corrections due to electromagnetic fluctuations are included
\citep{Gali2022}. Their impact on the onset of vestigial phases deserve
further investigations.

From a broader theoretical standpoint, our work reveals that there
is a larger and relatively unexplored landscape of vestigial orders
that can potentially be realized in systems whose symmetry group $\mathcal{G}=\mathcal{G}_{\mathrm{int}}\otimes\mathcal{G}_{s}$
is the product of a space group and an internal group. Here, we focused
on the complex-valued charge-$4e$ bilinears of superconductors, which
transform non-trivially under the internal group $\mathcal{G}_{\mathrm{int}}=U(1)$.
In magnetic systems, with internal group $\mathcal{G}_{\mathrm{int}}=SU(2)$,
the bilinears that transform non-trivially would be vector and tensorial
composite order parameters. A well-known example of the latter is
the spin-nematic order parameter, which has been proposed to be realized
in certain frustrated magnets \citep{Blume1969,Andreev1984}. While
other types of vestigial tensorial spin orders were briefly discussed
in Ref. \citep{Fernandes2019}, a systematic investigation has not
been performed. Of course, the independent classification of magnetic
bilinears in terms of IRs of $\mathcal{G}_{\mathrm{int}}=SU(2)$ and
$\mathcal{G}_{s}$ is only meaningful if the spin-orbit coupling is
weak, which may constrain their realization in actual materials. These
considerations for non-trivial vestigial phases are also relevant
for systems with emergent continuous symmetries \textendash{} such
as twisted bilayer graphene, which under certain conditions is described
by a model with emergent spin-valley $SU(4)$ or $U(4)\otimes U(4)$
symmetry \citep{Kang2019,Bultinck2020,Bernevig2021,Chichinadze2022}.

Because vestigial phases are fluctuation-driven phenomena, they are
most likely to be observed in low-dimensional and/or unconventional
superconductors, since the fluctuation regime of conventional BCS
superconductors is very narrow. In this regard, several materials
have been recently reported or proposed to be multi-component nematic
or chiral unconventional superconductors, making them natural candidates
to search for vestigial orders. This is the case for the doped topological
insulator $A_{\mathrm{x}}\mathrm{Bi_{2}Se_{3}}$, with $A=\mathrm{Cu},\,\mathrm{Nb},\,\mathrm{Sr}$,
which has a nematic superconducting ground state \citep{Fu2014,Matano2015,Pan2016,Asaba2016,Venderbos2016a,Hecker2018}.
Recent experiments have found strong evidence for a vestigial nematic
order preceding the superconducting phase \citep{Tamegai2019,Cho2019}.
Twisted bilayer graphene has also been shown to display nematic superconductivity
\citep{Cao2021}, and hints of a possible vestigial nematic phase
were observed in anisotropic transport measurements. $\mathrm{CaSn_{3}}$
\citep{Siddiquee2022} and few-layer $\mathrm{NbSe_{2}}$ \citep{Hamill2021,Cho2022}
are other examples of materials whose pairing states are accompanied
by broken lattice rotational symmetry; however, at least in the latter,
the data does not favor an interpretation in terms of a multi-component
superconductor. The heavy-fermion material $\mathrm{UPt_{3}}$ is
a well-established candidate for chiral two-component $f$-wave superconductivity
\citep{Sauls1994,Schemm2014,Avers2020}, which could host vestigial
orders as well. The same holds for other compounds where time-reversal
symmetry-breaking (TRSB) superconductivity has been reported, most
notably $\mathrm{Sr_{2}RuO_{4}}$ \citep{Luke1998,Jia2006}, $\mathrm{URu_{2}Si_{2}}$
\citep{Kasahara2009,Balicas2013,Schemm2015}, pressurized $\mathrm{KV_{3}Sb_{5}}$
\citep{Guguchia2023}, and 4Hb-$\mathrm{TaS_{2}}$ \citep{Ribak2020}
(for a more comprehensive list, see Ref. \citep{Ghosh2020}). In these
cases, however, it remains unsettled whether the observed TRSB arises
from a symmetry-enforced two-component superconducting order parameter.
In 4Hb-$\mathrm{TaS_{2}}$, recent Little-Parks \citep{Almoalem2022}
and critical field \citep{Silber2022} experiments provide strong
support for such a scenario. On the other hand, in $\mathrm{Sr_{2}RuO_{4}}$,
which has been recently proposed to be a two-component singlet superconductor
\citep{Benhabib2021,Ghosh2021}, the observation of nodal quasi-particles
and the lack of specific heat signatures across the second superconducting
transition have been interpreted in terms of an accidental degeneracy
between two one-component superconductors transforming as different
IRs \citep{Kivelson2020,Willa_Hecker2021}. It is important to emphasize
that vestigial phases can emerge even in cases where the degeneracy
between two superconducting orders is not symmetry-enforced, but accidental,
as discussed in Ref. \citep{Willa2020}. One superconductor where
this might be the case is $\mathrm{UTe_{2}}$, which was also reported
to spontaneously break time-reversal symmetry \citep{Hayes2021}.
Since its orthorhombic $\mathsf{D_{2h}}$ point group does not support
two-dimensional IRs, a TRSB superconducting state would require two
nearly degenerate states; however, whether $\mathrm{UTe_{2}}$ is
a single or two-component superconductor remains unsettled \citep{Rosa2022}.
Signatures consistent with a vestigial TRSB order have been recently
reported in K-doped $\mathrm{BaFe_{2}As_{2}}$ \citep{Grinenko2021},
whose ground state has been proposed to be a TRSB $s+\mathsf{i}s$
state \citep{Stanev2010,Maiti2010}.

Overall, our work significantly expands the class of systems where
the elusive charge-$4e$ condensates may be realized \citep{Nozieres1982,Korshunov1985,Nozieres1998,Volovik1992,Wu2005,Aligia2005,Berg2009,Radzihovsky2009,Babaev2010,Agterberg2011,Radzihovsky2011,Moon2012,Agterberg2020,HongYao2017,Fernandes2021,Jian2021,Zeng2021,Gnezdilov_Wang2022}.
Experimentally, recent magnetoresistance oscillation data in the kagome
superconductor $\mathrm{CsV_{3}Sb_{5}}$ have been interpreted as
signatures of charge-$4e$ and charge-$6e$ states above the onset
of charge-$2e$ order \citep{Ge2022}. Theoretically, it has been
previously shown that pair-density waves \citep{Berg2009,Agterberg2011},
coupled $U(1)\times U(1)$ superconductors \citep{Babaev2010}, and
hexagonal nematic superconductors \citep{Fernandes2021,Jian2021}
are good candidates to display vestigial $s$-wave charge-$4e$ order.
Our results reveal that, in fact, there are normal-state instabilities
in the charge-$4e$ channel in any two-component superconductor. Interestingly,
this instability is not restricted to the $s$-wave channel, but includes
also exotic $d$-wave charge-$4e$ states, whose properties deserve
further theoretical investigations. The main issue is that, except
for the case of a hexagonal nematic superconductor, the various vestigial
charge-$4e$ instabilities found here are subleading with respect
to the vestigial nematic or ferromagnetic instabilities. One way in
which this hierarchy of vestigial instabilities can be reversed is
via disorder, as previously discussed in Ref. \citep{Fernandes2021}
in the context of hexagonal nematic superconductors. Quite generally,
the type of disorder that is most detrimental for a given ordered
state is a random distribution of conjugate fields of the corresponding
order parameter \textendash{} also known as ``random-field'' disorder
\citep{Vojta2019}. Charge-$4e$ order parameters are generally protected
from random-field type of disorder, as their conjugate fields are
not present in crystals or devices. In contrast, random strain and,
to a lesser extent, diluted magnetic impurities, are present in many
realistic settings, acting as random-field disorder for the nematic
and ferromagnetic order parameters, respectively. The ability to control
these types of disorder could enable the stabilization of the subleading
vestigial charge-$4e$ states. Even in a perfectly clean system, the
presence of a subleading attractive charge-$4e$ instability should
be manifested in the collective excitations of the leading vestigial
phase, opening another route to access this elusive state of matter.
\begin{acknowledgments}
We thank T. Birol, M. Christensen, L. Fu, and P. Orth for valuable
discussions. M.H. and R.M.F. were supported by the U. S. Department
of Energy, Office of Science, Basic Energy Sciences, Materials Sciences
and Engineering Division, under Award No. DE-SC0020045. R.W. and J.S.
were supported by the German Research Foundation (DFG) through CRC
TRR 288 \textquotedblleft ElastoQMat\textquotedblright , project A07.
\end{acknowledgments}

\appendix

\section{Group-theoretical formalism\label{App:Real-and-complex}}

In this Appendix, we present the group-theoretical framework that
yields the results presented in Sec. \ref{sec:Classification-of-bilinears}
of the main text for the bilinears of the two-component superconducting
order parameters in Eqs. (\ref{eq:case1})-(\ref{eq:case2}), which
live in the product group $\mathcal{G}=U(1)\otimes\mathcal{G}_{p}$,
with $\mathcal{G}_{p}=\mathsf{D_{4h}}$ or $\mathcal{G}_{p}=\mathsf{D_{6h}}$.
The approach is the same as in Ref. \citep{Fernandes2019}, but generalized
to include complex-valued bilinears.

We start by studying a standard one-component superconductor with
order parameter $\Delta$. All bilinears are trivial under the operations
of the point group, since the product of two one-dimensional IRs always
yields the trivial IR $A_{1g}$. Thus, it is enough to focus on the
transformation properties of the unitary group $\mathcal{G}=U(1)$,
whose IRs we denote by $\Gamma_{m}^{U}$ , with $m=\{0,\pm1,\pm2,\dots\}$.
If the order parameter $\Delta$ transforms according to the IR $\Gamma_{+1}^{U}$,
then its complex conjugate $\bar{\Delta}$ transforms as $\Gamma_{-1}^{U}$.
Thus, the reasonable representation of the order parameter is given
through the ``Nambu'' vector $\hat{\boldsymbol{\Delta}}=(\Delta,\bar{\Delta})$,
which transforms according to the two-dimensional representation $\Gamma_{\Delta}=\Gamma_{+1}^{U}\oplus\Gamma_{-1}^{U}$.
This notation becomes more transparent if we consider the $U(1)$
symmetry operations as rotations in the complex plane:
\begin{align}
\left(\begin{array}{c}
\Re\Delta^{\prime}\\
\Im\Delta^{\prime}
\end{array}\right) & =\left(\begin{array}{cc}
\cos\varphi & -\sin\varphi\\
\sin\varphi & \cos\varphi
\end{array}\right)\left(\begin{array}{c}
\Re\Delta\\
\Im\Delta
\end{array}\right).\label{eq:U1_rot}
\end{align}
The transformation relation (\ref{eq:U1_rot}) can be (block-) diagonalized
upon application of the unitary matrix 
\begin{align*}
U & =\frac{1}{\sqrt{2}}\left(\begin{array}{cc}
1 & 1\\
-\mathsf{i} & \mathsf{i}
\end{array}\right),
\end{align*}
which leads to the relation
\begin{align}
\hat{\boldsymbol{\Delta}}^{\prime} & =\mathcal{R}_{\Delta}(\varphi)\hat{\boldsymbol{\Delta}}.\label{eq:Delta_prime_rot}
\end{align}
Here, $\mathcal{R}_{\Delta}(\varphi)=\mathcal{R}_{+1}(\varphi)\oplus\mathcal{R}_{-1}(\varphi)$
and $\mathcal{R}_{m}(\varphi)=e^{\mathsf{i}m\varphi}$ is the transformation
matrix associated with the IR $\Gamma_{m}^{U}$, see Table \ref{tab:U1_group}.
The transformation relation (\ref{eq:Delta_prime_rot}) demonstrates
that the rotation of the two real components $(\Re\Delta,\Im\Delta)$
in the complex plane is properly described by means of the Nambu vector
$\hat{\boldsymbol{\Delta}}$ transforming as $\Gamma_{\Delta}$. 

Next, we consider bilinear combinations of $\hat{\boldsymbol{\Delta}}$.
From the decomposition of the product representation (\ref{eq:U(1)}),
$\Gamma_{\Delta}\otimes\Gamma_{\Delta}=2\Gamma_{0}^{U}\oplus\left(\Gamma_{+2}^{U}\oplus\Gamma_{-2}^{U}\right)$,
there are two bilinears associated with the trivial sector and two
with the $m=\pm2$ (i.e. charge-$4e$) sector. The bilinears can be
written as
\begin{align}
C^{(m)} & =\frac{1}{2}\hat{\boldsymbol{\Delta}}^{T}\lambda^{m}\hat{\boldsymbol{\Delta}},\label{eq:B_U}
\end{align}
with the $2\times2$ matrices $\lambda^{m}$ acting in Nambu space.
These matrices $\lambda^{m}$ are defined implicitly through the transformation
condition 
\begin{align}
\mathcal{R}_{\Delta}^{T}(\varphi)\lambda^{m}\mathcal{R}_{\Delta}(\varphi) & =\mathcal{R}_{m}(\varphi)\lambda^{m}, & \forall\varphi & \in[0,2\pi).\label{eq:SC_tau_cond}
\end{align}
Applying this condition, we find the matrices shown in Table \ref{tab:U1_group}.
Inserting these matrices into Eq. (\ref{eq:B_U}), we obtain the $N_{\Gamma_{\Delta}}=3$
bilinear components
\begin{align}
C^{(0)} & =|\Delta|^{2}, & C^{(+2)} & =\Delta^{2}, & C^{(-2)} & =\bar{\Delta}^{2},\label{eq:Phi_U}
\end{align}
transforming according to $\Gamma_{0}^{U}$, $\Gamma_{+2}^{U}$, and
$\Gamma_{-2}^{U}$, respectively. While the antisymmetric matrix associated
with the trivial IR $\Gamma_{0}^{U}\big|_{a}$ yields a vanishing
bilinear (\ref{eq:B_U}) in the present case, it plays a role when
multiple groups are involved, as it can be paired with another antisymmetric
matrix.
\begin{table}
\begin{centering}
\begin{tabular}{ccc}
\toprule 
$U(1)$ & $E$ & $\mathcal{R}_{m}(\varphi)$\tabularnewline
\midrule
\midrule 
$\Gamma_{m}^{U}$ & $1$ & $e^{\mathsf{i}m\varphi}$\tabularnewline
\bottomrule
\end{tabular}$\quad$%
\begin{tabular}{cc}
\toprule 
$U(1)$ & $\lambda^{m}$\tabularnewline
\midrule
\midrule 
$\Gamma_{0}^{U}\big|_{s}$ & $\sigma^{x}$\tabularnewline
\midrule 
$\Gamma_{0}^{U}\big|_{a}$ & $-\mathsf{i}\,\sigma^{y}$\tabularnewline
\midrule 
$\Gamma_{+2}^{U}$ & $\sigma^{0}+\sigma^{z}$\tabularnewline
\midrule 
$\Gamma_{-2}^{U}$ & $\sigma^{0}-\sigma^{z}$\tabularnewline
\bottomrule
\end{tabular}$\hfill$%
\begin{tabular}{cc}
\toprule 
$\mathsf{D_{4h}}$ & $\tau^{n,l}$\tabularnewline
\midrule
\midrule 
$A_{1g}$ & $\tau^{0}$\tabularnewline
\midrule 
$A_{2g}$ & $\tau^{y}$\tabularnewline
\midrule 
$B_{1g}$ & $\tau^{z}$\tabularnewline
\midrule 
$B_{2g}$ & $\tau^{x}$\tabularnewline
\bottomrule
\end{tabular}$\hfill$%
\begin{tabular}{cc}
\toprule 
$\mathsf{D_{6h}}$ & $\tau^{n,l}$\tabularnewline
\midrule
\midrule 
$A_{1g}$ & $\tau^{0}$\tabularnewline
\midrule 
$A_{2g}$ & $\tau^{y}$\tabularnewline
\midrule 
$E_{2g}$ & $(\tau^{z},-\tau^{x})$\tabularnewline
\bottomrule
\end{tabular}
\par\end{centering}
\caption{(left) The character table of the unitary group $U(1)$ and its IRs
$\Gamma_{m}^{U}$, $m\in\mathbb{Z}$. (middle-left to right) Matrices
associated with the bilinear decomposition in the cases of the one-component
SC {[}Eqs. (\ref{eq:SC_tau_cond}) and (\ref{eq:U(1)}){]} and of
two-component real-valued order parameters transforming as two-dimensional
IRs of the point groups $\mathsf{D_{4h}}$ and $\mathsf{D_{6h}}$
{[}Eqs. (\ref{eq:trans_cond_D4h}), (\ref{eq:D4h}) and (\ref{eq:D3d}){]}.
\label{tab:U1_group}}
\end{table}

We now proceed by constructing the bilinears of a real-valued two-component
order parameter $\boldsymbol{\eta}=(\eta_{1},\eta_{2})$ that transforms
according to the IRs $E_{g}$ and $E_{u}$ of the point group $\mathcal{G}_{p}=\mathsf{D_{4h}}$
or the IRs $E_{1g}$, $E_{2g}$, $E_{1u}$, and $E_{2u}$ of the point
group $\mathcal{G}_{p}=\mathsf{D_{6h}}$. The bilinears are defined
as
\begin{align}
C^{n,l} & =\boldsymbol{\eta}^{T}\tau^{n,l}\boldsymbol{\eta},\label{eq:B_D4h}
\end{align}
where $n$ denotes the IR within the product decompositions in Eqs.
(\ref{eq:D4h}) and (\ref{eq:D3d}) (see also Table \ref{tab:U1_group})
and $l=1,\dots,\dim n$. Like in the $U(1)$ case, the associated
$2\times2$ matrices $\tau^{n,l}$ are defined implicitly through
the transformation condition 
\begin{align}
\mathcal{R}_{E_{g}}^{T}(g)\tau^{n,l}\mathcal{R}_{E_{g}}(g) & =\sum_{l^{\prime}}\mathcal{R}_{n}(g)_{ll^{\prime}}\tau^{n,l^{\prime}}, & \forall g & \in\mathcal{G}_{p},\label{eq:trans_cond_D4h}
\end{align}
where $\mathcal{R}_{n}(g)$ denotes the transformation matrix for
the group element $g$ within the IR $n$. Solving this equation gives
the matrices shown in Table \ref{tab:U1_group}. Consequently, the
bilinears (\ref{eq:B_D4h}) become 
\begin{align}
\mathsf{D_{4h}} & : & \!\!\!\!C^{A_{1g}} & =\eta_{1}^{2}+\eta_{2}^{2}, & \!\!\!C^{B_{1g}} & =\eta_{1}^{2}-\eta_{2}^{2},\;\;C^{B_{2g}}=2\eta_{1}\eta_{2},\label{eq:bil_real_D4h}\\
\mathsf{D_{6h}} & : & \!\!\!\!C^{A_{1g}} & =\eta_{1}^{2}+\eta_{2}^{2}, & \!\!\!\boldsymbol{C}^{E_{2g}} & =\big(\eta_{1}^{2}-\eta_{2}^{2},-2\eta_{1}\eta_{2}\big).\label{eq:bil_real_D3d}
\end{align}
Note that the antisymmetric matrix associated with the $A_{2g}$ bilinear
yields zero, i.e. $C^{A_{2g}}=0$.

We are now in position to derive the bilinears of the two-component
superconducting order parameter $\boldsymbol{\Delta}=(\Delta_{1},\Delta_{2})$
of Eqs. (\ref{eq:case1})-(\ref{eq:case2}) by combining the bilinear
decompositions obtained for the groups $U(1)$ and $\mathsf{\mathcal{G}_{p}}$
studied above. Let us start with the $\mathcal{G}_{p}=\mathsf{D_{4h}}$
case; as explained before, the superconducting order parameter is
given in the Nambu representation by $\hat{\boldsymbol{\Delta}}=(\boldsymbol{\Delta},\bar{\boldsymbol{\Delta}})^{T}$,
where the four-component ``vector'' $\hat{\boldsymbol{\Delta}}$
transforms as the representation $\Gamma=\Gamma_{\Delta}\otimes E_{i}$
of the symmetry group $\mathcal{G}=U(1)\otimes\mathsf{D_{4h}}$. The
bilinears are given by 
\begin{align}
C^{(m),n} & =\frac{1}{2}\hat{\boldsymbol{\Delta}}^{T}\lambda^{m}\tau^{n}\hat{\boldsymbol{\Delta}},\label{eq:bilinear_U_Eu}
\end{align}
where $\lambda^{m}$ and $\tau^{n}$ act on the Nambu (i.e. gauge)
and the $E_{i}$ (i.e. lattice) sectors, respectively. Since these
matrices are defined implicitly by the conditions (\ref{eq:SC_tau_cond})
and (\ref{eq:trans_cond_D4h}), they are the same matrices shown before
in Table \ref{tab:U1_group}. Thus, to identify the bilinear components
$C^{(m),n}$, all we need to do is construct a ``multiplication table''
according to the bilinear decompositions in the two sectors:
\begin{align*}
\Gamma\otimes\Gamma & =\left(\Gamma_{\Delta}\otimes\Gamma_{\Delta}\right)\otimes\left(E_{i}\otimes E_{i}\right)\\
 & =\big(2\Gamma_{0}^{U}\oplus\Gamma_{+2}^{U}\oplus\Gamma_{-2}^{U}\big)\otimes\left(A_{1g}\oplus A_{2g}\oplus B_{1g}\oplus B_{2g}\right).
\end{align*}
Such a multiplication table is given in Table \ref{tab:bil_components}
of the main text. Out of the $16$ possible bilinear combinations,
only $N_{\Gamma}=10$ components are non-zero. Inserting the matrices
into Eq. (\ref{eq:bilinear_U_Eu}) gives

\begin{align}
\Psi^{A_{1g}} & \!=|\Delta_{1}|^{2}+|\Delta_{2}|^{2}, & \!\!\psi^{\!A_{1g}} & \!=\Delta_{1}^{2}+\Delta_{2}^{2}, & \!\!\Psi^{A_{2g}} & \!=\mathsf{i}\bar{\Delta}_{2}\Delta_{1}\!-\mathsf{i}\bar{\Delta}_{1}\Delta_{2},\nonumber \\
\Psi^{B_{1g}} & \!=|\Delta_{1}|^{2}-|\Delta_{2}|^{2}, & \!\!\psi^{\!B_{1g}} & \!=\Delta_{1}^{2}-\Delta_{2}^{2},\nonumber \\
\Psi^{B_{2g}} & \!=\bar{\Delta}_{1}\Delta_{2}\!+\bar{\Delta}_{2}\Delta_{1}, & \!\!\psi^{\!B_{2g}} & \!=2\Delta_{1}\Delta_{2},\label{eq:bilinears-1-1}
\end{align}
where we have employed the same notation as in the main text, i.e.
real-valued bilinears ($m=0$) are labeled as $\Psi^{n}$ and complex-valued
ones ($m=\pm2$), as $(\psi^{n},\bar{\psi}^{n})$. The expressions
(\ref{eq:bilinears-1-1}) are identical to those in Eq. (\ref{eq:bilinears_exp})
of the main text. Alternatively, one can exploit the property $\hat{\boldsymbol{\Delta}}^{T}=\hat{\boldsymbol{\Delta}}^{\dagger}\sigma^{x}$
to rewrite the bilinears (\ref{eq:bilinear_U_Eu}) as 
\begin{align}
\Psi^{n} & =\hat{\boldsymbol{\Delta}}^{\dagger}M^{n}\hat{\boldsymbol{\Delta}}, & \psi^{n} & =\hat{\boldsymbol{\Delta}}^{\dagger}m^{n}\hat{\boldsymbol{\Delta}}.\label{eq:Cn_zetan_App}
\end{align}
Here, the matrices $M^{n}$, $m^{n}$ are defined as:
\begin{align}
M^{A_{1g}} & =\tau^{0}\sigma^{0}/2, & m^{A_{1g}} & =\tau^{0}\sigma^{-}, & M^{A_{2g}} & =\tau^{y}\sigma^{z}/2,\nonumber \\
M^{B_{1g}} & =\tau^{z}\sigma^{0}/2, & m^{B_{1g}} & =\tau^{z}\sigma^{-},\nonumber \\
M^{B_{2g}} & =\tau^{x}\sigma^{0}/2, & m^{B_{2g}} & =\tau^{x}\sigma^{-},\label{eq:M_m}
\end{align}
with $\sigma^{\pm}=(\sigma^{x}\pm\mathsf{i}\sigma^{y})/2$, which
gives Eq. (\ref{eq:M_m-1}) in the main text. This is the notation
used in Secs. \ref{sec:Large--approach} and \ref{sec:Variational-Approach}
of the main text. The case $\mathcal{G}_{p}=\mathsf{D_{6h}}$ can
be treated in the same way. The only change is that the bilinears
denoted by $\Psi^{B_{1g}}$, $\Psi^{B_{2g}}$ and $\psi^{B_{1g}}$,
$\psi^{B_{2g}}$, which in the $\mathsf{D_{4h}}$ case transform as
two separate one-dimensional IRs, combine to transform as the same
two-dimensional IR, $\boldsymbol{\Psi}^{E_{2g}}=(\Psi^{B_{1g}},-\Psi^{B_{2g}})$
and $\boldsymbol{\psi}^{E_{2g}}=(\psi^{B_{1g}},-\psi^{B_{2g}})$.
All bilinears of the $\mathsf{D_{6h}}$ case are also displayed in
Table \ref{tab:bil_components}. 

\section{Derivation of the Variational free energy \label{App:variational-details}}

In this Appendix, we derive the expression for the variational free
energy (\ref{eq:f_phi}) and the corresponding self-consistent equations
presented in Sec. \ref{sec:Variational-Approach}. The Gaussian trial
action (\ref{eq:trial_action}) is given by
\begin{align}
\mathcal{S}_{0} & =\frac{1}{2}\frac{V}{T}\sum_{k}\hat{\boldsymbol{\Delta}}_{k}^{\dagger}\,\mathcal{G}_{k}^{-1}\,\hat{\boldsymbol{\Delta}}_{k}\,,\label{eq:trial_action_App}
\end{align}
with the inverse Green's function $\mathcal{G}_{k}^{-1}$ introduced
in Eq. (\ref{eq:var_Greens_fcn}) and repeated here for convenience:
\begin{align}
\mathcal{G}_{k}^{-1} & =2\Big(\!R_{0}+\!f_{\boldsymbol{k}}^{A_{1g}}\!\Big)M^{A_{1g}}+2\!\!\sum_{n\in\mathbb{G}_{\mathbb{R}}}\!\!\!\Phi^{n}M^{n}+\!\!\sum_{n\in\mathbb{G}_{\mathbb{C}}}\!\!\!\left(\bar{\phi}^{n}m^{n}\!+\mathrm{H.c.}\right).\label{eq:G_ansatz_app}
\end{align}
Our goal is to compute the variational free energy (\ref{eq:var_free_energy}),
or equivalently, the variational free energy density:

\begin{equation}
f_{v}=\frac{F_{v}}{V}=-\frac{T}{V}\,\log\mathcal{Z}_{0}+\frac{T}{V}\left\langle \mathcal{S}-\mathcal{S}_{0}\right\rangle _{0}\,,\label{eq:app_fv}
\end{equation}
where the expectation values are taken with respect to the trial action
$\mathcal{S}_{0}$ as introduced in Eq. (\ref{eq:expect_value_A}),
and $\mathcal{Z}_{0}\equiv\int\mathcal{D}(\boldsymbol{\Delta},\bar{\boldsymbol{\Delta}})\,e^{-\mathcal{S}_{0}}$.
For completeness, we also reproduce the initial action $\mathcal{S}=\mathcal{S}^{(2)}+\mathcal{S}^{\mathrm{int}}$,
Eq. (\ref{eq:action1}), with the second- and fourth-order contributions:
\begin{align}
\mathcal{S}^{(2)} & =\frac{V}{T}\sum_{k}\hat{\boldsymbol{\Delta}}_{k}^{\dagger}\left(f_{\boldsymbol{k}}^{A_{1g}}+r_{0}\right)M^{A_{1g}}\hat{\boldsymbol{\Delta}}_{k},\label{eq:S_2_app}\\
\mathcal{S}^{\mathrm{int}} & =\int_{\mathsf{x}}\Big[u\,\big(\Psi^{A_{1g}}\big)^{2}+v\,\big(\Psi^{A_{2g}}\big)^{2}+w\,\big(\Psi^{B_{1g}}\big)^{2}\Big],\label{eq:S_int_app}
\end{align}
where, in line with the ansatz (\ref{eq:G_ansatz_app}), the non-trivial
fluctuations $f_{\boldsymbol{k}}^{B_{1g}}=f_{\boldsymbol{k}}^{B_{2g}}=0$
have been set to zero. In the following, we compute separately the
three contributions to the free energy density (\ref{eq:app_fv}):
\begin{align}
f_{v}^{(0)} & =-\frac{T}{V}\,\log\mathcal{Z}_{0}, & f_{v}^{(2)} & =\frac{T}{V}\left\langle \mathcal{S}^{(2)}-\mathcal{S}_{0}\right\rangle _{0},\nonumber \\
f_{v}^{(4)} & =\frac{T}{V}\left\langle \mathcal{S}^{\mathrm{int}}\right\rangle _{0}.\label{eq:fv_s}
\end{align}

Before doing so, we emphasize the absence of the ambiguity caused
by the Fierz identities (\ref{eq:Fierz_ids_D4h}) which posed a problem
to the large-$N$ method. The interaction action (\ref{eq:S_int_app})
only enters into the variational free energy through the contribution
$f_{v}^{(4)}$. Here, however, because the expectation value is a
linear map, the Fierz relations are still intact. More explicitly,
if we would choose the interaction representation as in Eq. (\ref{eq:alt_Sint}),
we would compute 
\begin{align}
f_{v}^{(4)} & =\frac{T}{V}\int_{\mathsf{x}}\Big[\left(u+w\right)\left\langle \big(\Psi^{A_{1g}}\big)^{2}\right\rangle _{0}+\left(v-w\right)\left\langle \big(\Psi^{A_{2g}}\big)^{2}\right\rangle _{0}\nonumber \\
 & \qquad\quad-w\,\left\langle \big(\Psi^{B_{2g}}\big)^{2}\right\rangle _{0}\Big].\label{eq:f4_alt_app}
\end{align}
Meanwhile, the insertion of the Fierz relation 
\begin{equation}
\big(\Psi^{B_{2g}}\big)^{2}=\big(\Psi^{A_{1g}}\big)^{2}-\big(\Psi^{B_{1g}}\big)^{2}-\big(\Psi^{A_{2g}}\big)^{2},\label{eq:aux_Fierz-1}
\end{equation}
reduces the expression (\ref{eq:f4_alt_app}) to the free energy contribution
that follows from the representation (\ref{eq:S_int_app}). The same
is true for any other interaction representation. In other words,
all interaction representations lead to the same result, and, for
convenience, we choose to work with the representation (\ref{eq:S_int_app}).

The evaluation of the Gaussian integral in the partition function
gives

\begin{align}
\mathcal{Z}_{0} & =\prod_{k}\,\Big[\det\big(VT\mathcal{G}_{k}^{-1}\big)\Big]^{-1/2},\label{eq:Z0_eval}
\end{align}
and thus, the first free energy contribution becomes 
\begin{align}
f_{v}^{(0)} & =\frac{T}{2V}\sum_{k}\mathrm{tr}\,\log\left(\mathcal{G}_{k}^{-1}\right).\label{eq:fv_0_eval}
\end{align}
Here, we dropped an unimportant constant and used the identity $\log\,\det\left(\mathcal{G}_{k}^{-1}\right)=\mathrm{tr}\,\log\left(\mathcal{G}_{k}^{-1}\right)$. 

To derive the second- and fourth-order contributions in (\ref{eq:fv_s}),
it remains to evaluate the expectation values
\begin{align}
\left\langle \!\Psi_{q=0}^{n}\right\rangle _{0} & \!=\!\sum_{k}\!\left\langle \hat{\boldsymbol{\Delta}}_{k}^{\dagger}M^{n}\hat{\boldsymbol{\Delta}}_{k}\right\rangle _{0},\qquad\!\left\langle \psi_{q=0}^{n}\right\rangle _{0}\!=\!\sum_{k}\!\left\langle \hat{\boldsymbol{\Delta}}_{k}^{\dagger}m^{n}\hat{\boldsymbol{\Delta}}_{k}\right\rangle _{0},\nonumber \\
\left\langle \Psi_{q}^{n}\Psi_{-q}^{n}\right\rangle _{0} & =\sum_{k,k^{\prime}}\left\langle \big(\hat{\boldsymbol{\Delta}}_{k}^{\dagger}M^{n}\hat{\boldsymbol{\Delta}}_{k+q}\big)\big(\hat{\boldsymbol{\Delta}}_{k^{\prime}}^{\dagger}M^{n}\hat{\boldsymbol{\Delta}}_{k^{\prime}-q}\big)\right\rangle _{0}.\label{eq:psis_expc_vals}
\end{align}
Such expectation values containing products of $\hat{\boldsymbol{\Delta}}_{k}$
can conveniently be computed by means of a conjugate field $\hat{\boldsymbol{j}}_{k}=\left(\boldsymbol{j}_{k},\bar{\boldsymbol{j}}_{-k}\right)$
linearly coupled to the gap function via 
\begin{align}
\mathcal{S}_{j} & =-\sum_{k}\hat{\boldsymbol{j}}_{k}^{T}\hat{\boldsymbol{\Delta}}_{k}.\label{eq:S_j_aux}
\end{align}
Then, using the new partition function 
\begin{align}
\mathcal{Z}_{0}\left[\boldsymbol{j}\right] & =\int\mathcal{D}\left(\boldsymbol{\Delta},\bar{\boldsymbol{\Delta}}\right)\,e^{-\mathcal{S}_{0}-\mathcal{S}_{j}},\label{eq:part_func_hj}
\end{align}
the expectation value of any function $\mathcal{F}\big(\hat{\Delta}_{i,k}\big)$
of the gap components $\hat{\Delta}_{i,k}$ is given by
\begin{align}
\left\langle \mathcal{F}\big(\hat{\Delta}_{i,k}\big)\right\rangle _{0} & =\mathcal{F}\left(\frac{\delta}{\delta\hat{j}_{i,k}}\right)\left.\frac{\mathcal{Z}_{0}\left[\boldsymbol{j}\right]}{\mathcal{Z}_{0}\left[0\right]}\right|_{\boldsymbol{j}=0}.\label{eq:expt_A2}
\end{align}
The Gaussian integral evaluation of (\ref{eq:part_func_hj}) can be
performed in a straightforward way. Upon exploiting the Nambu-space
identities $\hat{\boldsymbol{\Delta}}_{k}=S^{x}\hat{\boldsymbol{\Delta}}_{-k}^{*}$,
$\hat{\boldsymbol{j}}_{k}=S^{x}\hat{\boldsymbol{j}}_{-k}^{*}$, and
$\mathcal{G}_{k}^{-1}=S^{x}\left(\mathcal{G}_{k}^{-1}\right)^{T}S^{x}$
, with the $4\times4$ matrix $S^{x}=\tau^{0}\sigma^{x}$, we find:

\begin{equation}
\frac{\mathcal{Z}_{0}\left[\boldsymbol{j}\right]}{\mathcal{Z}_{0}\left[0\right]}=\exp\Big(\frac{1}{2}\frac{T}{V}\sum_{k}\hat{\boldsymbol{j}}_{k}^{T}\mathcal{G}_{k}\hat{\boldsymbol{j}}_{k}^{*}\Big)\,,\label{eq:Z_j_eval}
\end{equation}
with $\mathcal{Z}_{0}\left[0\right]=\mathcal{Z}_{0}$ given in Eq.
(\ref{eq:Z0_eval}). Then, exploiting the aforementioned Nambu-space
identities one finds the expectation value 
\begin{align}
\langle\hat{\Delta}_{i,k}^{*}\hat{\Delta}_{j,k^{\prime}}\rangle_{0} & =\frac{\delta}{\delta\hat{j}_{i,k}^{*}}\frac{\delta}{\delta\hat{j}_{j,k^{\prime}}}\left.\frac{\mathcal{Z}_{0}\left[\boldsymbol{j}\right]}{\mathcal{Z}_{0}\left[0\right]}\,\right|_{\boldsymbol{j}=0}=\frac{T}{V}\delta_{k,k^{\prime}}\mathcal{G}_{k}^{ji},\label{eq:expc_DeltaDelta}\\
\langle\hat{\Delta}_{i,k}\hat{\Delta}_{j,k^{\prime}}\rangle_{0} & =\frac{T}{V}\delta_{k,-k^{\prime}}\mathcal{G}_{k}^{jl}S_{li}^{x},\label{eq:expc_DeltaDelta2}
\end{align}
and similarly, 
\begin{align}
\langle\hat{\Delta}_{ik}^{*}\hat{\Delta}_{jk^{\prime}}\hat{\Delta}_{i^{\prime}p}^{*}\hat{\Delta}_{j^{\prime}p^{\prime}}\rangle_{0} & =\frac{T^{2}}{V^{2}}\Big\{\delta_{k,-p}\delta_{k^{\prime},-p^{\prime}}S_{il^{\prime}}^{x}\mathcal{G}_{k}^{l^{\prime}i^{\prime}}\mathcal{G}_{k^{\prime}}^{j^{\prime}l}S_{lj}^{x}\nonumber \\
 & \!\!\!\!\!\!\!\!\!\!\!\!\!\!\!\!+\delta_{k,k^{\prime}}\delta_{p,p^{\prime}}\mathcal{G}_{k}^{ji}\mathcal{G}_{p}^{j^{\prime}i^{\prime}}+\delta_{k,p^{\prime}}\delta_{k^{\prime},p}\mathcal{G}_{k}^{j^{\prime}i}\mathcal{G}_{k^{\prime}}^{ji^{\prime}}\Big\}.\label{eq:expc_DDDD}
\end{align}
The summation over doubly occurring indices is implied. As the trial
action (\ref{eq:trial_action_App}) is chosen to describe the system
above the superconducting regime only, one obtains $\langle\hat{\boldsymbol{\Delta}}_{k}\rangle_{0}=0$.
Note that the fourth-order expectation value (\ref{eq:expc_DDDD})
could alternatively be computed using Wick's theorem 
\begin{align*}
\langle ABCD\rangle & =\langle AB\rangle\langle CD\rangle+\langle AC\rangle\langle BD\rangle+\langle AD\rangle\langle BC\rangle,
\end{align*}
and the expressions (\ref{eq:expc_DeltaDelta})-(\ref{eq:expc_DeltaDelta2}).

For the bilinear expectation values in (\ref{eq:psis_expc_vals}),
one directly obtains 
\begin{align}
\left\langle \Psi_{q=0}^{n}\right\rangle _{0} & =2\Pi^{n}, & \left\langle \psi_{q=0}^{n}\right\rangle _{0} & =2\pi^{n},\label{eq:bil_expc_vals_0}
\end{align}
with the integrals defined in Eq. (\ref{eq:Pi_n}), or explicitly
repeated 
\begin{align}
\Pi^{n} & =\frac{T}{2V}\sum_{k}\mathrm{tr}\left[\mathcal{G}_{k}M^{n}\right], & \pi^{n} & =\frac{T}{2V}\sum_{k}\mathrm{tr}\left[\mathcal{G}_{k}m^{n}\right].\label{eq:Pi_n-2}
\end{align}
The corresponding second-order contribution to the free energy density
in (\ref{eq:fv_s}) becomes 
\begin{align}
f_{v}^{(2)} & =2\left(r_{0}-R_{0}\right)\Pi^{A_{1g}}-2\sum_{n\in\mathbb{G}_{\mathbb{R}}}\Phi^{n}\Pi^{n}-\sum_{n\in\mathbb{G}_{\mathbb{C}}}\left(\phi^{n}\bar{\pi}^{n}+\bar{\phi}^{n}\pi^{n}\right).\label{eq:fv2_final}
\end{align}
The fourth-order free energy contribution in momentum space is explicitly
given by 
\begin{align}
f_{v}^{(4)} & =\frac{V}{T}\sum_{q}\Big[u\,\left\langle \Psi_{q}^{A_{1g}}\Psi_{-q}^{A_{1g}}\right\rangle _{0}+v\,\left\langle \Psi_{q}^{A_{2g}}\Psi_{-q}^{A_{2g}}\right\rangle _{0}\nonumber \\
 & \qquad+w\,\left\langle \Psi_{q}^{B_{1g}}\Psi_{-q}^{B_{1g}}\right\rangle _{0}\Big].\label{eq:f4_explicit}
\end{align}
Using the derived expressions (\ref{eq:expc_DDDD}) and (\ref{eq:psis_expc_vals}),
an individual term in (\ref{eq:f4_explicit}) can be simplified:
\begin{align}
\sum_{q}\langle\Psi_{q}^{n}\Psi_{-q}^{n}\rangle_{0} & =4\left(\Pi^{n}\right)^{2}+2\,\mathrm{tr}\left[M^{n}\text{\ensuremath{\underbar{\ensuremath{\mathcal{G}}}}}M^{n}\text{\ensuremath{\underbar{\ensuremath{\mathcal{G}}}}}\right].\label{eq:PsiPsi_expc_val}
\end{align}
Here, we used the relation $S^{x}\left(M^{n}\right)^{T}S^{x}=M^{n}$,
and defined $\text{\ensuremath{\underbar{\ensuremath{\mathcal{G}}}}}=\frac{T}{V}\sum_{k}\mathcal{G}_{k}$.
To further simplify the above relation, we invert the Green's function
matrix (\ref{eq:G_ansatz_app}),
\begin{align}
\mathcal{G}_{k} & =2\!\!\!\sum_{n\in\mathbb{G}_{\mathbb{R}}^{0}}\!\!\!G_{k}^{n}M^{n}+\!\!\sum_{n\in\mathbb{G}_{\mathbb{C}}}\!\![g_{k}^{n}(m^{n})^{\dagger}+\bar{g}_{k}^{n}m^{n}],\label{eq:inverted_inv_Greens_fcn}
\end{align}
with the elements defined through
\begin{align}
G_{k}^{n} & =\mathrm{tr}\left[M^{n}\mathcal{G}_{k}\right]/2, & g_{k}^{n} & =\mathrm{tr}\left[m^{n}\mathcal{G}_{k}\right]/2.\label{eq:G_g}
\end{align}
Note that the matrices are orthogonal with 
\begin{align}
\mathrm{tr}\left[M^{n_{1}}M^{n_{2}}\right]=\frac{1}{2}\mathrm{tr}\left[(m^{n_{1}})^{\dagger}m^{n_{2}}\right] & =\delta_{n_{1}n_{2}},\nonumber \\
\mathrm{tr}\left[M^{n_{1}}m^{n_{2}}\right]=\mathrm{tr}\left[m^{n_{1}}m^{n_{2}}\right] & =0.\label{eq:mats_ortho}
\end{align}
After momentum summation the Green's function matrix (\ref{eq:inverted_inv_Greens_fcn})
is expressed in terms of the integrals (\ref{eq:Pi_n-2}),
\begin{align}
\text{\ensuremath{\underbar{\ensuremath{\mathcal{G}}}}} & =\frac{T}{V}\sum_{k}\mathcal{G}_{k}=2\!\!\!\sum_{n\in\mathbb{G}_{\mathbb{R}}^{0}}\!\!\!\Pi^{n}M^{n}+\!\!\sum_{n\in\mathbb{G}_{\mathbb{C}}}\!\![\pi^{n}(m^{n})^{\dagger}+\bar{\pi}^{n}m^{n}].\label{eq:inverted_inv_Greens_fcn-1}
\end{align}
Because the matrices $M^{A_{1g}}$, $M^{A_{2g}}$ and $M^{B_{1g}}$
either commute or anti-commute with all other matrices $M^{n}$, $m^{n}$,
the matrix combination $M^{n}\text{\ensuremath{\underbar{\ensuremath{\mathcal{G}}}}}M^{n}$
inside the trace in Eq. (\ref{eq:PsiPsi_expc_val}) simplifies to
$M^{n}\text{\ensuremath{\underbar{\ensuremath{\mathcal{G}}}}}M^{n}=\frac{1}{4}\text{\ensuremath{\underbar{\ensuremath{\tilde{\mathcal{G}}}}}}.$
The Green's function $\text{\ensuremath{\underbar{\ensuremath{\tilde{\mathcal{G}}}}}}$
is still of the same type as Eq. (\ref{eq:inverted_inv_Greens_fcn-1})
but certain symmetry channels have acquired a relative minus sign,
dependent on which particular matrix $M^{n}$ was at play. This has
two important consequences. First, because of the orthogonality of
the matrices (\ref{eq:mats_ortho}), the trace in Eq. (\ref{eq:PsiPsi_expc_val})
only generates non-mixed terms of the sort $(\Pi^{n})^{2}$ or $|\pi^{n}|^{2}$.
Second, the relative minus signs are responsible for the eventual
interaction parameter combinations within the given symmetry channels. 

While $M^{A_{1g}}$ commutes with all matrices, $M^{A_{2g}}$ and
$M^{B_{1g}}$ commute and anti-commute according to

\begin{align*}
\Big[M^{A_{2g}},\left(\begin{array}{c}
m^{B_{1g}}\\
m^{B_{2g}}
\end{array}\right)\Big]_{-} & =\boldsymbol{0}, & \Big[M^{A_{2g}},\left(\begin{array}{c}
M^{B_{1g}}\\
M^{B_{2g}}\\
m^{A_{1g}}
\end{array}\right)\Big]_{+} & =\boldsymbol{0},\\
\Big[M^{B_{1g}},\left(\begin{array}{c}
m^{A_{1g}}\\
m^{B_{1g}}
\end{array}\right)\Big]_{-} & =\boldsymbol{0}, & \Big[M^{B_{1g}},\left(\begin{array}{c}
M^{A_{2g}}\\
M^{B_{2g}}\\
m^{B_{2g}}
\end{array}\right)\Big]_{+} & =\boldsymbol{0}.
\end{align*}
Let us exemplify the outlined ideas on one of the fourth-order free
energy terms in (\ref{eq:f4_explicit}):
\begin{align}
v\!\sum_{q}\!\left\langle \Psi_{q}^{A_{2g}}\Psi_{-q}^{A_{2g}}\right\rangle _{0} & \!=4v\big(\Pi^{A_{2g}}\big)^{2}+2v\,\mathrm{tr}\left[M^{A_{2g}}\text{\ensuremath{\underbar{\ensuremath{\mathcal{G}}}}}M^{A_{2g}}\text{\ensuremath{\underbar{\ensuremath{\mathcal{G}}}}}\right]\nonumber \\
 & \!\!\!\!\!\!\!\!\!\!\!\!\!\!\!\!\!\!\!\!\!\!\!\!\!\!\!\!=6v\big(\Pi^{A_{2g}}\big)^{2}\!+2v\big(\Pi^{A_{1g}}\big)^{2}\!-2v\big(\Pi^{B_{1g}}\big)^{2}\!-2v\big(\Pi^{B_{2g}}\big)^{2}\nonumber \\
 & \!\!\!\!\!\!\!\!\!\!\!\!\!\!\!\!\!\!\!\!\!\!\!\!\!\!\!\!-2v\left|\pi^{A_{1g}}\right|^{2}+2v\left|\pi^{B_{1g}}\right|^{2}+2v\left|\pi^{B_{2g}}\right|^{2}.\label{eq:v_fin_expr}
\end{align}
Finally, inserting the expression (\ref{eq:v_fin_expr}), and the
respective two other terms into the fourth-order free energy density
(\ref{eq:f4_explicit}) we obtain 
\begin{align}
f_{v}^{(4)} & =2\sum_{n\in\mathbb{G}_{\mathbb{R}}^{0}}U_{n}\left(\Pi^{n}\right)^{2}+\sum_{n\in\mathbb{G}_{\mathbb{C}}}2u_{n}|\pi^{n}|^{2}\,,\label{eq:fv4_final}
\end{align}
with the effective interaction parameters
\begin{align}
U_{A_{1g}} & =3u+v+w, & u_{A_{1g}} & =u-v+w, & U_{A_{2g}} & =u+3v-w,\nonumber \\
U_{B_{1g}} & =u-v+3w, & u_{B_{1g}} & =u+v+w,\nonumber \\
U_{B_{2g}} & =u-v-w, & u_{B_{2g}} & =u+v-w,\label{eq:int_parameters-1}
\end{align}
repeated in Eq. (\ref{eq:int_parameters}) of the main text. Combining
Eqs. (\ref{eq:fv_0_eval}), (\ref{eq:fv2_final}), and (\ref{eq:fv4_final})
then gives the variational free energy (\ref{eq:f_phi}) of the main
text, which we repeat here for convenience:

\begin{align}
f_{v} & =\!\frac{T}{2V}\!\!\sum_{k}\!\mathrm{tr}\log\left(\mathcal{G}_{k}^{-1}\right)+2\left[r_{0}\!-\!R_{0}\!+\!U_{A_{1g}}\Pi^{A_{1g}}\right]\!\Pi^{A_{1g}}\nonumber \\
 & \quad-\!2\!\!\sum_{n\in\mathbb{G}_{\mathbb{R}}}\left[\Phi^{n}\!-\!U_{n}\Pi^{n}\right]\Pi^{n}-\!\!\!\sum_{n\in\mathbb{G}_{\mathbb{C}}}\!\!\!\left[\left(\phi^{n}\!-\!u_{n}\pi^{n}\right)\bar{\pi}^{n}\!+\mathrm{c.c.}\right].\label{eq:f_phi-1}
\end{align}

Now, let us briefly prove the relation (\ref{eq:partial_fv}) of the
main text,
\begin{align}
\frac{\partial f_{v}}{\partial X_{i}}\Big|_{\Pi^{n},\pi^{n}} & =0\,,\label{eq:partial_fv-1}
\end{align}
stating that the partial derivative of the variational free energy
(\ref{eq:f_phi-1}) with respect to any of its variational parameters
$X_{i}\in\left\{ R_{0},\Phi^{n},\phi^{n},\bar{\phi}^{n}\right\} $
vanishes if $\Pi^{n}$ and $\pi^{n}$ are kept constant. The relation
(\ref{eq:partial_fv-1}) can directly be read off using the derivative
\begin{equation}
\frac{\partial}{\partial X_{i}}\frac{T}{2V}\!\!\sum_{k}\!\mathrm{tr}\log\left(\mathcal{G}_{k}^{-1}\right)=\begin{cases}
2\Pi^{A_{1g}} & ,X_{i}=R_{0}\\
2\Pi^{n} & ,X_{i}\in\left\{ \Phi^{n}\right\} \\
\bar{\pi}^{n} & ,X_{i}\in\left\{ \phi^{n}\right\} \\
\pi^{n} & ,X_{i}\in\left\{ \bar{\phi}^{n}\right\} 
\end{cases}\,.
\end{equation}
Therefore, we obtain Eq. (\ref{eq:partial_fv-1}), and the minimization
of the variational free energy follows the steps explained in Sec.
\ref{subsec:Free-energy-minimum}.

We finish this Appendix by demonstrating that the expectation value
of a bilinear reduces to its expectation value with respect to the
trial action within the variational approach, i.e. we derive Eq. (\ref{eq:expc_Cn_zetan_main})
of the main text. The expectation values of the (uniform) bilinears
are given by:
\begin{align}
\left\langle \Psi_{q=0}^{n}\right\rangle  & =\!\sum_{k}\left\langle \!\hat{\boldsymbol{\Delta}}_{k}^{\dagger}M^{n}\hat{\boldsymbol{\Delta}}_{k}\!\right\rangle , & \left\langle \psi_{q=0}^{n}\right\rangle  & =\!\sum_{k}\left\langle \!\hat{\boldsymbol{\Delta}}_{k}^{\dagger}m^{n}\hat{\boldsymbol{\Delta}}_{k}\!\right\rangle .
\end{align}
To proceed, we introduce the external conjugate fields $Y_{0}^{n}$
and $(y_{0}^{n},\bar{y}_{0}^{n})$ that couple linearly to the bilinear
combinations via:

\begin{align}
\mathcal{S}_{Y} & =-\sum_{n\in\mathbb{G}_{\mathbb{R}}}\Psi^{n}Y_{0}^{n}-\sum_{n\in\mathbb{G}_{\mathbb{C}}}\left(\psi^{n}\bar{y}_{0}^{n}+\bar{\psi}^{n}y_{0}^{n}\right)\,.\label{eq:S_Sigma}
\end{align}
The new partition function 
\begin{align}
\mathcal{Z}^{Y} & =\int\mathcal{D}(\boldsymbol{\Delta},\bar{\boldsymbol{\Delta}})\,e^{-\mathcal{S}-\mathcal{S}_{Y}},\label{eq:part_fcn_Y}
\end{align}
allows for a direct computation of the above expectation values

\begin{align}
\left\langle \Psi_{q=0}^{n}\right\rangle  & =\frac{\partial\log\mathcal{Z}^{Y}}{\partial Y_{0}^{n}}\Big|_{Y_{0}^{n}=y_{0}^{n}=0}, & \left\langle \psi_{q=0}^{n}\right\rangle  & =\frac{\partial\log\mathcal{Z}^{Y}}{\partial\bar{y}_{0}^{n}}\Big|_{Y_{0}^{n}=y_{0}^{n}=0},\label{eq:expc_Cn_zetan}
\end{align}
for $n\in\mathbb{G}_{\mathbb{R}}$ and $n\in\mathbb{G}_{\mathbb{C}}$,
respectively. We rewrite the new partition function (\ref{eq:part_fcn_Y})
to systematically correctly embed it into the framework of the variational
approach, cf. Eq. (\ref{eq:part_func1}), 
\begin{align}
\mathcal{Z}^{Y} & =\mathcal{Z}_{0}^{Y}\langle e^{-\left(\mathcal{S}-\mathcal{S}_{0}\right)}\rangle_{0}^{Y},\label{eq:part_fcn_Y-1}
\end{align}
where $\langle\mathcal{O}\rangle_{0}^{Y}$ denotes the expectation
value with respect to $\mathcal{S}_{0}+\mathcal{S}_{Y}$,
\begin{align}
\langle\mathcal{O}\rangle_{0}^{Y} & \equiv\big[\mathcal{Z}_{0}^{Y}\big]^{-1}\,\int\mathcal{D}(\boldsymbol{\Delta},\bar{\boldsymbol{\Delta}})\,\,\mathcal{O}\,\,e^{-\mathcal{S}_{0}-\mathcal{S}_{Y}},\label{eq:expect_value_A-2}
\end{align}
and $\mathcal{Z}_{0}^{Y}\equiv\int\mathcal{D}(\boldsymbol{\Delta},\bar{\boldsymbol{\Delta}})\,e^{-\mathcal{S}_{0}-\mathcal{S}_{Y}}$
is the associated partition function. Now, we employ the convexity
inequality (\ref{eq:convex_ineq}) on the expression (\ref{eq:part_fcn_Y-1}),
to derive the corresponding variational free energy in the presence
of the external field,
\begin{align}
f_{v}^{Y} & =-\frac{T}{V}\log\mathcal{Z}_{0}^{Y}+\frac{T}{V}\left\langle \mathcal{S}-\mathcal{S}_{0}\right\rangle _{0}^{Y}.\label{eq:Fv_Sigmasigma}
\end{align}
The derivative of the free energy (\ref{eq:Fv_Sigmasigma}) with respect
to the conjugate field gives the bilinear expectation values (\ref{eq:expc_Cn_zetan}).
To compute these derivatives, we note that $\mathcal{S}_{0}+\mathcal{S}_{Y}$
only depends on the variable combinations $\tilde{\Phi}^{n}=\Phi^{n}-\frac{T}{V}Y_{0}^{n}$
and $\tilde{\phi}^{n}=\phi^{n}-2\frac{T}{V}y_{0}^{n}$. Correspondingly,
we define $\tilde{f}_{v}^{Y}=f_{v}^{Y}-\frac{T}{V}\left\langle \mathcal{S}_{Y}\right\rangle _{0}^{Y}$
such that $\tilde{f}_{v}^{Y}$ also only depends on $\tilde{\Phi}^{n}$
and $\tilde{\phi}^{n}$. Then, derivatives of the type (\ref{eq:expc_Cn_zetan})
vanish by construction, for example:
\begin{align*}
\frac{\partial\tilde{f}_{v}^{Y}}{\partial Y_{0}^{n}}\Big|_{Y_{0}^{n}=y_{0}^{n}=0} & =\frac{\partial\tilde{f}_{v}^{Y}}{\partial\tilde{\Phi}^{n}}\frac{\partial\tilde{\Phi}^{n}}{\partial Y_{0}^{n}}\Big|_{Y_{0}^{n}=y_{0}^{n}=0}=-\frac{T}{V}\frac{\partial f_{v}}{\partial\Phi^{n}}=0.
\end{align*}
Thus, rewriting the free energy (\ref{eq:Fv_Sigmasigma}) as $f_{v}^{Y}=\tilde{f}_{v}^{Y}+\frac{T}{V}\left\langle \mathcal{S}_{Y}\right\rangle _{0}^{Y}$
is convenient as only its second term contributes to the expectation
values (\ref{eq:expc_Cn_zetan}). With $\mathcal{S}_{Y}$ being already
linear in $(Y_{0}^{n},y_{0}^{n},\bar{y}_{0}^{n})$, see Eq. (\ref{eq:S_Sigma}),
the derivatives are straightforwardly evaluated: 
\begin{align}
\left\langle \Psi_{q=0}^{n}\right\rangle  & =\left\langle \Psi_{q=0}^{n}\right\rangle _{0}=2\Pi^{n}, & \left\langle \psi_{q=0}^{n}\right\rangle  & =\left\langle \psi_{q=0}^{n}\right\rangle _{0}=2\pi^{n}.\label{eq:expc_Cn_zetan_2}
\end{align}
Here, we inserted the previously derived expressions (\ref{eq:bil_expc_vals_0}).
As expected, within the Gaussian variational approach, the expectation
values of the bilinears reduce to their trial expectation values. 

\section{One-component superconductor in the variational approach\label{App:One-component-superconductor-in}}

As discussed in the main text, the straightforward extension of the
variational ansatz to include the possibility of long-range superconducting
order yields unreasonable results. The issue is not particular to
vestigial phases or to multi-component superconductivity, as it emerges
already in the simpler case of a one-component superconductor, indicating
that this is likely a drawback of the variational ansatz itself. In
this Appendix, we show that the variational ansatz applied to a one-component
superconductor gives a first-order superconducting transition.

The action describing this system is given by 
\begin{align}
\mathcal{S} & =\frac{V}{T}\sum_{k}\bar{\Delta}_{k}\left(r_{0}+f_{\boldsymbol{k}}^{A_{1g}}\right)\Delta_{k}+\int_{x}u\,|\Delta|^{4},\label{eq:action2=00005D}
\end{align}
with the gradient term the same as before. For a one-component superconductor,
the Nambu basis $\hat{\boldsymbol{\Delta}}_{k}=(\Delta_{k},\bar{\Delta}_{-k})$
has only two components. As an additional variational parameter, we
introduce the non-zero expectation value of the superconducting order
parameter, $\delta=|\delta|e^{\mathsf{i}\varphi}$, which in the Nambu
basis becomes $\hat{\boldsymbol{\delta}}=(\delta,\bar{\delta})$.
The modified variational trial action is given by:

\begin{equation}
\mathcal{S}_{0}=\frac{1}{2}\frac{V}{T}\sum_{k}\left(\hat{\boldsymbol{\Delta}}_{k}^{\dagger}-\hat{\boldsymbol{\delta}}^{\dagger}\delta_{k,0}\right)\,\mathcal{G}_{k}^{-1}\,\left(\hat{\boldsymbol{\Delta}}_{k}-\hat{\boldsymbol{\delta}}\delta_{k,0}\right).\label{eq:SC_trial_action}
\end{equation}
Since there are only two bilinears allowed in this case, $\Phi^{A_{1g}}=R_{0}-r_{0}$
and $\phi^{A_{1g}}$, the Green's function simplifies to 
\begin{align}
\mathcal{G}_{k}^{-1} & =2\left(R_{0}+f_{\boldsymbol{k}}^{A_{1g}}\right)M^{A_{1g}}+\left[\phi^{A_{1g}}(m^{A_{1g}})^{\dagger}+\mathrm{H.c.}\right],\label{eq:inv_Greens_mat_1_comp}
\end{align}
with the $2\times2$ matrices $M^{A_{1g}}=\frac{1}{2}\sigma^{0}$
and $m^{A_{1g}}=\sigma^{-}$. For convenience, we parameterize the
inverted matrix by 
\begin{align}
\mathcal{G}_{k} & =2G_{k}^{A_{1g}}M^{A_{1g}}+\left[g_{k}^{A_{1g}}(m^{A_{1g}})^{\dagger}+\mathrm{H.c.}\right],\label{eq:inverted_inv_Greens_1comp}
\end{align}
and introduce $D^{A_{1g}}\equiv\hat{\boldsymbol{\delta}}^{\dagger}M^{A_{1g}}\hat{\boldsymbol{\delta}}=|\delta|^{2}$
and $d^{A_{1g}}\equiv\hat{\boldsymbol{\delta}}^{\dagger}m^{A_{1g}}\hat{\boldsymbol{\delta}}=\delta^{2}$.
To obtain the variational free energy, we follow the same steps as
detailed in App. \ref{App:variational-details}. The only two technical
differences are the fact that now the superconducting order parameter
$\boldsymbol{\delta}$ is non-zero, and that we want to derive an
expression for the superconducting susceptibility $\chi$. To accomplish
the latter, we introduce an external conjugate field $\hat{\boldsymbol{h}}_{k}=(h_{k},\bar{h}_{-k})$
that couples to the uniform superconducting order parameter in the
action via:

\begin{align}
\mathcal{S}_{h} & =-\hat{\boldsymbol{\Delta}}_{0}^{\dagger}\hat{\boldsymbol{h}}_{0}\,.\label{eq:Sh-1-1}
\end{align}
Then, the partition function subjected to the external field,
\begin{align}
\mathcal{Z}[\boldsymbol{h}] & =\int\mathcal{D}(\boldsymbol{\Delta},\bar{\boldsymbol{\Delta}})\,e^{-\mathcal{S}-\mathcal{S}_{h}},\label{eq:Z_h}
\end{align}
allows for a direct evaluation of the superconducting susceptibility:
\begin{align}
\chi_{ij} & =\frac{\partial^{2}\log\mathcal{Z}[\boldsymbol{h}]}{\partial\hat{h}_{i,0}^{*}\partial\hat{h}_{j,0}}\Bigg|_{\boldsymbol{h}=0}=\left\langle \hat{\Delta}_{i,0}\hat{\Delta}_{j,0}^{*}\right\rangle -\Big\langle\hat{\Delta}_{i,0}\Big\rangle\left\langle \hat{\Delta}_{j,0}^{*}\right\rangle .\label{eq:chi_formula-1-1}
\end{align}
The incorporation of the new partition function (\ref{eq:Z_h}) into
the variational framework follows the same logic outlined in the previous
section, around Eq. (\ref{eq:part_fcn_Y-1}). Correspondingly, the
variational free energy\textemdash in the presence of the external
field $\boldsymbol{h}$\textemdash becomes
\begin{align}
f_{v}[\boldsymbol{h}] & =-\frac{T}{V}\log\mathcal{Z}_{0}^{h}+\frac{T}{V}\left\langle \mathcal{S}-\mathcal{S}_{0}\right\rangle _{0}^{h},\label{eq:Fv_Sigmasigma-1}
\end{align}
with expectation values defined as
\begin{align}
\langle\mathcal{O}\rangle_{0}^{h} & \equiv\big[\mathcal{Z}_{0}^{h}\big]^{-1}\,\int\mathcal{D}(\boldsymbol{\Delta},\bar{\boldsymbol{\Delta}})\,\,\mathcal{O}\,\,e^{-\mathcal{S}_{0}-\mathcal{S}_{h}},\label{eq:expect_value_A-2-1}
\end{align}
and the associated partition function $\mathcal{Z}_{0}^{h}\equiv\int\mathcal{D}(\boldsymbol{\Delta},\bar{\boldsymbol{\Delta}})\,e^{-\mathcal{S}_{0}-\mathcal{S}_{h}}$.

The calculation of the free energy density (\ref{eq:Fv_Sigmasigma-1})
is tedious but straightforward. For convenience, we show the intermediate
results
\begin{align}
\mathcal{Z}_{0}^{h} & =\exp\left(\frac{T}{2V}\hat{\boldsymbol{h}}_{0}^{\dagger}\mathcal{G}_{0}\hat{\boldsymbol{h}}_{0}+\hat{\boldsymbol{h}}_{0}^{\dagger}\hat{\boldsymbol{\delta}}\right)\prod_{k}\Big[\!\det\left(VT\mathcal{G}_{k}^{-1}\right)\!\Big]^{-\frac{1}{2}}\!,\nonumber \\
\frac{\mathcal{Z}_{0}^{h}[\boldsymbol{j}]}{\mathcal{Z}_{0}^{h}} & =\exp\left(\frac{T}{2V}\sum_{k}\hat{\boldsymbol{j}}_{k}^{T}\mathcal{G}_{k}\hat{\boldsymbol{j}}_{k}^{*}+\hat{\boldsymbol{j}}_{k}^{T}\Big[\hat{\boldsymbol{\delta}}+\frac{T}{V}\mathcal{G}_{0}\hat{\boldsymbol{h}}_{0}\Big]\right),\label{eq:Z_j_oneComp}
\end{align}
where we introduced an auxiliary field $\boldsymbol{j}_{k}$ similar
as in Eq. (\ref{eq:S_j_aux}) that allows for the direct computation
of expectation values. For example, one obtains 
\begin{align}
\langle\hat{\boldsymbol{\Delta}}_{k}\rangle_{0}^{h} & =\frac{\delta}{\delta\hat{\boldsymbol{j}}_{k}}\left.\frac{\mathcal{Z}_{0}^{h}[\boldsymbol{j}]}{\mathcal{Z}_{0}^{h}}\right|_{\boldsymbol{j}=0}=\delta_{k,0}\left(\hat{\boldsymbol{\delta}}+\frac{T}{V}\mathcal{G}_{0}\hat{\boldsymbol{h}}_{0}\right),\label{eq:expc_Delta_OneC}\\
\langle\hat{\Delta}_{i,k}^{*}\hat{\Delta}_{j,k^{\prime}}\rangle_{0}^{h} & =\frac{\delta}{\delta\hat{j}_{i,k}^{*}}\frac{\delta}{\delta\hat{j}_{j,k^{\prime}}}\left.\frac{\mathcal{Z}_{0}^{h}[\boldsymbol{j}]}{\mathcal{Z}_{0}^{h}}\right|_{\boldsymbol{j}=0}\nonumber \\
 & =\frac{T}{V}\delta_{k,k^{\prime}}\mathcal{G}_{k}^{ji}+\delta_{k,0}\delta_{k^{\prime},0}\hat{\delta}_{h,i}^{*}\hat{\delta}_{h,j},\label{eq:expc_DD_OneC}
\end{align}
with $\hat{\boldsymbol{\delta}}_{h}=\hat{\boldsymbol{\delta}}+\frac{T}{V}\mathcal{G}_{0}\hat{\boldsymbol{h}}_{0}$,
and similarly for higher order terms. Eventually, one arrives at the
expression for the variational free energy (\ref{eq:Fv_Sigmasigma-1}).
For convenience, we separate the field-dependent and field-independent
parts, $f_{v}[\boldsymbol{h}]=f_{v}^{0}+\delta f_{v}^{h}$ with $f_{v}^{0}=f_{v}[\boldsymbol{0}]$,
and $\delta f_{v}^{h}=f_{v}[\boldsymbol{h}]-f_{v}[\boldsymbol{0}]$.
The first part is given by 
\begin{align}
f_{v}^{0} & =\frac{1}{2}\frac{T}{V}\sum_{k}\mathrm{tr}\log\left(\mathcal{G}_{k}^{-1}\right)+\left[r_{0}-R_{0}+2u\Pi^{A_{1g}}\right]\Pi^{A_{1g}}\nonumber \\
 & -\frac{1}{2}\left[\left(\phi^{A_{1g}}-u\pi^{A_{1g}}\right)\bar{\pi}^{A_{1g}}+\mathrm{c.c.}\right]+\frac{1}{2}\hat{\boldsymbol{\delta}}^{\dagger}\Big(\mathcal{K}_{0}^{-1}+\frac{1}{6}\mathcal{D}_{0}^{-1}\Big)\hat{\boldsymbol{\delta}},\label{eq:f_phi_1_comp}
\end{align}
with the integrals: 
\begin{align}
\Pi^{A_{1g}} & =\frac{T}{V}\sum_{k}\mathrm{tr}\left[\mathcal{G}_{k}M^{A_{1g}}\right]=\frac{T}{V}\sum_{k}G_{k}^{A_{1g}},\label{eq:1comp_PiA1g}\\
\pi^{A_{1g}} & =\frac{T}{V}\sum_{k}\mathrm{tr}\left[\mathcal{G}_{k}m^{A_{1g}}\right]=\frac{T}{V}\sum_{k}g_{k}^{A_{1g}},\label{eq:1comp_piA1g}
\end{align}
and the $2\times2$ matrices:
\begin{align}
\mathcal{K}_{0}^{-1} & =\left(r_{0}+4u\Pi^{A_{1g}}\right)2M^{A_{1g}}+2u\left[\bar{\pi}^{A_{1g}}m^{A_{1g}}+\mathrm{H.c.}\right],\label{eq:K0_m1_one_comp}\\
\mathcal{D}_{0}^{-1} & =8uD^{A_{1g}}M^{A_{1g}}+2u\left[\bar{d}^{A_{1g}}m^{A_{1g}}+\mathrm{H.c.}\right].\label{eq:D0_m1_one_comp}
\end{align}
The field-dependent part becomes
\begin{align}
\delta f_{v}^{h} & =-\frac{T^{2}}{V^{2}}\hat{\boldsymbol{h}}_{0}^{\dagger}\Big[\mathcal{G}_{0}-\frac{1}{2}\mathcal{G}_{0}\left(\mathcal{K}_{0}^{-1}+4uD^{A_{1g}}M^{A_{1g}}\right)\mathcal{G}_{0}\Big]\hat{\boldsymbol{h}}_{0}\nonumber \\
 & +\frac{T^{2}}{V^{2}}u\left(\hat{\boldsymbol{h}}_{0}^{\dagger}\mathcal{G}_{0}\hat{\boldsymbol{\delta}}\right)^{2}-\frac{T}{V}\hat{\boldsymbol{h}}_{0}^{\dagger}\Big[\mathbbm{1}-\mathcal{G}_{0}\Big(\mathcal{K}_{0}^{-1}+\frac{1}{3}\mathcal{D}_{0}^{-1}\Big)\Big]\hat{\boldsymbol{\delta}}.\label{eq:dfh_OneC}
\end{align}
Here, terms of the order $\mathcal{O}\big(\boldsymbol{h}^{3}\big)$
are neglected as they do not contribute to the susceptibility (\ref{eq:chi_formula-1-1}).
In between, we used the identity $\hat{\boldsymbol{\delta}}^{\dagger}\mathcal{D}_{0}^{-1}\hat{\boldsymbol{\delta}}=12u\left(D^{A_{1g}}\right)^{2}$. 

Let us now minimize the free energy (\ref{eq:f_phi_1_comp}) with
respect to the variational parameters. The corresponding derivatives
have the same structure as Eq. (\ref{eq:dfdx}). The saddle-point
equations for the variables $X_{i}\in\left\{ R_{0},\,\phi^{A_{1g}},\bar{\phi}^{A_{1g}}\right\} $
becomes:

\begin{equation}
\frac{df_{v}^{0}}{dX_{i}}=V_{A_{1g}}\frac{\partial\Pi^{A_{1g}}}{\partial X_{i}}+v_{A_{1g}}\frac{\partial\pi^{A_{1g}}}{\partial X_{i}}+\!\bar{v}_{A_{1g}}\frac{\partial\bar{\pi}^{A_{1g}}}{\partial X_{i}}=0,\label{eq:df_dX_app2}
\end{equation}
with:

\begin{align}
V_{A_{1g}} & =\frac{\partial f_{v}^{0}}{\partial\Pi^{A_{1g}}}=r_{0}-R_{0}+4u\left(\Pi^{A_{1g}}+D^{A_{1g}}\right),\label{eq:VA1_1comp}\\
v_{A_{1g}} & =\frac{\partial f_{v}^{0}}{\partial\pi^{A_{1g}}}=-\frac{1}{2}\bar{\phi}^{A_{1g}}+u\left(\bar{\pi}^{A_{1g}}+\bar{d}^{A_{1g}}\right).\label{eq:vA1_1comp}
\end{align}
The saddle-point equation related to the variational parameter $\delta$
is given by:
\begin{align}
\frac{df_{v}}{d\hat{\boldsymbol{\delta}}^{*}} & =\left[\mathcal{K}_{0}^{-1}+\frac{1}{3}\mathcal{D}_{0}^{-1}\right]\,\hat{\boldsymbol{\delta}}=0.\label{eq:SC_sadd_1comp}
\end{align}
Above $T_{c}$, we can set $\delta=0$, which automatically solves
the saddle-point equation (\ref{eq:SC_sadd_1comp}). Moreover, since
the matrix encapsulated by Eq. (\ref{eq:df_dX_app2}) is non-singular,
the saddle-point equations above $T_{c}$ give $V_{A_{1g}}=v_{A_{1g}}=0$.
Interestingly, $v_{A_{1g}}=0$ can only be solved by $\phi^{A_{1g}}=0$,
since $u>0$. Thus, a single-component superconductor does not support
a vestigial charge-$4e$ order. 

Below $T_{c}$, while the matrix encoded in Eq. (\ref{eq:df_dX_app2})
remains non-singular, there is an additional constraint imposed by
Goldstone's theorem, which reduces the number of variational parameters
from 3 to 2 by imposing a condition $R_{0}=R_{0}\left(\Phi^{A_{1g}},\phi^{A_{1g}}\right)$
. To see that, we need to first calculate the superconducting susceptibility.
From the field-dependent free energy density (\ref{eq:dfh_OneC}),
we can directly compute both the superconducting expectation value,
as well as the superconducting susceptibility (\ref{eq:chi_formula-1-1}).
One obtains 
\begin{align}
\langle\hat{\boldsymbol{\Delta}}_{0}\rangle & =\Big[1-\mathcal{G}_{0}\Big(\mathcal{K}_{0}^{-1}+\frac{1}{3}\mathcal{D}_{0}^{-1}\Big)\Big]\,\hat{\boldsymbol{\delta}}=\hat{\boldsymbol{\delta}}\,,\label{eq:expc_val_D_OneC1}\\
\frac{V}{T}\chi & =\mathcal{G}_{0}-\mathcal{G}_{0}\tilde{\mathcal{D}}_{0}^{-1}\mathcal{G}_{0}\,,\label{eq:susceptibility_1comp_app}
\end{align}
where, in the first line, we inserted the saddle-point equation (\ref{eq:SC_sadd_1comp})
and, in the second line, we defined the matrix 

\begin{align}
\tilde{\mathcal{D}}_{0}^{-1} & =-\mathcal{G}_{0}^{-1}+\mathcal{K}_{0}^{-1}+\mathcal{D}_{0}^{-1}\nonumber \\
 & =V_{A_{1g}}2M^{A_{1g}}+2\left[v_{A_{1g}}m^{A_{1g}}+\mathrm{H.c.}\right].\label{eq:D0_tilde_m1_one_comp}
\end{align}
Note that, above the superconducting transition temperature, where
the saddle-point equation gives $V_{A_{1g}}=v_{A_{1g}}=0$, we have
$\tilde{\mathcal{D}}_{0}^{-1}=0$, thus recovering the expected result
$\frac{V}{T}\chi=\mathcal{G}_{0}$, where the susceptibility only
diverges if $\det\mathcal{G}_{0}^{-1}=0$. For later convenience,
we ``rotate'' the susceptibility to the basis of real and imaginary
components of $\Delta_{k}$. This is accomplished by performing a
unitary transformation 
\begin{align}
\chi_{b} & =U_{b}^{\dagger}\,\chi\,U_{b}, & \text{with} &  & U_{b} & =\frac{1}{\sqrt{2}}\left(\begin{array}{cc}
1 & \mathsf{i}\\
1 & -\mathsf{i}
\end{array}\right).\label{eq:chi_b}
\end{align}

Inside the superconducting phase, Goldstone's theorem requires the
transverse component of the susceptibility (\ref{eq:susceptibility_1comp_app})
to be divergent at all temperatures, which translates into one eigenvalue
of $\mathcal{G}_{0}^{-1}$ being constrained to be zero. Diagonalizing
the inverse Green's function matrix gives
\begin{align}
U^{\dagger}\mathcal{G}_{0}^{-1}U & =\left(\begin{array}{cc}
\Lambda_{-} & 0\\
0 & \Lambda_{+}
\end{array}\right), & U & =\frac{1}{\sqrt{2}}\left(\begin{array}{cc}
1 & e^{\mathsf{i}\varphi_{A}}\\
-e^{-\mathsf{i}\varphi_{A}} & 1
\end{array}\right),\label{eq:G0_m1_diag_1comp}
\end{align}
where $\Lambda_{\pm}\equiv R_{0}\pm\left|\phi^{A_{1g}}\right|$ are
the eigenvalues and the phase $\varphi_{A}$ is defined according
to $\phi^{A_{1g}}=|\phi^{A_{1g}}|e^{\mathsf{i}\varphi_{A}}$. Since
$R_{0}\geq0$, only the eigenvalue $\Lambda_{-}$ can vanish, leading
to the constraint $R_{0}=|\phi^{A_{1g}}|$ inside the superconducting
state. As a result, $R_{0}$ and $\phi^{A_{1g}}$ are no longer independent,
which must be taken into account when minimizing the variational free
energy. The saddle-point equation is then given by: 
\begin{align}
\frac{df_{v}^{0}}{d\phi^{A_{1g}}} & =\frac{\partial f_{v}^{0}}{\partial\phi^{A_{1g}}}\Big|_{R_{0}}+\frac{\partial f_{v}^{0}}{\partial R_{0}}\frac{\partial R_{0}}{\partial\phi^{A_{1g}}}=0,
\end{align}
with a similar expression for $\bar{\phi}^{A_{1g}}$. Using the fact
that $\partial R_{0}/\partial\phi^{A_{1g}}=\frac{1}{2}e^{-\mathsf{i}\varphi_{A}}$,
we can recast the saddle point equation as:
\begin{align}
e^{\mathsf{i}\varphi_{A}}\frac{df_{v}^{0}}{d\phi^{A_{1g}}}\pm e^{-\mathsf{i}\varphi_{A}}\frac{df_{v}^{0}}{d\bar{\phi}^{A_{1g}}} & =-\frac{T}{2V}\mathrm{tr}\left[\tilde{\mathcal{D}}_{0}^{-1}Z^{\pm}\right]=0,\label{eq:KpKm_1comp}
\end{align}
where we introduced 
\begin{align}
Z^{\pm} & =\sum_{k}\mathcal{G}_{k}\Big[2M^{A_{1g}}\frac{1\pm1}{2}\!+e^{\mathsf{i}\varphi_{A}}(m^{A_{1g}})^{\dagger}\!\pm e^{-\mathsf{i}\varphi_{A}}m^{A_{1g}}\Big]\mathcal{G}_{k}.
\end{align}
Using the Green's function parametrization (\ref{eq:inverted_inv_Greens_1comp})
and the property $g_{k}^{A_{1g}}=-\big|g_{k}^{A_{1g}}\big|e^{\mathsf{i}\varphi_{A}}$,
the expressions above simplify to:
\begin{align}
Z^{+} & =\!\left[2M^{A_{1g}}\!+\!e^{\mathsf{i}\varphi_{A}}(m^{A_{1g}})^{\dagger}\!+\!e^{-\mathsf{i}\varphi_{A}}m^{A_{1g}}\right]\!\sum_{k}\!\!\Big(\!G_{k}^{A_{1g}}\!\!-\!|g_{k}^{A_{1g}}|\Big)^{\!2},\label{eq:Zp_1comp}\\
Z^{-} & =\!\left[e^{\mathsf{i}\varphi_{A}}(m^{A_{1g}})^{\dagger}\!-\!e^{-\mathsf{i}\varphi_{A}}m^{A_{1g}}\right]\!\sum_{k}\!\Big[(G_{k}^{A_{1g}})^{2}\!-\!|g_{k}^{A_{1g}}|^{2}\Big].\label{eq:Zm_1comp}
\end{align}
Substituting them back in Eq. (\ref{eq:KpKm_1comp}) gives:
\begin{align}
0 & =V_{A_{1g}}+v_{A_{1g}}e^{\mathsf{i}\varphi_{A}}+\bar{v}_{A_{1g}}e^{-\mathsf{i}\varphi_{A}},\label{eq:SC_sadd1_1comp}\\
0 & =v_{A_{1g}}e^{\mathsf{i}\varphi_{A}}-\bar{v}_{A_{1g}}e^{-\mathsf{i}\varphi_{A}},\label{eq:SC_sadd2_1comp}
\end{align}
which, together with Eq. (\ref{eq:SC_sadd_1comp}), defines the saddle-point
equations below $T_{c}$. 

Before proceeding, let us compute explicitly the superconducting susceptibility
$\chi$. First, we note that the unitary matrix $U$ in Eq. (\ref{eq:G0_m1_diag_1comp})
also diagonalizes the matrix $\tilde{\mathcal{D}}_{0}^{-1}$ in Eq.
(\ref{eq:D0_tilde_m1_one_comp}), 
\begin{align}
U^{\dagger}\tilde{\mathcal{D}}_{0}^{-1}U & =\mathrm{diag}\left(\tilde{V},0\right),\label{eq:D0tm1_diag_1comp}
\end{align}
where we used Eqs. (\ref{eq:SC_sadd1_1comp})-(\ref{eq:SC_sadd2_1comp})
and defined 
\begin{align}
\tilde{V} & =V_{A_{1g}}-v_{A_{1g}}e^{\mathsf{i}\varphi_{A}}-\bar{v}_{A_{1g}}e^{-\mathsf{i}\varphi_{A}}.\label{eq:V_bar_one_comp}
\end{align}
Using Eq. (\ref{eq:D0tm1_diag_1comp}), the application of the same
unitary matrix $U$ (\ref{eq:G0_m1_diag_1comp}) simplifies the superconducting
susceptibility given in Eq. (\ref{eq:susceptibility_1comp_app}),
\begin{align*}
\frac{V}{T}\chi & =U\,\mathrm{\,diag}\Big(\frac{\Lambda_{-}-\tilde{V}}{\Lambda_{-}^{2}},\frac{1}{\Lambda_{+}}\Big)\,\,U^{\dagger}.
\end{align*}
Upon rotating it to the basis of real and imaginary components of
the gap, i.e. upon computing $\chi_{b}$ in Eq. (\ref{eq:chi_b}),
we find:
\begin{align*}
\frac{V}{T}\chi_{b} & =\left(\!\!\frac{\Lambda_{-}\!\!-\!\tilde{V}}{2\Lambda_{-}^{2}}\!+\!\frac{1}{2\Lambda_{+}}\!\right)\!\sigma^{0}\!-\!\left(\!\!\frac{\Lambda_{-}\!\!-\!\tilde{V}}{2\Lambda_{-}^{2}}\!-\!\frac{1}{2\Lambda_{+}}\!\right)\!\left(\!\!\begin{array}{cc}
\cos\varphi_{A} & \!\!\sin\varphi_{A}\\
\sin\varphi_{A} & \!\!\text{-}\cos\varphi_{A}
\end{array}\!\!\right)\!.
\end{align*}
We now choose to condense the superconducting order parameter along
the real axis, i.e. we choose a real $\delta$. Then, according to
Goldstone's theorem, the susceptibility associated with the real component
(longitudinal susceptibility) has to be finite, whereas the susceptibility
associated with the imaginary component (transverse susceptibility)
has to be divergent at all temperatures below $T_{c}$. Of course,
the Goldstone mode will end up gapped via the Anderson-Higgs mechanism,
which is not included in our model since there is no coupling to the
electromagnetic fields in the action. Because $\Lambda_{-}=0$ inside
the superconducting state, a vanishing transverse susceptibility can
only be achieved by setting the charge-$4e$ order-parameter phase
$\varphi_{A}=0$. In this case, we find 
\begin{align}
\frac{V}{T}\chi_{b}\left(\varphi_{A}=0\right) & =\mathrm{diag}\Big(\frac{1}{\Lambda_{+}},\frac{\Lambda_{-}-\tilde{V}}{\Lambda_{-}^{2}}\Big).\label{eq:chi_b_fin_1comp}
\end{align}
Note that inclusion of the charge-$4e$ parameter $\phi^{A_{1g}}$
is essential to ensure that the longitudinal and transverse susceptibilities
are different. 

Moving on to the superconducting saddle-point equation (\ref{eq:SC_sadd_1comp})
for the case of a real $\delta$, we note that the unitary matrix
$U$ (\ref{eq:G0_m1_diag_1comp}) also diagonalizes the matrix $\mathcal{D}_{0}^{-1}$
(\ref{eq:D0_m1_one_comp}), yielding $U^{\dagger}\mathcal{D}_{0}^{-1}U=2u\,\delta^{2}\left(2\sigma^{0}-\sigma^{z}\right)$.
Since $\mathcal{K}_{0}^{-1}=\tilde{\mathcal{D}}_{0}^{-1}+\mathcal{G}_{0}^{-1}-\mathcal{D}_{0}^{-1}$,
and using the fact that $U$ diagonalizes all three matrices $\tilde{\mathcal{D}}_{0}^{-1}$,
$\mathcal{G}_{0}^{-1}$, and $\mathcal{D}_{0}^{-1}$, the saddle-point
equation (\ref{eq:SC_sadd_1comp}) simplifies to:
\begin{align}
\frac{df_{v}}{d\hat{\boldsymbol{\delta}}^{*}}=0 & =\Big[\mathcal{G}_{0}^{-1}+\tilde{\mathcal{D}}_{0}^{-1}-\frac{2}{3}\mathcal{D}_{0}^{-1}\Big]\hat{\boldsymbol{\delta}}\nonumber \\
0 & =\mathrm{diag}\Big(\Lambda_{-}+\tilde{V}-\frac{4}{3}u\,\delta^{2},\Lambda_{+}-4u\delta^{2}\Big)\,\,\hat{\boldsymbol{\delta}}_{d},\label{eq:SC_sadd_diag_1comp}
\end{align}
where $\hat{\boldsymbol{\delta}}_{d}=U^{\dagger}\hat{\boldsymbol{\delta}}=\sqrt{2}\delta\,(0,1)^{T}.$
Thus, a non-trivial solution exists only when the second diagonal
matrix element of Eq. (\ref{eq:SC_sadd_diag_1comp}) vanishes. Imposing
this condition, we find:

\begin{equation}
\delta^{2}=\frac{\Lambda_{+}}{4u}=\frac{R_{0}}{2u}\,,
\end{equation}
where, in the last step, we used the facts that $\Lambda_{+}\equiv R_{0}+\left|\phi^{A_{1g}}\right|$
and $R_{0}=\left|\phi^{A_{1g}}\right|$. The parameter $R_{0}$ is
given by the remaining two saddle-point equations (\ref{eq:SC_sadd1_1comp})
and (\ref{eq:SC_sadd2_1comp}). Since $\varphi_{A}=0$, $\phi^{A_{1g}}$
is real, and thus the saddle-point equation (\ref{eq:SC_sadd2_1comp})
is automatically satisfied. As for the first saddle-point equation
(\ref{eq:SC_sadd1_1comp}), we obtain:
\begin{align}
r_{0} & =-R_{0}-2u\left(2\Pi^{A_{1g}}-\big|\pi^{A_{1g}}\big|\right).\label{eq:r0_eq_one_comp}
\end{align}
where

\begin{align}
\Pi^{A_{1g}}\left(R_{0}\right) & =\frac{T}{V}\!\!\sum_{k}\!\!\frac{R_{0}+f_{\boldsymbol{k}}^{A_{1g}}}{\big(2R_{0}\!+\!f_{\boldsymbol{k}}^{A_{1g}}\big)\!f_{\boldsymbol{k}}^{A_{1g}}}\,,\\
\big|\pi^{A_{1g}}\left(R_{0}\right)\big| & =\frac{T}{V}\!\!\sum_{k}\!\!\frac{R_{0}}{\big(2R_{0}\!+\!f_{\boldsymbol{k}}^{A_{1g}}\big)\!f_{\boldsymbol{k}}^{A_{1g}}}\,.
\end{align}
Equation (\ref{eq:r0_eq_one_comp}) is an implicit equation for $R_{0}$
as a function of the temperature $r_{0}$. To proceed, we note that,
according to Eq. (\ref{eq:r0_eq_one_comp}), a second-order superconducting
transition is signaled by the vanishing of $R_{0}$. It is therefore
convenient to define the reference temperature $r_{0}^{c}$ by setting
$R_{0}=0$ in Eq. (\ref{eq:r0_eq_one_comp}), yielding $r_{0}^{c}=-4u\,\Pi^{A_{1g}}\left(R_{0}=0\right)$,
where $\Pi^{A_{1g}}\left(R_{0}=0\right)=\frac{T}{V}\sum_{k}\big(f_{\boldsymbol{k}}^{A_{1g}}\big)^{-1}$.
Note that the last integral is infrared-divergent in $d=2$, reflecting
Mermin-Wagner theorem. Introducing $\delta r_{0}=r_{0}-r_{0}^{c}$
, we rewrite Eq. (\ref{eq:r0_eq_one_comp}) as
\begin{align}
\delta r_{0} & =-R_{0}-2u\left[2\Pi^{A_{1g}}-2\Pi^{A_{1g}}\left(R_{0}=0\right)-\big|\pi^{A_{1g}}\big|\right],\label{eq:delta_r0_one_comp}
\end{align}
where it holds $\delta r_{0}\left(R_{0}=0\right)=0$, by construction.
In order for the superconducting transition to be second-order, it
must hold that $\frac{\partial R_{0}}{\partial\delta r_{0}}\big|_{\delta r_{0}=0}<0$,
which implies $\frac{\partial\delta r_{0}}{\partial R_{0}}\big|_{R_{0}=0}<0$.
A straightforward calculation gives
\begin{align}
\frac{\partial\delta r_{0}}{\partial R_{0}}\Big|_{R_{0}=0} & =-1+6u\frac{T}{V}\sum_{k}\frac{1}{\big(f_{\boldsymbol{k}}^{A_{1g}}\big)^{2}},\label{eq:der_ddr_1comp}
\end{align}
which contains an infrared divergent integral for $d\leq4$. This
means that for $d\leq4$, the derivative (\ref{eq:der_ddr_1comp})
is always positive and, thus, the superconducting transition is not
second-order, but first-order for this variational ansatz.

\section{Two-component superconductor in the variational approach\label{App:Two-component-superconductor}}

In this Appendix, we show how the formulas presented in App. \ref{App:One-component-superconductor-in}
are modified in the case of a two-component superconductor. We demonstrated
in App. \ref{App:One-component-superconductor-in} that the Gaussian
variational ansatz does not correctly capture the superconducting
transition, as it always leads to a first-order transition. Nonetheless,
for the sake of completeness, here we summarize the description of
the two-component superconductor below $T_{c}$. 

The derivation follows the same steps outlined in App. \ref{App:variational-details},
and in particular, App. \ref{App:One-component-superconductor-in}.
Extending the trial action (\ref{eq:trial_action_App}) as done in
Eq. (\ref{eq:SC_trial_action}), we include the superconducting variational
parameter $\hat{\boldsymbol{\delta}}=(\boldsymbol{\delta},\bar{\boldsymbol{\delta}})^{T}$
with $\boldsymbol{\delta}=(\delta_{1},\delta_{2})$. For later convenience,
we introduce the symmetry-classified bilinear combinations: 
\begin{align}
D^{n} & =\hat{\boldsymbol{\delta}}^{\dagger}\,M^{n}\,\hat{\boldsymbol{\delta}}, & d^{n} & =\hat{\boldsymbol{\delta}}^{\dagger}\,m^{n}\,\hat{\boldsymbol{\delta}},\label{eq:SC_bilinears}
\end{align}
for $n\in\mathbb{G}_{\mathbb{R}}^{0}$ and $n\in\mathbb{G}_{\mathbb{C}}$,
respectively. Because the superconducting order parameter is now non-zero,
the variational free-energy density (\ref{eq:f_phi-1}), acquires
an additional contribution $f_{\delta}$, i.e. $f_{v}\rightarrow f_{v}+f_{\delta}$.
In close analogy to the one-component superconducting case {[}cf.
Eq. (\ref{eq:f_phi_1_comp}){]}, this extra contribution evaluates
to 
\begin{align}
f_{\delta} & =\!\frac{1}{2}\hat{\boldsymbol{\delta}}^{\dagger}\big[\mathcal{K}_{0}^{-1}+\frac{1}{6}\mathcal{D}_{0}^{-1}\big]\hat{\boldsymbol{\delta}},\label{eq:f_delta}
\end{align}
where we introduced the $4\times4$ matrices 
\begin{align}
\mathcal{D}_{0}^{-1} & =\!\!\!\sum_{n\in\mathbb{G}_{\mathbb{R}}^{0}}\!\!\!2U_{n}D^{n}M^{n}+\sum_{n\in\mathbb{G}_{\mathbb{C}}}u_{n}\left[\bar{d}^{n}m^{n}+d^{n}(m^{n})^{\dagger}\right],\label{eq:D0_m1_appendix}\\
\mathcal{K}_{0}^{-1} & =\left(r_{0}+2U_{A_{1g}}\Pi^{A_{1g}}\right)2M^{A_{1g}}+\sum_{n\in\mathbb{G}_{\mathbb{R}}}4U_{n}\Pi^{n}M^{n}\nonumber \\
 & \quad+\sum_{n\in\mathbb{G}_{\mathbb{C}}}2u_{n}\left[\bar{\pi}^{n}m^{n}+\pi^{n}(m^{n})^{\dagger}\right].\label{eq:K0_minus1}
\end{align}
Note the relationship: 
\begin{align}
u\big(D^{A_{1g}}\big)^{\!2}\!+\!v\big(D^{A_{2g}}\big)^{\!2}\!+\!w\big(D^{B_{1g}}\big)^{\!2} & =\frac{1}{12}\hat{\boldsymbol{\delta}}^{\dagger}\mathcal{D}_{0}^{-1}\hat{\boldsymbol{\delta}}.\label{eq:D_relation}
\end{align}
Due to Eq. (\ref{eq:f_delta}), the partial derivatives in the free
energy minimization (\ref{eq:dfdx}) acquire an explicit $\boldsymbol{\delta}$
dependence:
\begin{align}
\!\!V_{A_{1g}} & =\frac{\partial f_{v}}{\partial\Pi^{A_{1g}}}=2\left(r_{0}\!-\!R_{0}\right)\!+\!2U_{A_{1g}}\!\left(2\Pi^{A_{1g}}\!+\!D^{A_{1g}}\right),\!\!\!\!\label{eq:VA1g_App}\\
\!\!V_{n} & =\frac{\partial f_{v}}{\partial\Pi^{n}}=-2\Phi^{n}+2U_{n}\left(2\Pi^{n}+D^{n}\right),\label{eq:Vn_App}\\
\!\!v_{n} & =\frac{\partial f_{v}}{\partial\pi^{n}}=-\bar{\phi}^{n}+u_{n}\left(2\bar{\pi}^{n}+\bar{d}^{n}\right),\label{eq:vn_App}
\end{align}
for $n\in\mathbb{G}_{\mathbb{R}}$ and $n\in\mathbb{G}_{\mathbb{C}}$,
respectively. Variation of the free energy with respect to $\boldsymbol{\delta}$
yields:
\begin{align}
\frac{df_{v}}{d\hat{\boldsymbol{\delta}}^{*}} & =\big(\mathcal{K}_{0}^{-1}+\frac{1}{3}\mathcal{D}_{0}^{-1}\big)\hat{\boldsymbol{\delta}}=0.\label{eq:dfv_ddelta}
\end{align}

For the superconducting susceptibility, after a somewhat tedious derivation
similar to that in App. \ref{App:One-component-superconductor-in},
we find
\begin{align}
\frac{V}{T}\chi & =\mathcal{G}_{0}-\mathcal{G}_{0}\tilde{\mathcal{D}}_{0}^{-1}\mathcal{G}_{0},\label{eq:SC_susceptibility}
\end{align}
where
\begin{align}
\tilde{\mathcal{D}}_{0}^{-1} & =-\mathcal{G}_{0}^{-1}+\mathcal{K}_{0}^{-1}+\mathcal{D}_{0}^{-1}\label{eq:D0tm1_relation}\\
 & =\sum_{n\in\mathbb{G}_{\mathbb{R}}^{0}}V_{n}M^{n}+\sum_{n\in\mathbb{G}_{\mathbb{C}}}\left(\bar{v}_{n}(m^{n})^{\dagger}+v_{n}m^{n}\right),\label{eq:D0tm1_relation_b}
\end{align}
and with $V_{n}$, $v_{n}$ given in Eqs. (\ref{eq:VA1g_App})-(\ref{eq:vn_App}).

As explained in App. \ref{App:One-component-superconductor-in}, to
properly describe the SC state, one has to ensure that the transverse
susceptibility remains divergent below $T_{c}$. Consequently, one
has to ensure that one eigenvalue of $\mathcal{G}_{0}^{-1}$ always
vanishes below $T_{c}$, which leads to an additional constraint $R_{0}=R_{0}\left(\Phi^{n},\phi^{n}\right)$.
Being subjected to this constraint, the free energy $f_{v}$ can be
minimized. For a given superconducting ground state, saddle-point
equations similar to Eqs. (\ref{eq:SC_sadd1_1comp})-(\ref{eq:SC_sadd2_1comp})
are obtained.

\bibliographystyle{apsrev4-1-titles}
\bibliography{Literature/bilinears,Literature/charge4e}

\end{document}